\documentclass{aa}
\usepackage{epsf}

\def\ave#1{\langle #1 \rangle}
\def\eck#1{\left\lbrack #1 \right\rbrack}

\def \la {\mathrel{\vcenter
     {\offinterlineskip \hbox{$<$}\hbox{$\sim$}}}}
\def \ga {\mathrel{\vcenter
     {\offinterlineskip \hbox{$>$}\hbox{$\sim$}}}}

\begin{document}

\thesaurus{06(13.07.1, 02.07.2, 02.05.1, 08.14.1, 08.02.1, 02.08.1)}

\title{Colliding neutron stars}
\subtitle{Gravitational waves, neutrino emission, and gamma-ray bursts}

\author{M.~Ruffert\inst{1}\thanks{e-mail: {\tt mruffert@ast.cam.ac.uk}}
\and H.-Th.~Janka\inst{2}\thanks{e-mail: {\tt thj@mpa-garching.mpg.de}}
}
\institute{Institute of Astronomy, Madingley Road, Cambridge~CB3~0HA,
United Kingdom
\and Max-Planck-Institut f\"ur Astrophysik, Postfach 1523,
85740~Garching, Germany}
\offprints{H.-Th.~Janka}



\maketitle

\begin{abstract}
Three-dimensional hydrodynamical simulations are presented for the
direct head-on or off-center collision of two neutron stars,
employing a basically Newtonian PPM code but including the 
emission of gravitational waves and their back-reaction on the
hydrodynamical flow. A physical nuclear equation of state
is used that allows us to follow the thermodynamical evolution 
of the stellar matter and to compute the emission of neutrinos.
Predicted gravitational wave signals, luminosities and 
waveforms, are presented. The models are evaluated for their
implications for gamma-ray burst scenarios. We find an
extremely luminous outburst of neutrinos with a peak
luminosity of more than $4\cdot 10^{54}\,{\rm erg/s}$ for several
milliseconds. This leads to an efficiency of about 1\% for
the annihilation of neutrinos with antineutrinos, corresponding
to an average energy deposition rate of more than $10^{52}\,{\rm erg/s}$
and a total energy of about $10^{50}\,{\rm erg}$ deposited in
electron-positron pairs around the collision site within 10$\,$ms.
Although these numbers seem very favorable for gamma-ray burst
scenarios, the pollution of the $e^\pm$ pair-plasma cloud with
nearly $10^{-1}\,M_{\odot}$ of dynamically ejected baryons is
5 orders of magnitude too large. Therefore the formation
of a relativistically expanding fireball that leads to a gamma-ray 
burst powered by neutrino emission from colliding neutron stars 
is definitely ruled out.

\end{abstract}

\section{Introduction}

The {\it merging} of two neutron stars that make up a binary system is
of interest both because it is a powerful source of gravitational
waves and because it might be the central engine of gamma-ray bursts.
These mergings have been studied intensely during the past few years, since
the occurrence rate is high enough (e.g.~Narayan et al.~1991, Phinney 1991,
Tutukov et al.~1992, Tutukov \& Yungelson 1993, Lipunov et al.~1995,
van den Heuvel \& Lorimer 1996, Lipunov et al.~1997, Prokhorov et al.~1997,
Bethe \& Brown 1998) that one
might be able to measure the observable consequences.
However, direct {\it collisions} of two neutron stars are much
rarer [the rate is $\ll 10^{-10}$ per year per galaxy 
(Centrella \& McMillan 1993) as compared to
the binary merger rate of  $10^{-6}$ to $10^{-5}$ per year per galaxy]
and thus have not received much attention.

Two previously published sets of hydrodynamical models of colliding
neutron stars were computed by Centrella \& McMillan (1993) and by 
Rasio \& Shapiro (1992).
Both used an SPH code to simulate the dynamics of two polytropic 
neutron stars of equal masses with adiabatic index $\gamma=2$ and 
started at an initial distance of about four neutron star radii on 
parabolic orbits. Using a relatively small number of particles (2048),
Centrella \& McMillan (1993) were able to produce a catalog of
gravitational wave luminosities and gravitational 
waveforms for different impact parameters. Rasio \& Shapiro (1992)
restricted themselves to fewer simulations with higher resolution 
(16000 particles) of neutron star coalescence as well as 
head-on collisions. They found that up to 5\% of the total mass 
can escape from the systems and that strong shocks occur.
Their results for the gravitational waveforms and luminosities 
reveal a smaller first maximum (caused by the initial free fall 
of the two stars),
followed by the main peak associated with the rapid deceleration
of the colliding matter during the propagation of the recoil shocks.

Neutron star collisions have repeatedly been suggested in the 
literature as possible sources of gamma-ray bursts
(e.g., Katz \& Canel 1996, Dokuchaev\& Eroshenko 1996; 
also Dokuchaev et al.~1998),
powered either by neutrino-antineutrino annihilation which produces
an $e^\pm$ pair-\-pho\-ton fireball, or by highly relativistic shocks which
are formed during the collision (or coalescence) and eject matter 
at relativistic velocities (Shaviv \& Dar 1995).
Katz \& Canel (1996), in particular, developed the idea that long
gamma-ray bursts might be explained by accretion induced collapse of
bare degenerate white dwarfs, while short bursts might originate from
neutron star {\it collisions}. They estimated the post-collision
temperature to be $kT \approx 100$~MeV, and expected a neutrino pulse
with a width of 2.5$\,$ms, and a total number of escaping neutrinos
of $10^{58}$ (refering to Dar et al.~1992). With a mean neutrino 
energy of about $6\,$MeV this corresponds to a total energy of
$10^{53}$~erg emitted in neutrinos. 
They sketched the picture that shortly after the collision,
dense lumps of hot matter expand and become optically thin to 
neutrinos which are released in a powerful outburst that is luminous
enough to produce the desired gamma-ray burst energy by 
neutrino-antineutrino annihilation if one assumes an efficiency
of 1\%. In order to obtain the observed gamma-ray burst rates they
had to invoke a total of $10^9$ hypothetical clusters within
$z\approx 1$ with $10^8$~neutron stars each. They stated that
``we cannot exclude the possibility that such clusters are
commonly found at the centers of galaxies''. 

Neutron star collisions might also be considered as interesting,
because they could be viewed at as more violent variants of the
merging process with a more extreme dynamical evolution. This
might support the formation of relativistic shocks and possibly
lead to a larger burst of gravitational waves and neutrinos.
More violence of the merging process can be expected from 
general relativistic effects which were so far neglected in the
large majority of simulations (see, however, 
Wilson et al.~1996 for an interesting step towards
fully general relativistic modeling). 
Oohara \& Nakamura (1997), using a grid-based TVD scheme, 
performed preliminary simulations to compare the coalescence 
with a Newtonian potential to the case with post-Newtonian
physics. Their neutron stars were modeled as two polytropes
with $\gamma=2$, $M=0.62\,M_\odot$, and $R=15\,$km.
The principal difference between the two cases was that the
post-Newtonian merging indeed turned out to be more violent: 
the impact is more central and shocks develop, because ``general
relativity effectively increases the gravitational force''.
Although the strong shock itself has little immediate effect on the 
gravitational waveform, the dynamics can be changed because of
the higher temperatures and densities. Wex (1995) and
Ogawaguchi \& Kojima (1996) showed that both spin-orbit
and spin-spin interactions appear formally to be {\it first}
order post-Newtonian corrections, just as gravitational potential
corrections are (gravitational waves are 2.5$\,$PN), but the inferred
magnitude of these corrections for known compact binary systems is
actually smaller.

\begin{table*}
\caption[]{
Characterizing parameters and some computed quantities for all models.
$N$ is the number of grid zones per dimension,
$L$ the size of the largest grid,
$l$ the size of the smallest zone,
$M_{\rho<11}$ the gas mass with density below $10^{11}$~g/cm$^3$
at the end of the simulation, and
$M_{\rm d}$ the mass with specific angular momentum larger than 
$j^*\equiv v_{\rm Kepler}(3R_{\rm s})3R_{\rm s}$ where 
$R_{\rm s}$ is the event horizon of a $\sim 3\,M_{\odot}$ black hole 
probably forming from the colliding neutron stars.
$M_{\rm g}$ denotes the mass swept off the grid,
$M_{\rm u}$ is the mass leaving the grid unbound, and
$T_{\rm ex}$ the maximum temperature (in energy units)
reached on the grid during the simulation of a model.
}

\begin{flushleft}
\tabcolsep=3mm
\begin{tabular}{lllcccccccc}
\hline\\[-3mm]
Model & impact & spin & $N$ & $L$ & $l$ &
   $M_{\rho<11}$ & 
   $M_{\rm d}$ & $M_{\rm g}$ & $M_{\rm u}$ & $T_{\rm ex}$ \\
 & direction & & & km & km &
   {\scriptsize$10^{-2}M_\odot$} & 
   {\scriptsize$10^{-2}M_\odot$} & {\scriptsize$10^{-2}M_\odot$} &
   {\scriptsize$10^{-2}M_\odot$} & {\scriptsize MeV}
\\[0.3ex] \hline\\[-3mm]
h & head-on       & none & ~32 & 328 & 1.28
    & 4.0 & 0.0 & 5.2 & 1.5 & 96.                     \\   
H & head-on       & none & ~64 & 328 & 0.64
    & 4.1 & 0.0 & 6.6 & 1.5 & 96.                     \\   
${\cal H}$& head-on  & none & 128 & 328 & 0.32
    & --- & --- & --- & --- & --- 
      \\[0.7ex]  
o & off-center    & none & ~32 & 328 & 1.28
    & 9.0 & 0.03 & 1.5 & 0.16 & 57.                     \\   
O & off-center    & none & ~64 & 328 & 0.64
    & 11.0& 0.03 & 1.8 & 0.19 & 58.            
      \\[0.7ex]  
\hline
\end{tabular}
\end{flushleft}

\label{tab:models1}
\end{table*}

\begin{table*}
\caption[]{
Gravitational-wave and neutrino emission properties for all models.
$\widehat{\cal L}$ is the maximum gravitational-wave luminosity,
${\cal E}$ the total energy emitted in gravitational waves,
$r\widehat{h}$ the maximum amplitude of the gravitational waves as
observed from a distance $r$,
$L_{\nu_e}$ the stationary value of the electron neutrino luminosity 
which is reached at about 6--10~ms after the start of the
simulations, $L_{\bar{\nu}_e}$ the
corresponding electron antineutrino luminosity, and
$L_{\nu_x}$ the luminosity of each individual species of $\nu_x$
($ = \nu_{\mu},\,\bar\nu_{\mu},\,\nu_{\tau},\,\bar\nu_{\tau}$).
$L_\nu$ represents the total neutrino luminosity at the end of the
simulation,
$\langle\epsilon_{\nu_e}\rangle$, $\langle\epsilon_{\bar{\nu}_e}\rangle$
and $\langle\epsilon_{\nu_x}\rangle$ are the mean energies of the
different neutrino and antineutrino flavors.
$\dot{E}_{\nu\bar{\nu}}$ denotes the integral rate of energy
deposition by neutrino-antineutrino annihilation, averaged over the
simulation time of 10$\,$ms. 
}

\begin{flushleft}
\tabcolsep=2mm
\begin{tabular}{llccccccccccc}
\hline\\[-3mm]
Model & impact & 
   $\widehat{\cal L}$ & ${\cal E}$ & $r\widehat{h}$ &
   $L_{\nu_e}$ & $L_{\bar{\nu}_e}$ & $L_{\nu_x}$ & $L_\nu$ & 
   $\langle\epsilon_{\nu_e}\rangle$ & 
   $\langle\epsilon_{\bar{\nu}_e}\rangle$ & 
   $\langle\epsilon_{\nu_x}\rangle$ & 
   $\dot{E}_{\nu\bar{\nu}}$  \\
 & direction & 
   {\scriptsize$10^{55}\frac{\rm erg}{\rm s}$} &
   {\scriptsize$10^{52}$ erg} & {\scriptsize$10^4$cm} &
   {\scriptsize$10^{53}\frac{\rm erg}{\rm s}$} &
   {\scriptsize$10^{53}\frac{\rm erg}{\rm s}$} &
   {\scriptsize$10^{53}\frac{\rm erg}{\rm s}$} &
   {\scriptsize$10^{53}\frac{\rm erg}{\rm s}$} &
   {\scriptsize MeV} & {\scriptsize MeV} & {\scriptsize MeV} & 
   {\scriptsize$10^{50}\frac{\rm erg}{\rm s}$}
\\[0.3ex] \hline\\[-3mm]
h & head-on        
    & 4.2  & 0.42 & 6.3 
    & 2.4 & 5.0  & 1.4  & 13. & 14. & 20. & 26. & --- \\   
H & head-on  
    & 3.61 & 0.39 & 6.2 
    & 2.0 & 4.0  & 1.0  & 10. & 13. & 20. & 25. & 100 \\   
${\cal H}$& head-on  
    & 3.55 & ---  & 6.2 
    & --- & ---  & ---  & --- & --- & --- & --- & --- 
      \\[0.7ex]  
o & off-center    
    & 6.6 & 4.1 & 12.3 
    & 1.3  & 2.8  & 0.8  & 7.3 & 11. & 18. & 25. & --- \\   
O & off-center    
    & 6.1 & 3.6 & 11.9 
    & 1.2  & 3.1  & 0.8  & 7.5 & 11. & 18. & 25. & --- 
      \\[0.7ex]  
\hline
\end{tabular}
\end{flushleft}

\label{tab:models2}
\end{table*}

Our project of simulating neutron star collisions with a Newtonian
PPM code was motivated by the aspects described in the last two
paragraphs. On the one hand, we intended to put potential
gamma-ray burst scenarios to a test, on the other hand we wanted
to study a situation that mimics the {\it merging} of two neutron 
stars with extreme violence and maximal parameters like pre-merging
kinetic energy and angular momentum in the system. Thus we hoped
not only to obtain an upper bound on the gravitational wave
emission to be expected from the merging of two neutron stars,
even with general relativistic effects included. We also wanted to
see whether the most extreme conditions during the collision of the
two stars lead to sufficiently large neutrino emission to explain
the gamma-ray burst energetics by the annihilation of neutrinos
and antineutrinos emitted during the dynamical event. The latter
seems impossible in case of the final stages of the coalescence 
of binary neutron stars because the prompt neutrino burst, although
very luminous, fails by several orders of magnitude to produce 
about $10^{51}\,{\rm erg}$ of gamma-rays within the short time
of only a few milliseconds that it takes the two neutron stars
to merge into one massive body that is most likely going to collapse
into a black hole on a dynamical timescale (Ruffert et al.~1996).


The paper is organized as follows. 
In Sect.~\ref{sec:numer} the basic aspects and new elements of our
computational method (in extension of Ruffert et al.~1996) and the 
chosen initial conditions for our simulations are given.
Our results are presented in the following sections, where we 
describe the hydrodynamical and thermodynamical 
evolution (Sect.~\ref{sec:hydevol}), the 
gravitational wave emission (Sect.~\ref{sec:gwem}), 
and the neutrino production (Sect.~\ref{sec:nuem}) in the
colliding stars together with the evaluation of our models for
the efficiency of neutrino-antineutrino 
annihilation (Sect.~\ref{sec:annihi}). 
Section~\ref{sec:end} contains a summary and conclusions.

\begin{figure*}
\tabcolsep=0.0mm
\begin{tabular}{cc}
  \epsfxsize=8.8cm \epsfclipon \epsffile{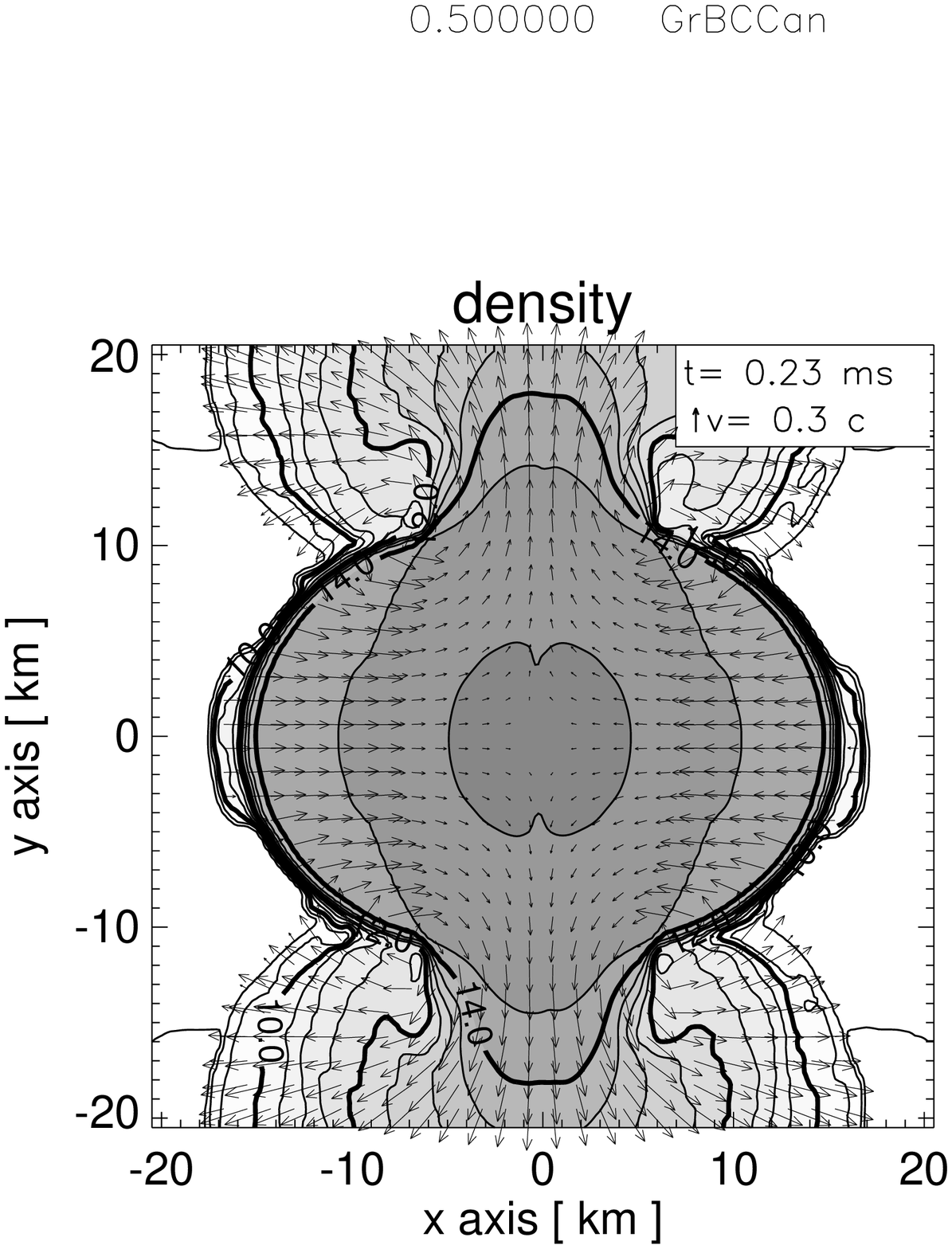}
  \put(-0.92,6.75){{\Large \bf \sf a}}   &
  \epsfxsize=8.8cm \epsfclipon \epsffile{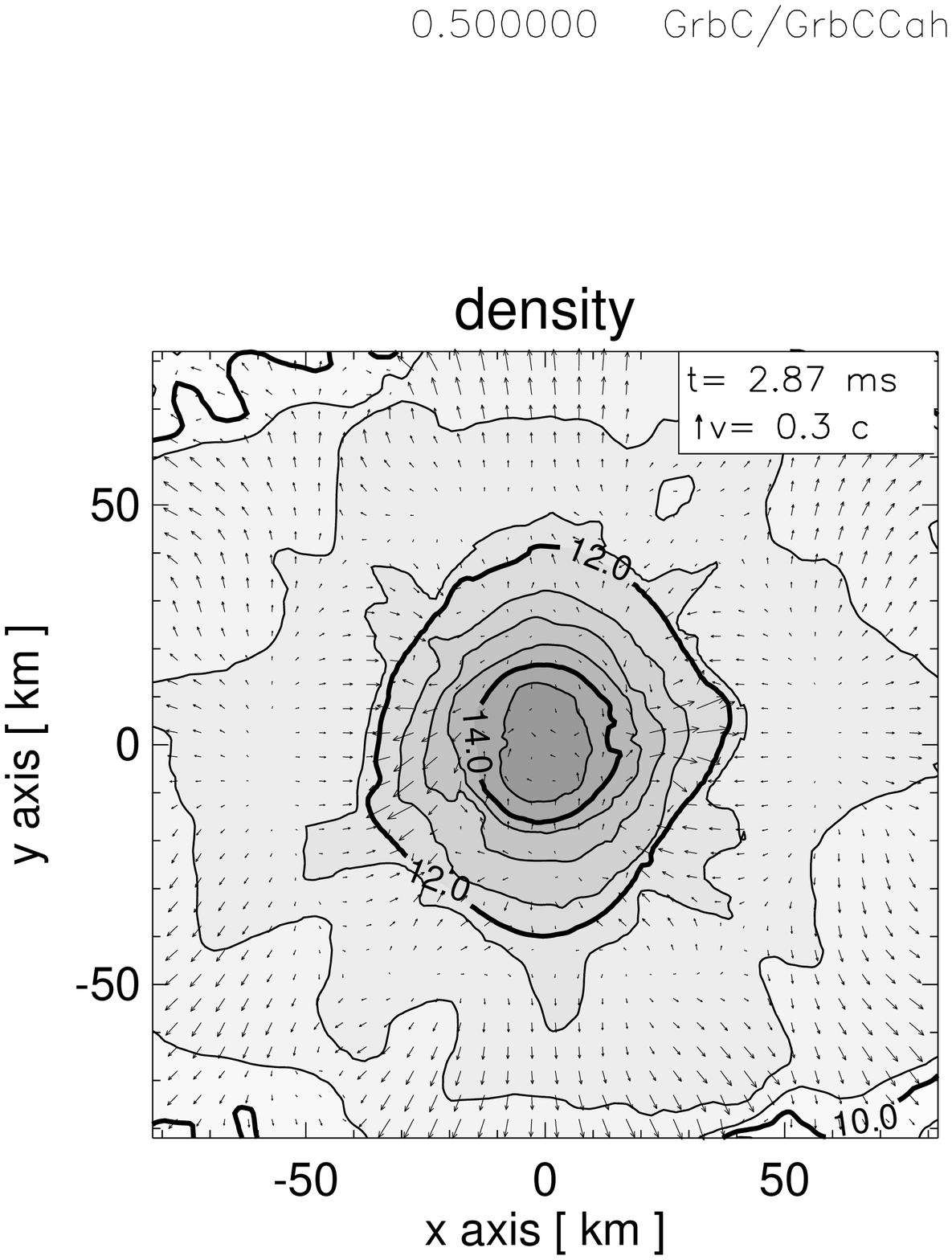}
  \put(-0.92,6.75){{\Large \bf \sf b}}  \\[-2ex]
  \epsfxsize=8.8cm \epsfclipon \epsffile{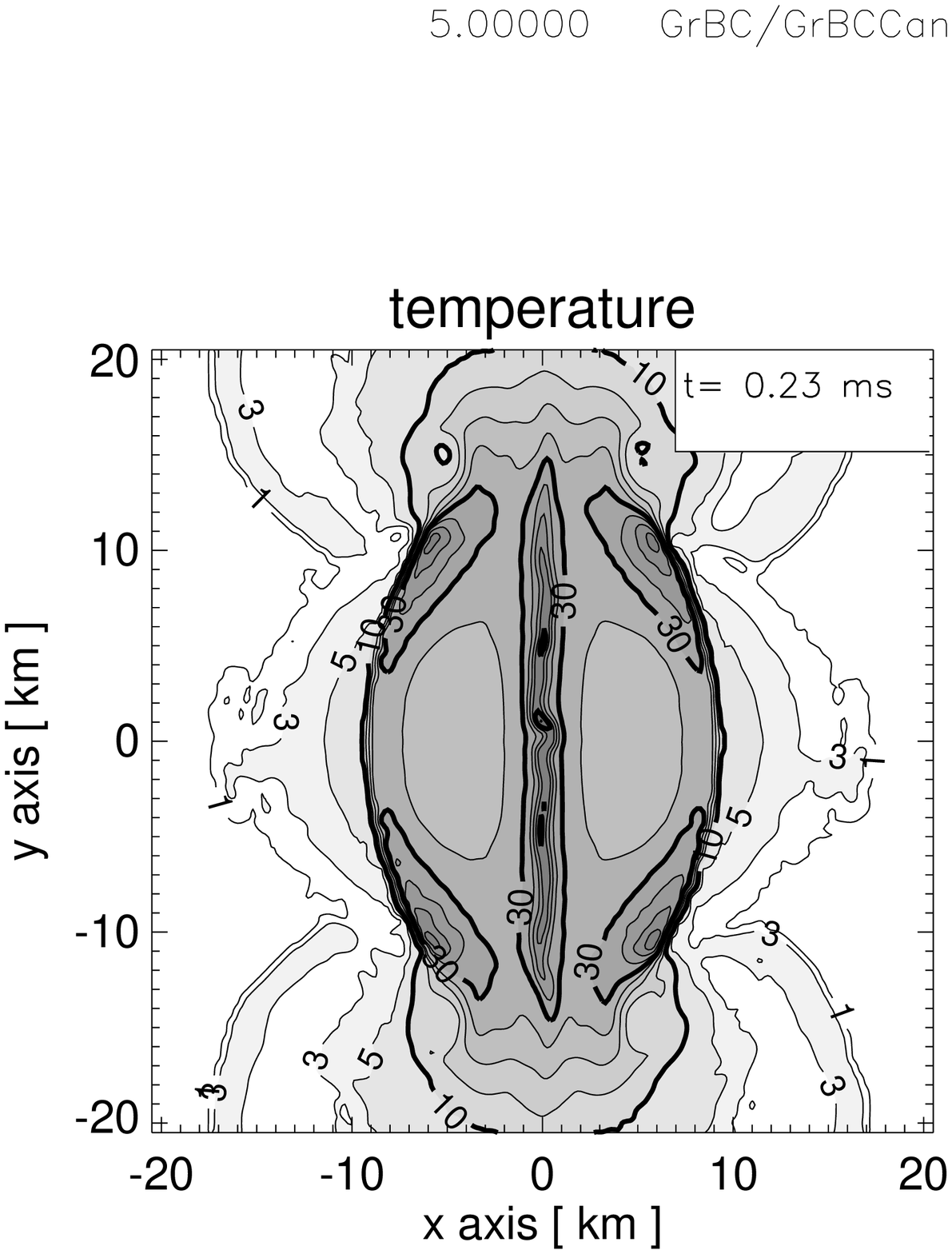}
  \put(-0.92,6.75){{\Large \bf \sf c}}  &
  \epsfxsize=8.8cm \epsfclipon \epsffile{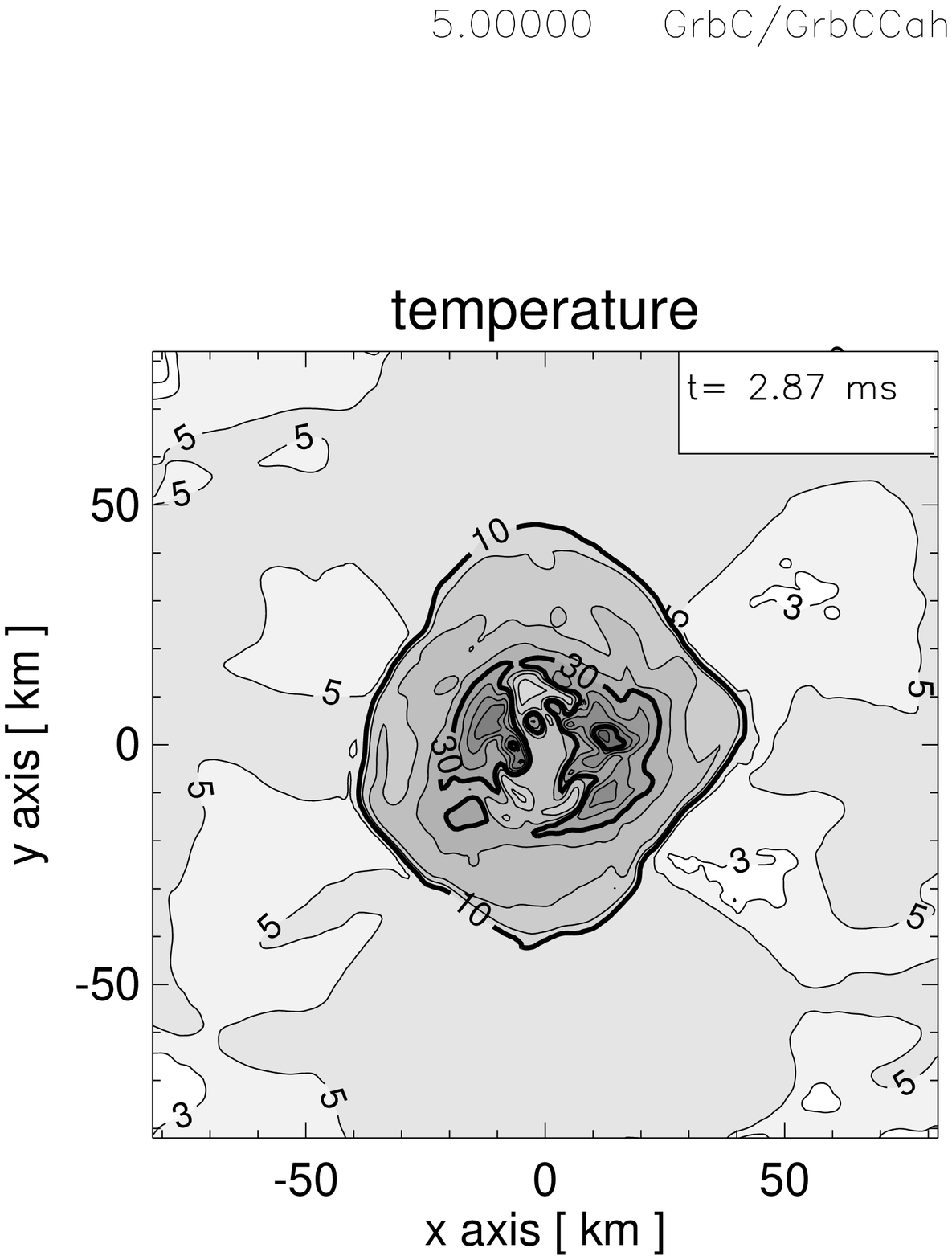}
  \put(-0.92,6.75){{\Large \bf \sf d}}  \\[-4ex]
\end{tabular}
\caption[]{Contour plots of Model~${\cal H}$ (left panels) and~H (right
panels) showing cuts in a plane containing the $x$-axis which is the 
symmetry axis of the initial model. The displayed physical quantities
are the density together with the velocity field (panels a and b) and the 
temperature (panels c and d). The density is measured in ${\rm g\,cm}^{-3}$,
the temperature in MeV. The density contours are spaced 
logarithmically with intervals of 0.5~dex, while the temperature contours
are linearly spaced, starting with 1$\,$MeV, 3$\,$MeV, 5$\,$MeV, and then 
continuing with an increment of 5~MeV. The bold contours are
labeled with their corresponding values ($10^{10}$, $10^{12}$, and
$10^{14}\,{\rm g\,cm}^{-3}$, and 10, 30, and 50$\,$MeV, respectively).
In the box in the upper right corner of each panel, the velocity vectors
and the time elapsed since the beginning of the simulation are given.
The mirror symmetry relative to the plane $x = 0$ and the rotational 
symmetry around the $x$-axis are broken during the 
evolution (panels b and d) because of an instability of the contact
layer of the two neutron stars against shear motions by which numerical
fluctuations (panel c) are amplified.
}
\label{fig:Hcont1}
\end{figure*}

\begin{figure*}
\tabcolsep=0.0mm
\begin{tabular}{cc}
  \epsfxsize=8.8cm \epsfclipon \epsffile{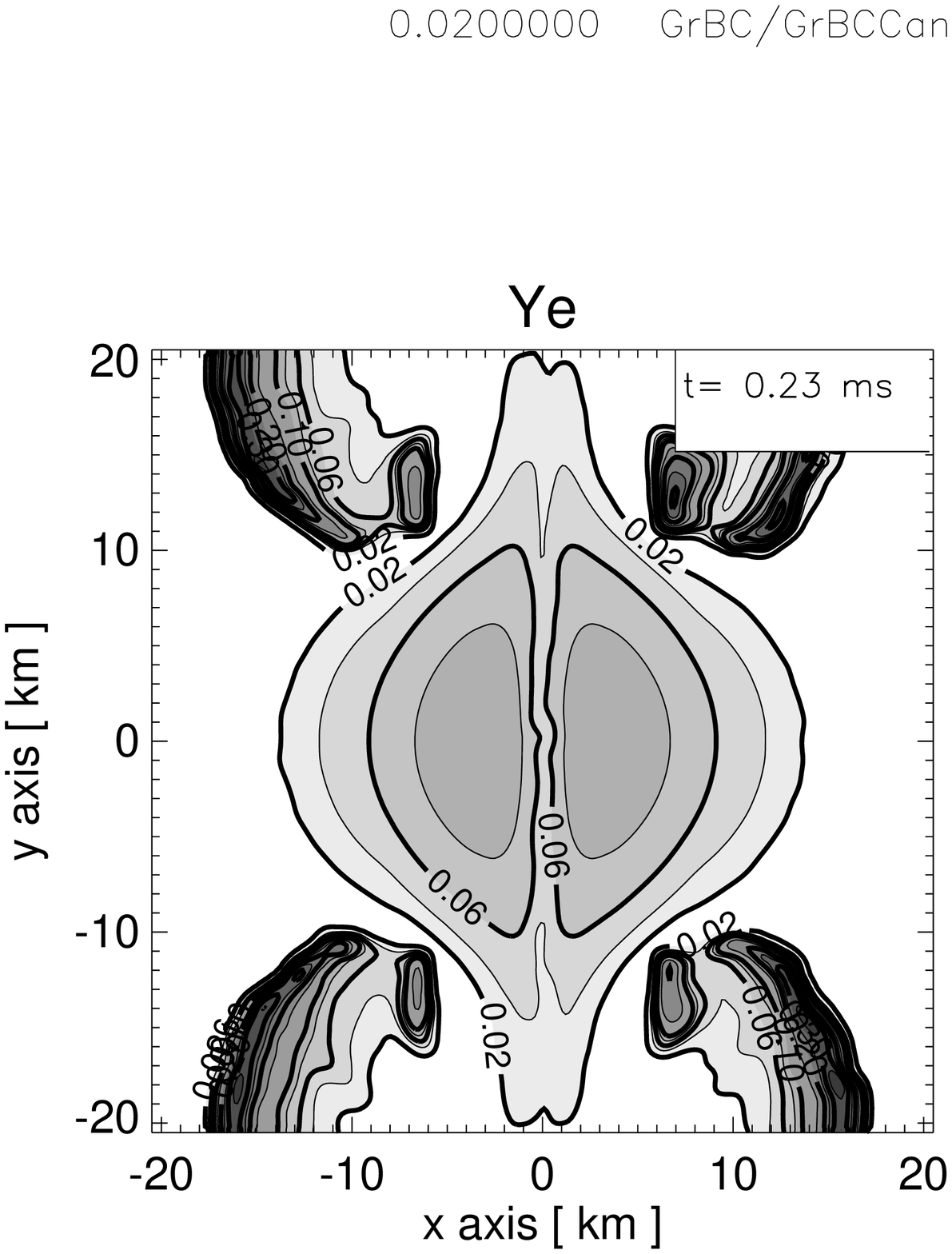}
  \put(-0.92,6.75){{\Large \bf \sf a}}   &
  \epsfxsize=8.8cm \epsfclipon \epsffile{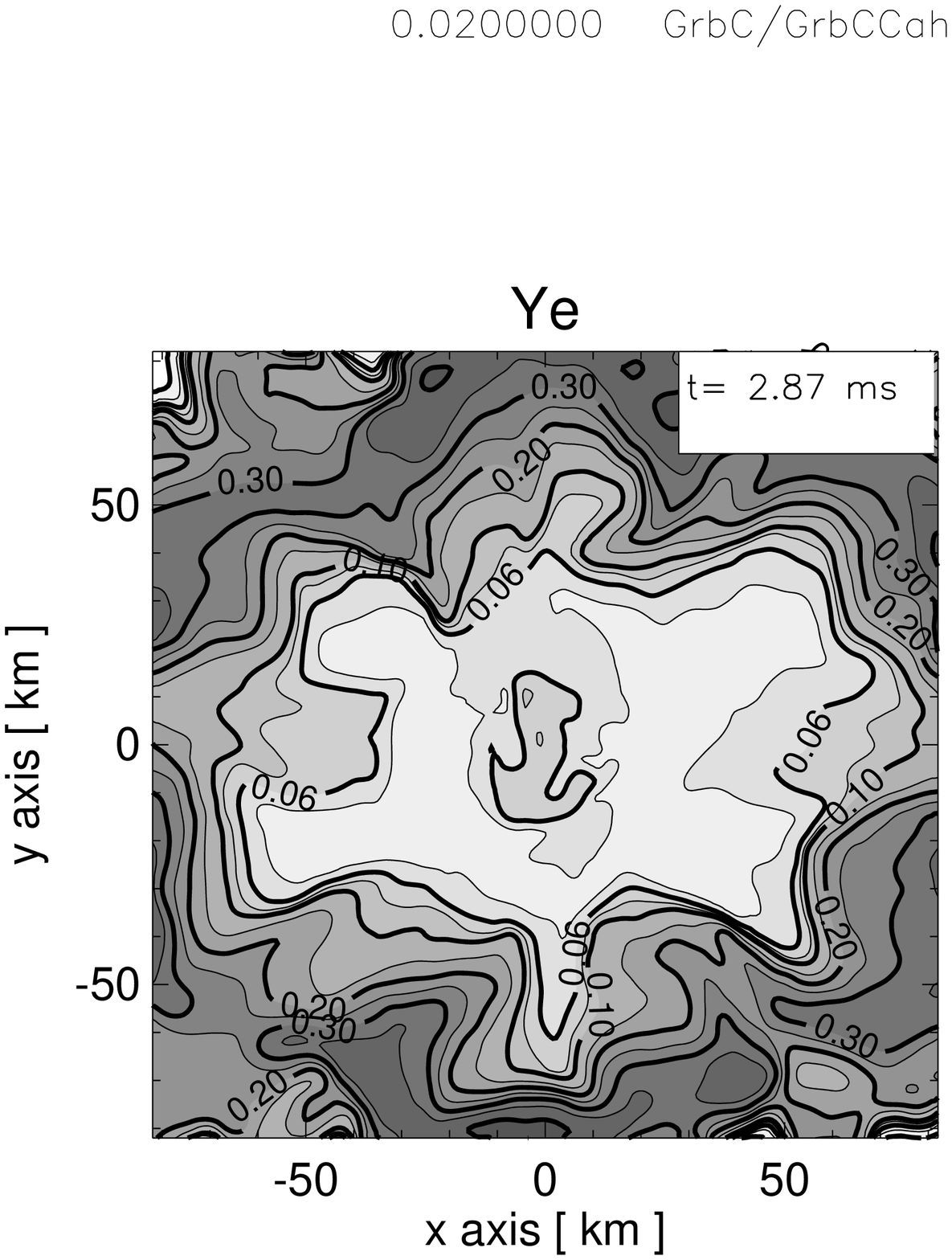}
  \put(-0.92,6.75){{\Large \bf \sf b}}  \\[-2ex]
  \epsfxsize=8.8cm \epsfclipon \epsffile{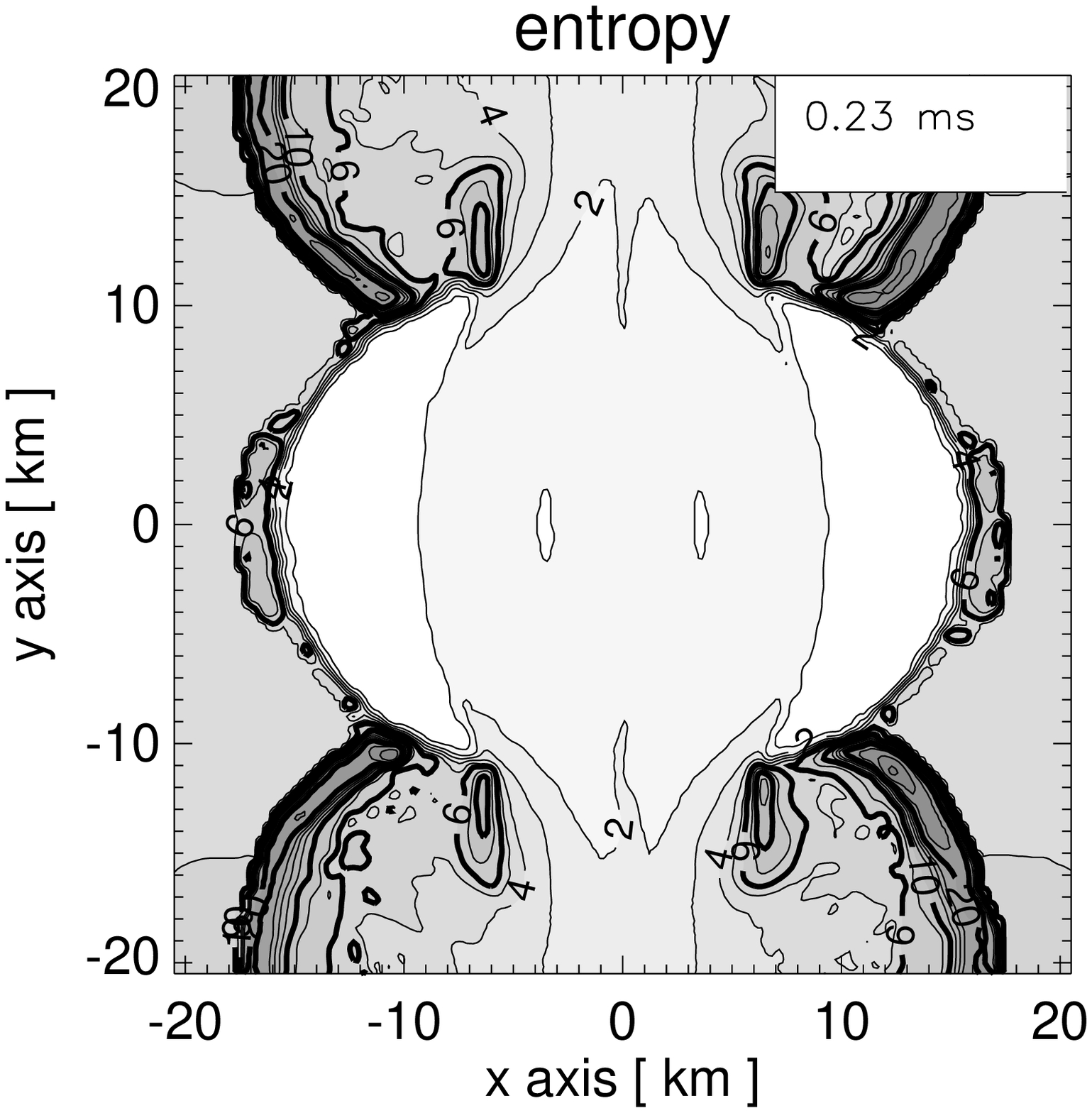}
  \put(-0.92,6.75){{\Large \bf \sf c}}  &
  \epsfxsize=8.8cm \epsfclipon \epsffile{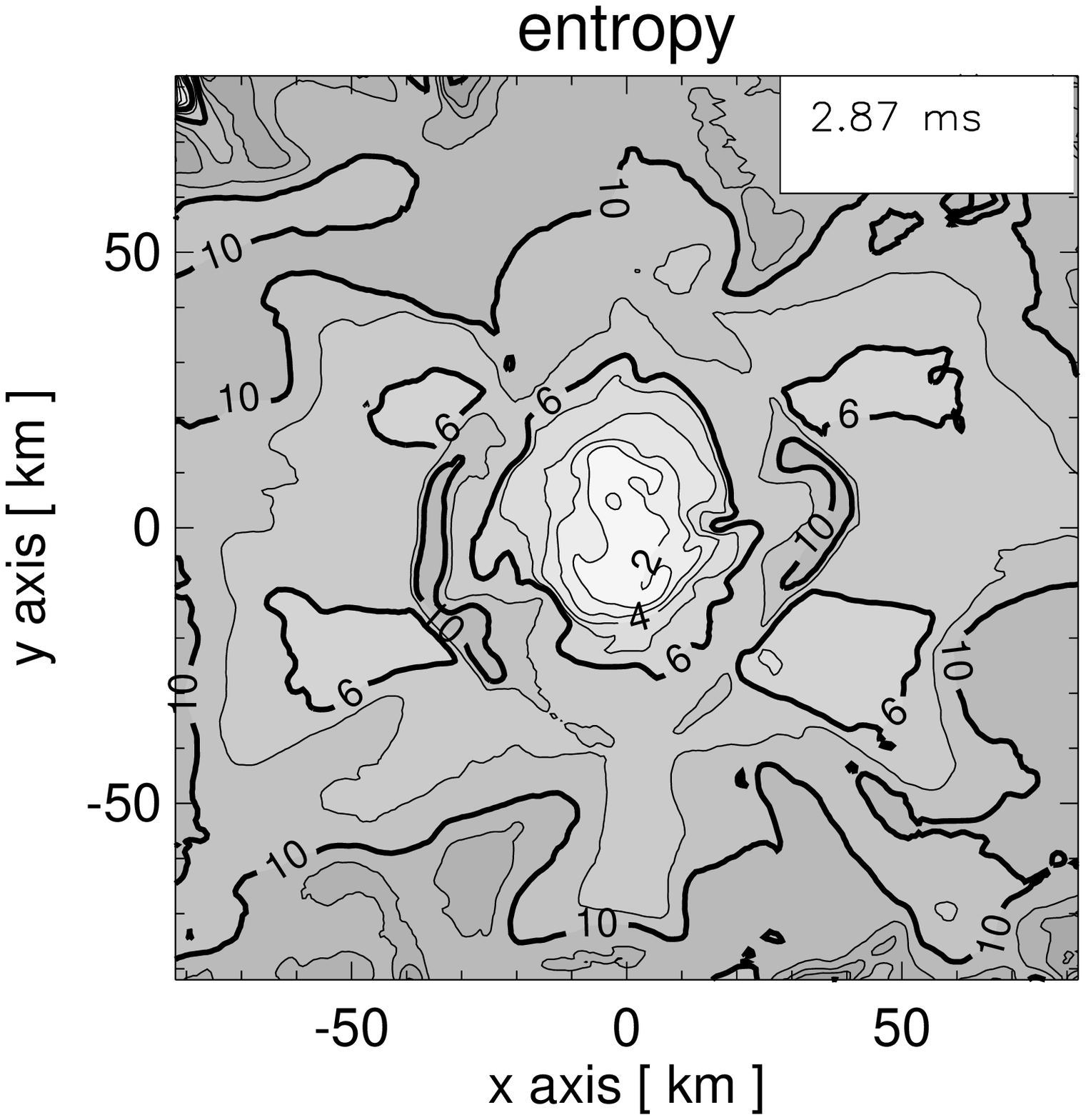}
  \put(-0.92,6.75){{\Large \bf \sf d}}  \\[-4ex]
\end{tabular}
\caption[]{Contour plots of Model~${\cal H}$ (left panels) and~H (right
panels) showing cuts in a plane containing the $x$-axis which is the 
symmetry axis of the initial model. The displayed physical quantities
are the electron fraction $Y_e$ (panels a and b) and the entropy 
(panels c and d), the latter measured in 
units of Boltzmann's constant $k$ per nucleon. The contours of the
electron fraction are linearly spaced with intervals of 0.02 below 0.1 
and with intervals of 0.05 above, the entropy contours are given in
steps of $1\,k$/nucleon from 1 to 6, in steps of $2\,k$/nucleon
between 6 and 16, and then for values of 20, 25, 30, and 40$\,k$/nucleon.
The bold contours are labeled with their corresponding values
(0.02, 0.06, 0.10, 0.20, 0.30, and 0.40 for $Y_e$, and 6, 10, and 20 for 
the entropy). Maximum values of $Y_e$ are above 0.4, 
of the entropy near $30\,k$/nucleon. 
In the box in the upper right corner of each panel, the
time elapsed since the beginning of the simulation is given. 
}
\label{fig:Hcont2}
\end{figure*}

\begin{figure*}
\tabcolsep=0.0mm
\begin{tabular}{cc}
  \epsfxsize=8.8cm \epsfclipon \epsffile{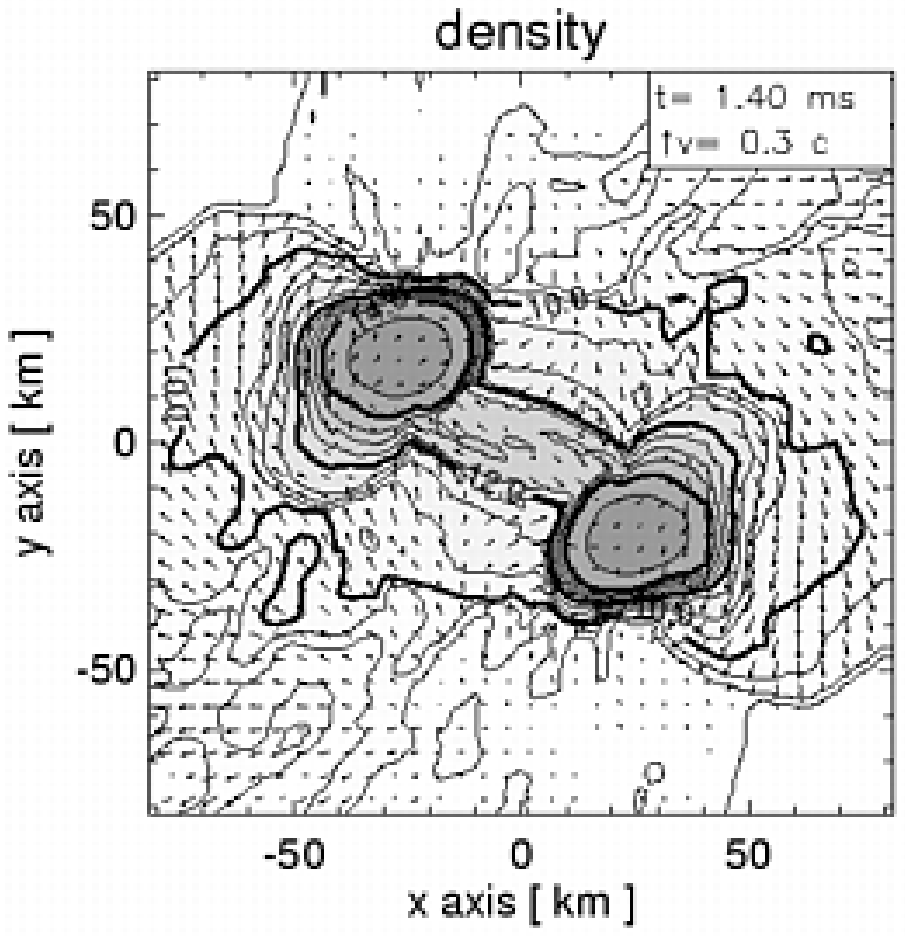} 
  \put(-0.92,6.75){{\Large \bf \sf a}}  &
  \epsfxsize=8.8cm \epsfclipon \epsffile{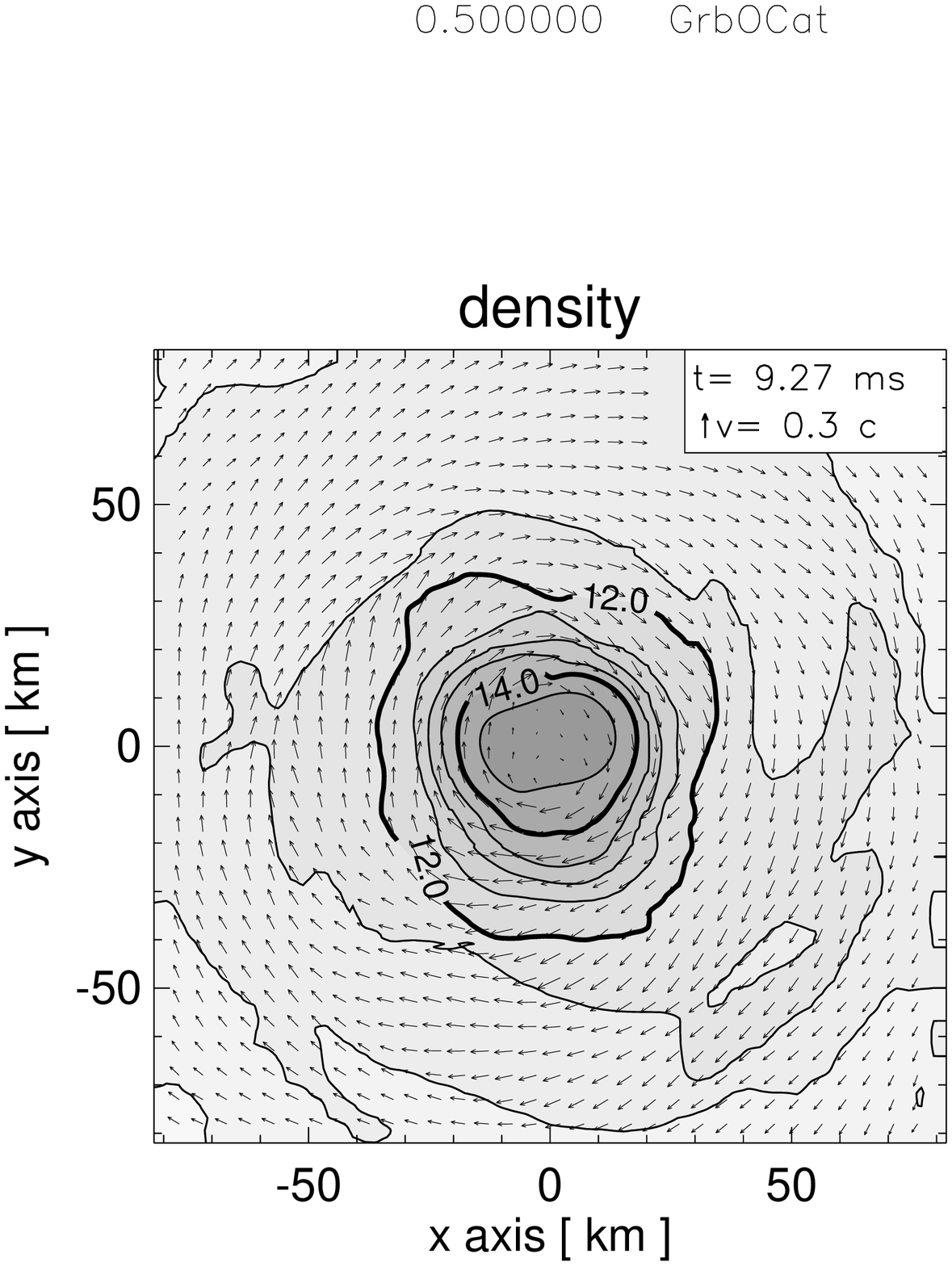} 
  \put(-0.92,6.75){{\Large \bf \sf b}} \\[-2ex]
  \epsfxsize=8.8cm \epsfclipon \epsffile{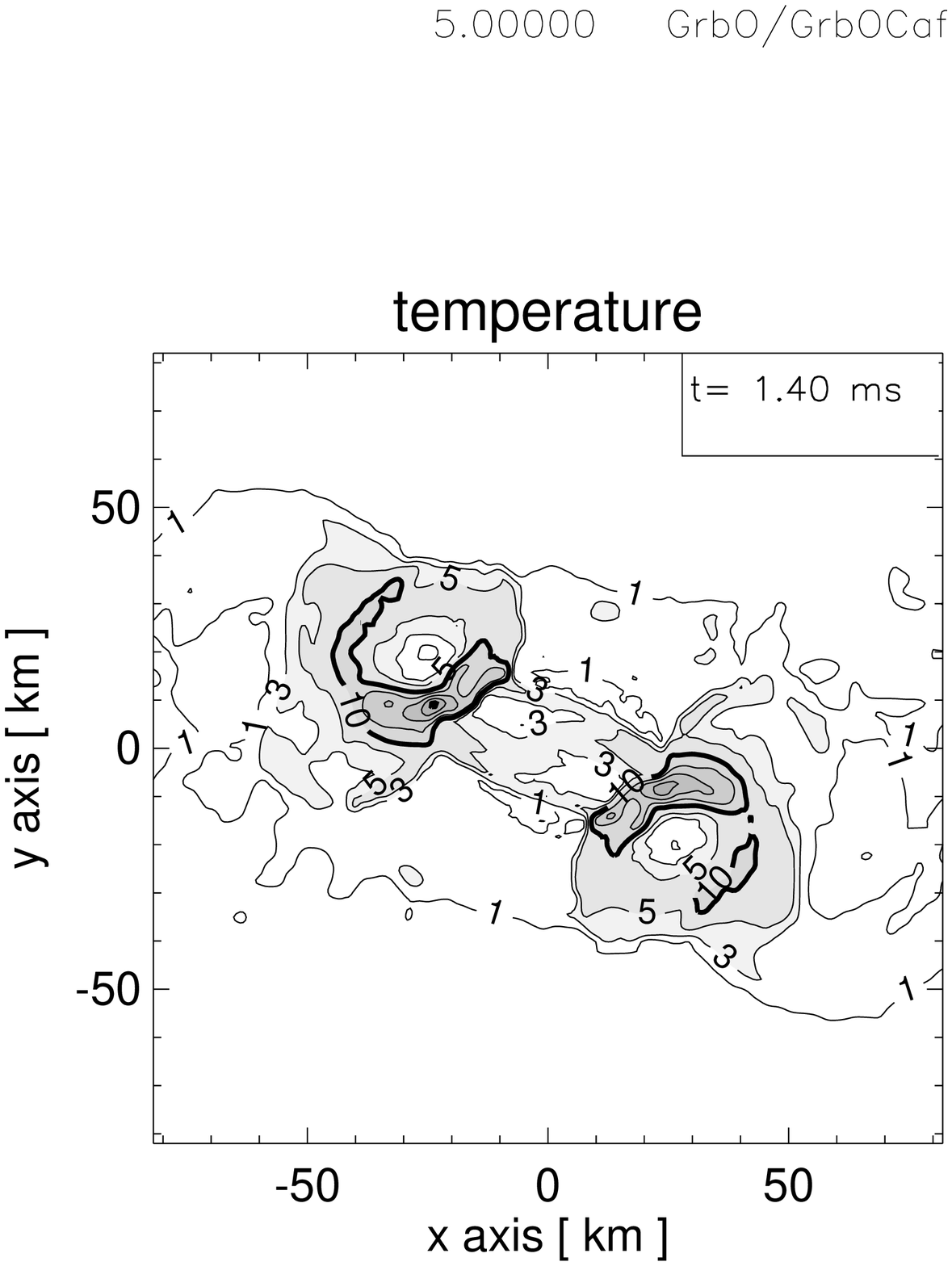}
  \put(-0.92,6.75){{\Large \bf \sf c}}  &
  \epsfxsize=8.8cm \epsfclipon \epsffile{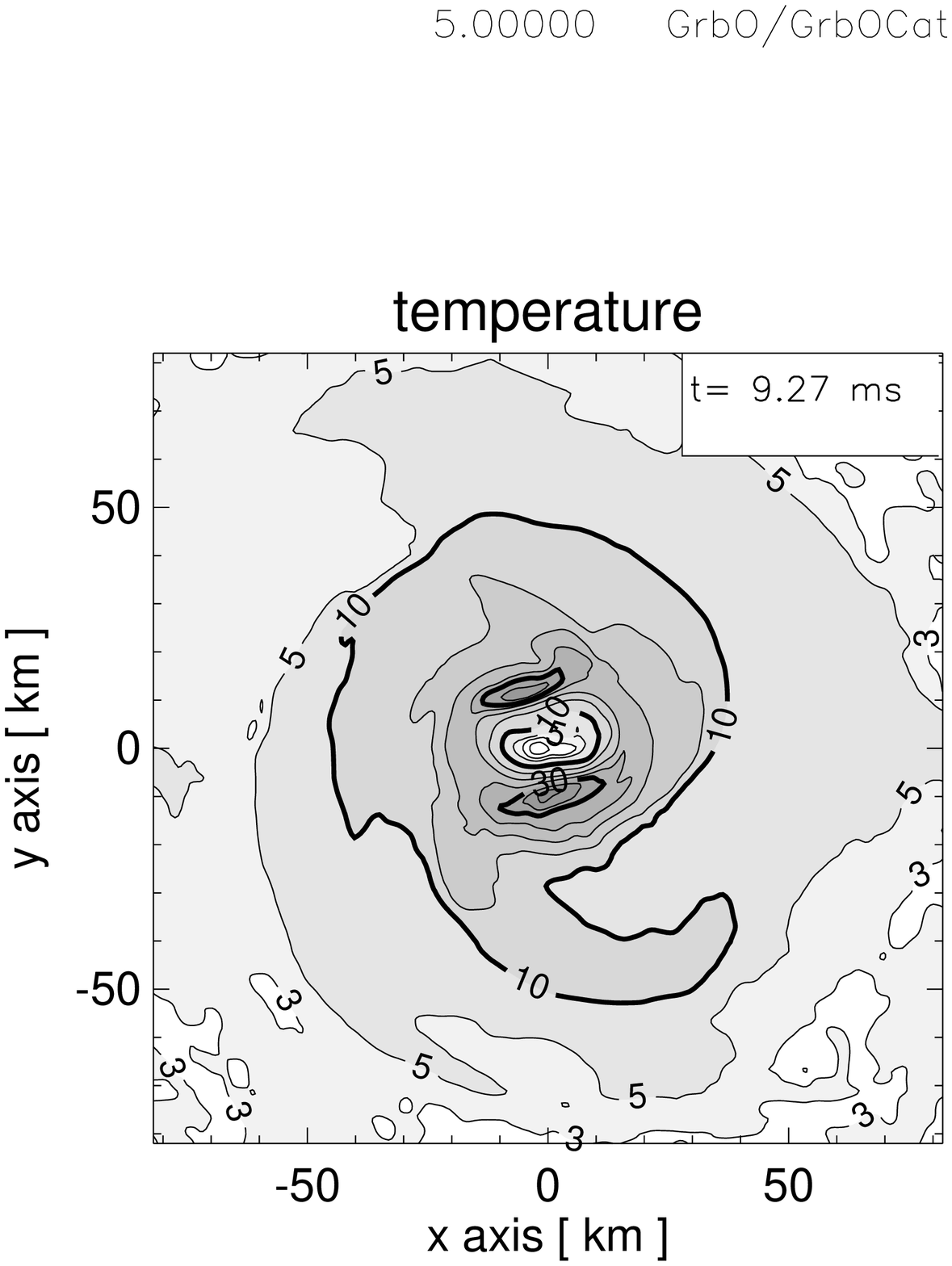} 
  \put(-0.92,6.75){{\Large \bf \sf d}} \\[-4ex]
\end{tabular}
\caption[]{Contour plots of Model~O showing cuts in the orbital plane for 
the density together with the velocity field (panels a and b) and for the 
temperature (panels c and d). The density is measured in ${\rm g\,cm}^{-3}$,
the temperature in MeV. The density contours are spaced
logarithmically with intervals of 0.5~dex, while the temperature contours
are linearly spaced, starting with 1$\,$MeV, 3$\,$MeV, 5$\,$MeV, and then
continuing with an increment of 5~MeV. The bold contours are
labeled with their corresponding values ($10^{10}$, $10^{12}$, and
$10^{14}\,{\rm g\,cm}^{-3}$, and 10, 30, and 50$\,$MeV, respectively).
In the box in the upper right corner of each panel, the velocity vectors
and the time elapsed since the beginning of the simulation are given.
}
\label{fig:Ocont1}
\end{figure*}

\begin{figure*}
\tabcolsep=0.0mm
\begin{tabular}{cc}
  \epsfxsize=8.8cm \epsfclipon \epsffile{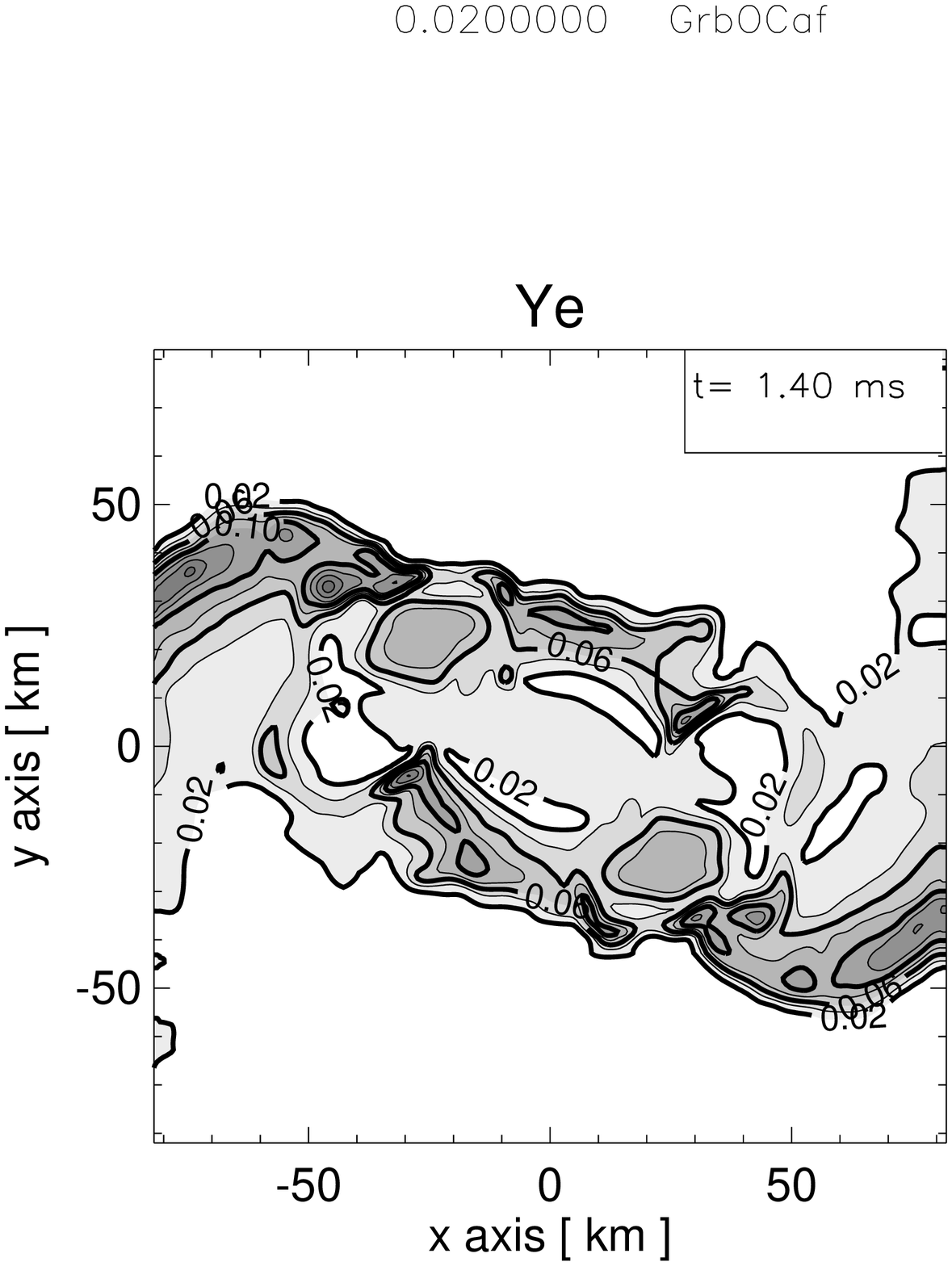} 
  \put(-0.92,6.75){{\Large \bf \sf a}}  &
  \epsfxsize=8.8cm \epsfclipon \epsffile{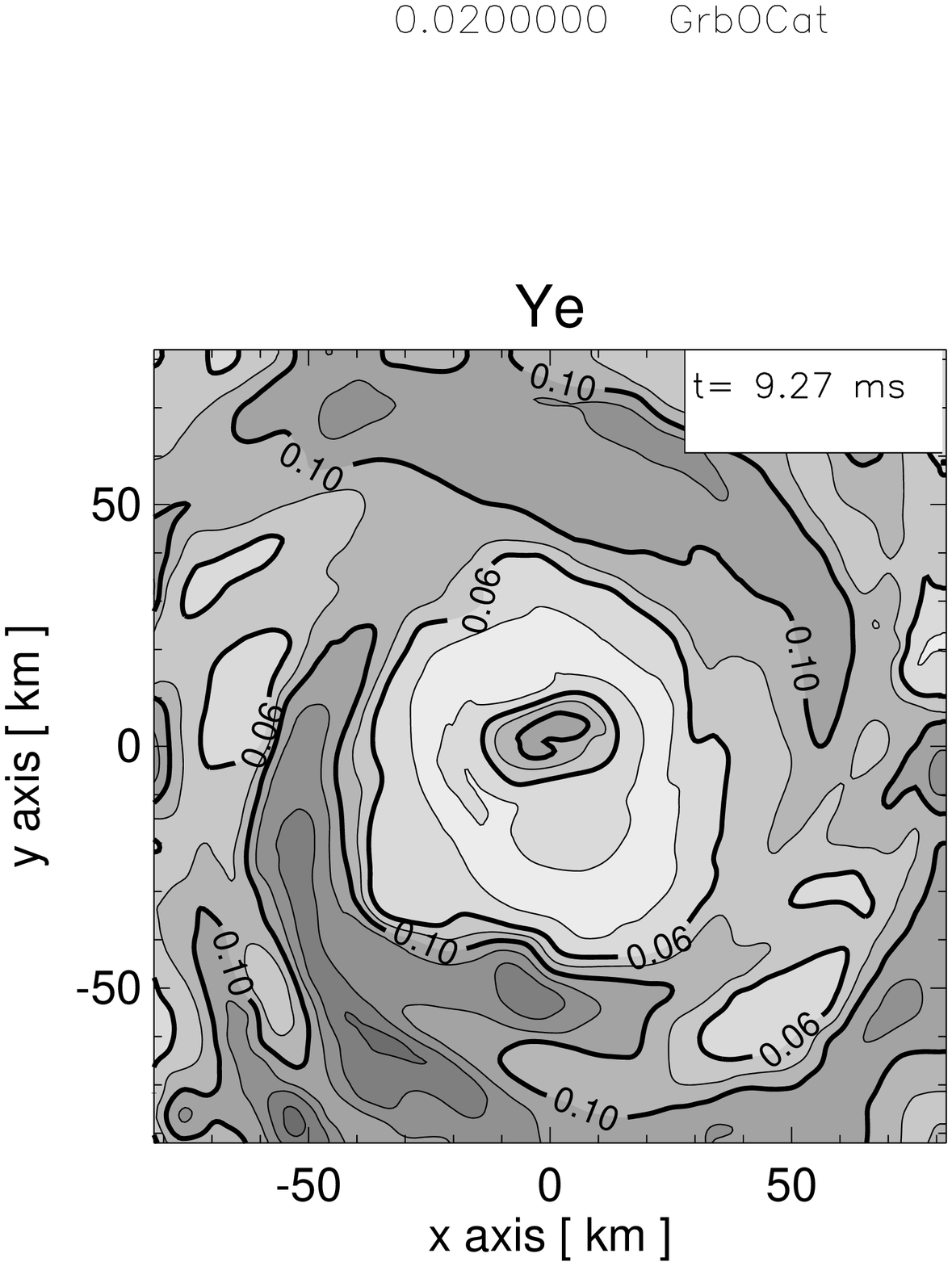} 
  \put(-0.92,6.75){{\Large \bf \sf b}} \\[-2ex]
  \epsfxsize=8.8cm \epsfclipon \epsffile{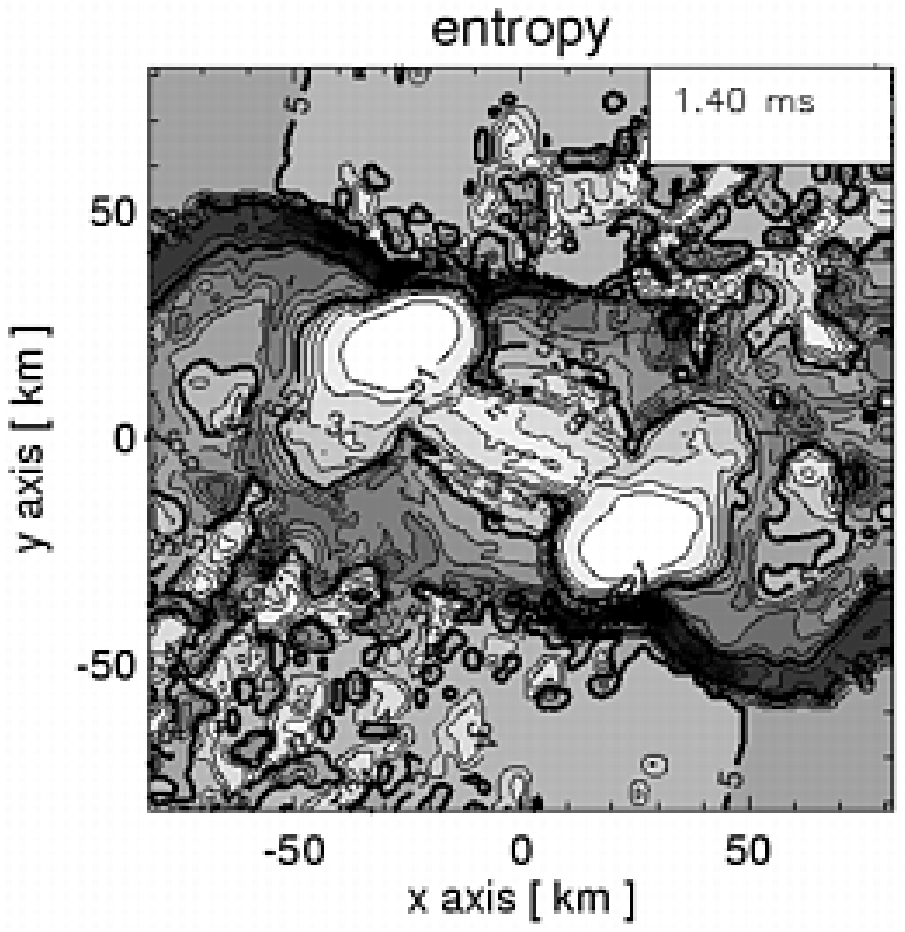}
  \put(-0.92,6.75){{\Large \bf \sf c}}  &
  \epsfxsize=8.8cm \epsfclipon \epsffile{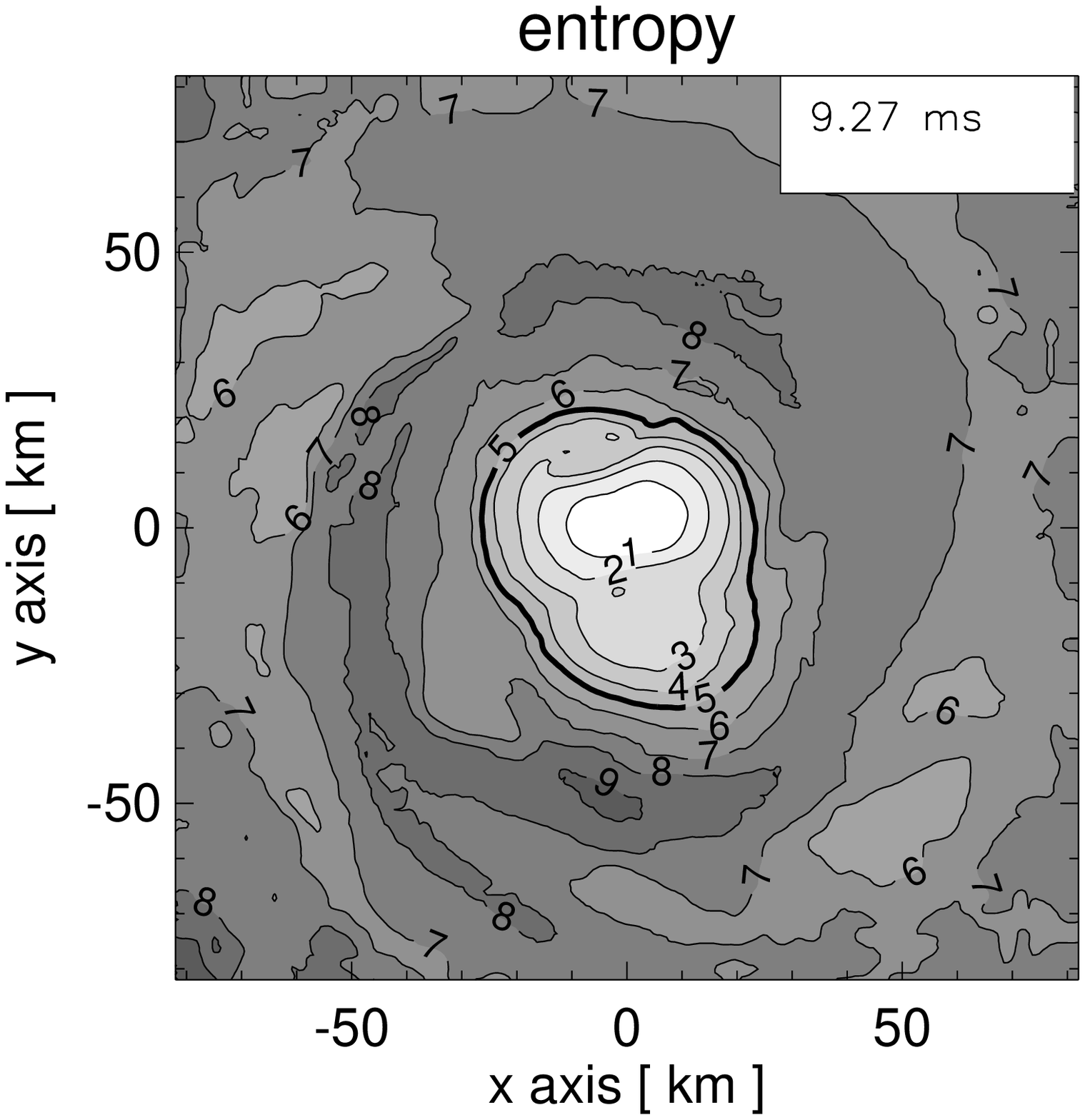} 
  \put(-0.92,6.75){{\Large \bf \sf d}} \\[-4ex]
\end{tabular}
\caption[]{Contour plots of Model~O showing cuts in the orbital plane 
for the electron fraction $Y_e$ (panels a and b) and the entropy 
(panels c and d), the latter quantity measured in units of
Boltzmann's constant $k$ per nucleon. The contours are
linearly spaced with intervals of 0.02 for the electron fraction and
$1\,k$/nucleon for the entropy. The bold contours are labeled with their
corresponding values (0.02, 0.06, 0.10, 0.20 for $Y_e$ and 5 and 10 for
the entropy). Maximum values of $Y_e$ are around 0.16, 
of the entropy about $10\,k$/nucleon.
In the box in the upper right corner of each panel, the
time elapsed since the beginning of the simulation is given.
}
\label{fig:Ocont2}
\end{figure*}

\section{Computational procedure and initial conditions\label{sec:numer}}

In this section we summarize the numerical methods and the
treatment of the input physics used for the presented
simulations. In addition, we specify the initial conditions by which
our different models are distinguished.
More detailed information about the employed numerical procedures
can be found in RJS (Ruffert et al.~1996) and RJTS (Ruffert et al.~1997).

\subsection{Hydrodynamical code}

The hydrodynamical simulations were done with a code based on the
Piecewise Parabolic Method (PPM) developed by Colella \& Woodward (1984).
The code is basically Newtonian, but contains the terms 
necessary to describe gravitational-wave emission and the corresponding
back-reaction on the hydrodynamical flow (Blanchet et al.~1990).
The modifications that follow from the gravitational
potential are implemented as source terms in the PPM algorithm.
The necessary spatial derivatives are evaluated as standard centered
differences on the grid. 

In order to describe the thermodynamics of the neutron star matter, 
we use the equation of state (EOS) of Lattimer \& Swesty (1991)
for a compressibility modulus of bulk nuclear matter of
$K = 180\,$MeV in tabular form. Energy loss and changes of the electron
abundance due to the emission of neutrinos and antineutrinos are 
taken into account by an elaborate ``neutrino leakage scheme''.
The energy source terms contain the production of all types of 
neutrino pairs by thermal processes and additionally of electron 
neutrinos and antineutrinos by lepton captures onto baryons. The latter
reactions act as sources or sinks of lepton number, too, and are
included as source terms in a continuity equation for the electron
lepton number. Matter is rendered optically thick to neutrinos due to
the main opacity producing reactions which are neutrino-nucleon
scattering and absorption of electron-type neutrinos onto nucleons.

More detailed information about the employed numerical procedures
can be found in RJS, in particular about the
implementation of the gravitational-wave radiation and back-reaction
terms and the treatment of the neutrino lepton number and energy loss
terms in the hydrodynamical code.

We have extended and improved the numerical treatment as compared 
to RJS in several aspects (a comparison of published results for
coalescing neutron stars obtained with the old code against results
from the improved one will be given in a separate, forthcoming
paper):

(a) Numerical resolution:
The presented simulations were done on multiply nested and refined grids.
With an only modest increase in CPU time, the nested grids allow one to
simulate a substantially larger computational volume while at the same
time they permit a higher local spatial resolution of the merged
object. The former is important to follow the fate of matter that is
flung out to distances far away from the collision site either to
become unbound or to eventually fall back. The latter is necessary to
adequately resolve the strong shock fronts and steep discontinuities 
of the plasma flow that develop during the collision.
The procedures used here are based on the algorithms that can be found
in Berger \& Colella (1989), Berger (1987) and Berger \& Oliger (1984).
Our version is described in detail in Section~4 of Ruffert (1992) so
only the most important features are to be summarized here. The 
individual grids are equidistant and Cartesian with each finer grid
having a factor of two smaller zone size and extent than the next coarser
one. Hereby the number of zones remains the same for all grids,
typically $32^3$ for low-resolution test calculations, $64^3$ for our
``standard'' simulations, and $128^3$ for models of special interest
where high resolution seems desirable. 

(b) Physics input:
The table for the Lattimer \& Swesty (1991) equation of state 
was extended to higher and lower temperatures and now spans
$0.01\,{\rm MeV}\le T\le 100\,{\rm MeV}$, and also the lower
density bound was moved down to now 
$\rho_{\rm min}=5\cdot 10^7\,{\rm g\,cm}^{-3}$ so that the density
range now covered by the table is 
$5\cdot 10^7\,{\rm g\,cm}^{-3}\le \rho\le 2.9\cdot 10^{15}\,{\rm g\,cm}^{-3}$.

\begin{figure}
\epsfxsize=8.8cm \epsfclipon \epsffile{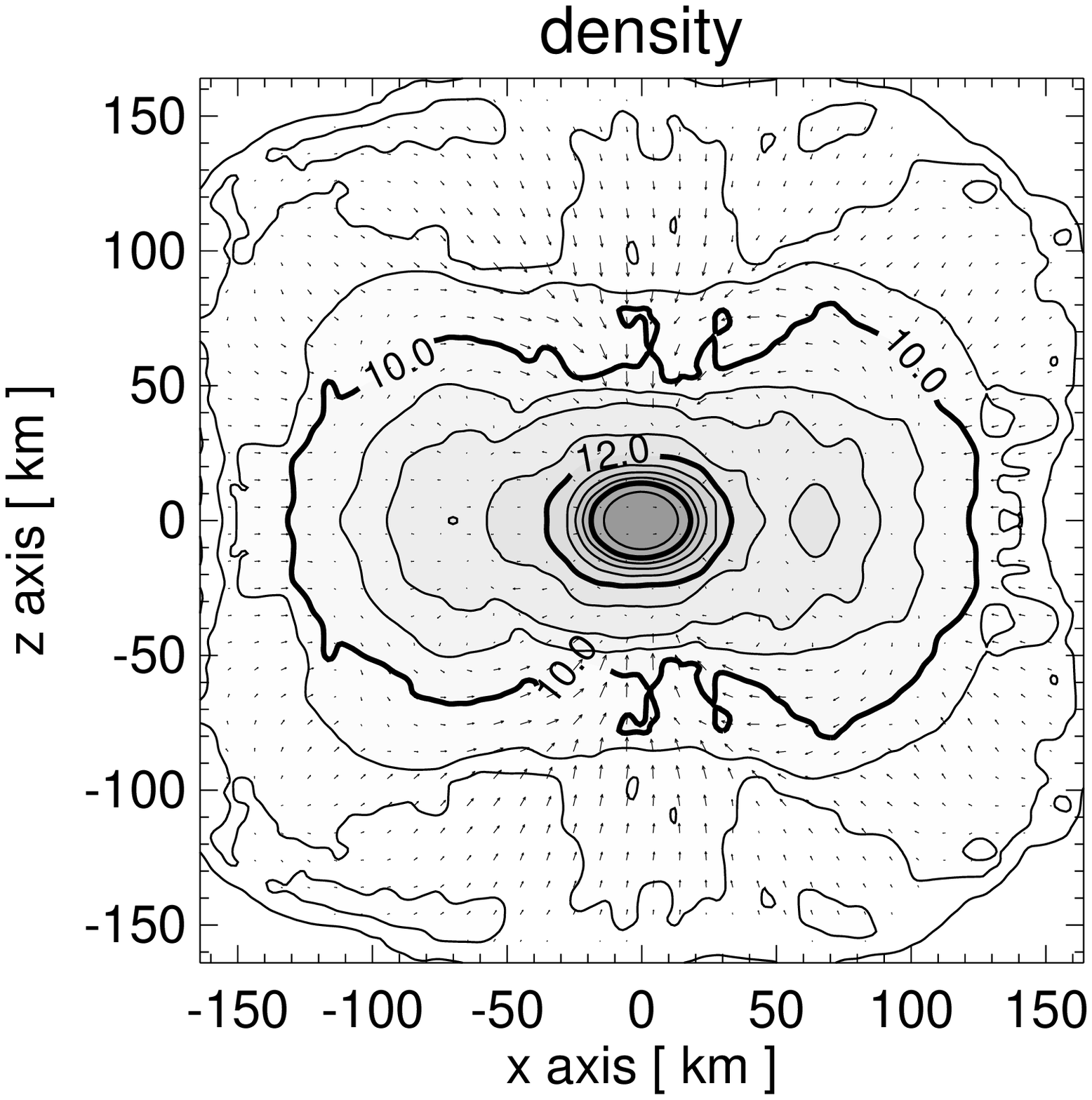}
\caption[]{Cut through Model~O showing the density contours together
with the velocity field in the plane $y = 0$ perpendicular to the
orbital plane near the end of the simulation ($t = 9.27\,{\rm ms}$). 
The contours are spaced logarithmically with intervals
of 0.5 dex, the bold contours correspond to density values of
$10^{10}$, $10^{12}$, and $10^{14}\,{\rm g\,cm}^{-3}$, respectively.
The velocity vectors are normalized as in Fig.~\ref{fig:Ocont1}b.
}
\label{fig:O2densv}
\end{figure}

\subsection{Evaluation of neutrino-antineutrino annihilation
\label{sec:evalann}}

In a post-processing step, performed after the hydrodynamical evolution
had been calculated, we evaluated our models for neutrino-antineutrino
($\nu\bar{\nu}$) annihilation in the surroundings of the collided stars
in order to construct a map showing the local energy deposition rates
per unit volume. Spatial integration finally yields the total rate of
energy deposition outside the neutrino emitting high-density regions.
The ``brute force'' approach,
however, which was applied by RJTS, is not feasible any longer because
it involved explicit summation of the contributions of neutrino and
antineutrino loss terms of all grid cells and at every location where
the local annihilation rate was to be determined. The computational
load for this procedure increases roughly with the third power of the
number of grid zones if the annihilation map has about the same spatial
resolution as the grid for the hydrodynamical simulation. With the larger
number of zones on several levels of the nested grid, such a strategy
is currently computationally impossible.

Therefore we resort to a different approach which involves five
distinct steps.
(1) First, the relevant physical quantities are mapped from only that
fractional volume of the nested grids where most of the neutrino emission
(and neutrino opacity)
comes from, to an equidistant Cartesian grid of fairly high resolution
(typically $140^3$). This mapping is done by tri-linear interpolation.
(2) Second, on this grid, the neutrinosphere for each flavor of neutrino
or antineutrino $\nu_i$ is determined. We define this two-dimensional 
hypersurface for $\nu_i$ by the set of those triples $(x,y,z)$ where, for 
each chosen pair of coordinates $x$ and $y$, the vertical optical depth
$\tau_{\nu_i}(z)$ satisfies the condition 
$\tau_{\nu_i}(z) = \Delta z\cdot\sum_{j = z}^\infty\kappa_i(j) = 1$
where $\Delta z$ is the cell size of the Cartesian grid and 
$\kappa_i(j)$ the local opacity of neutrino $\nu_i$ at position $j$.
(3) Thirdly, the local neutrino number and energy loss terms of neutrino 
$\nu_i$ are added up along the $z$-direction (for each fixed pair $(x,y)$),
and the total neutrino emissivity and the corresponding average energy
of the emitted neutrinos are projected to originate from the 
neutrinosphere of neutrino $\nu_i$ determined in step (2).
(4) Fourthly, and most importantly, the energy deposition rates by
$\nu\bar\nu$ annihilation are calculated by integrating (summing) only
over the two-dimensional neutrinospheres as neutrino and antineutrino 
sources instead of the three-dimensional stellar
volume as done in RJTS. Additional conditions
imposed on the construction of the annihilation map by RJTS are also
used here, i.e., the neutrino emission is assumed to occur isotropically
around the {\it outward pointing} local density gradients at the 
neutrinospheres, and the energy deposition by $\nu\bar\nu$ annihilation 
is evaluated only in those regions where the baryon density is below a 
certain threshold, typically $\rho < 10^{11}\,{\rm g\,cm}^{-3}$, and
on a cylindrical grid with coarser resolution than the Cartesian grid 
used to represent the neutrino sources, in order to limit the costs
of the numerically intense calculations.
(5) Finally, the local energy deposition rates per unit volume,
$\dot{E}_{\rm ann}(\varpi,\phi,z)$, are averaged over the $\phi$ direction
of the cylindrical grid:
\begin{equation}
\dot{E}_{\rm ann}(\varpi,z) =
  \frac{1}{2\pi}\int_0^{2\pi}\dot{E}_{\rm ann}(\varpi,\phi,z)
  \,{\rm d}\phi \,.
  \label{eq:doteann}
\end{equation}
From these average values two-dimensional maps like the one shown in
Sect.~\ref{sec:annihi} are plotted 
and integral numbers can be obtained by summation along radial or
vertical directions in the cylindrical grid.

The total energy deposition rate (``annihilation luminosity'') 
$L_{\rm ann}$ is obtained from the local values of the energy deposition
rate per unit volume, $\dot{E}_{\rm ann}(\varpi,z)$, 
by integration over the whole space
outside the neutrino emitting stellar source:
\begin{equation}
L_{\rm ann}=\int\dot{E}_{\rm ann}(\varpi,\phi,z) \, {\rm d}V
\label{eq:Lann}
\end{equation}
with ${\rm d}V=\varpi\,{\rm d}\varpi\,{\rm d}\phi\,{\rm d}z$.
Given the time dependent function $L_{\rm ann}(t)$ one can then
calculate the cumulative energy deposition by neutrino-antineutrino
annihilation according to
\begin{equation}
E_{\rm ann}=\int L_{\rm ann}(t) \, {\rm d}t \,.
\label{eq:Eexact}
\end{equation}
Because the computation is so expensive, however, $L_{\rm ann}(t)$
cannot be evaluated on a fine temporal grid, but only at a few 
discrete points in time, $t_i$. With the values $L_{\rm ann}(t_i)$
we therefore make the following approximation for $E_{\rm ann}$:
\begin{equation}
E_{\rm ann} \approx \frac{1}{N} \sum_{i=1}^{N} 
        \left( \frac{L_{\rm ann}(t_i)}{{\cal F}(t_i)} \right) \cdot
         \int {\cal F}(t) \, {\rm d}t
\label{eq:Eapprox}
\end{equation}
where ${\cal F}(t)$ is defined by
\begin{eqnarray}
{\cal F}(t)\,\equiv\, & &\frac{1}{R(t)}\Bigl\{ 
       L_{\nu_e}(t)\,L_{\bar{\nu}_e}(t)\,\eck{
      \ave{\epsilon_{\nu_e}(t)}  + 
      \ave{\epsilon_{\bar{\nu}_e}(t)} } + \nonumber \\
    &  & 2\cdot 
		  L_{\nu_{\mu}}(t)\,L_{\bar{\nu}_{\mu}}(t)\,\eck{
      \ave{\epsilon_{\nu_{\mu}}(t)}  +
      \ave{\epsilon_{\bar{\nu}_{\mu}}(t)} }
      \Bigr\} \,.
\label{eq:abbrf}
\end{eqnarray}
Here the term multiplied by the factor 2 accounts for the equal 
contributions from $\nu_{\mu}\bar\nu_{\mu}$ annihilation and
$\nu_{\tau}\bar\nu_{\tau}$ annihilation. The form of ${\cal F}(t)$
in Eq.~(\ref{eq:abbrf}) reflects the main dependences of the 
$\nu\bar\nu$ annihilation rate: The energy deposition rate increases 
proportional to the product of the neutrino and antineutrino 
luminosities times the sum of the mean energies of the annihilating
neutrinos and antineutrinos; in the denominator the characteristic
radial extent $R(t)$ of the neutrino source comes from the volume 
integral of Eq.~(\ref{eq:Lann}) when the latter is performed in spherical
coordinates (compare Eqs.~(3) and (10) in RJTS and references therein). 
The ratio appearing in Eq.~(\ref{eq:Eapprox}) in the sum in front of 
the time integral then contains geometrical effects which result
from the dependence of the
$\nu\bar\nu$ annihilation rate on the angular distributions of 
neutrinos and antineutrinos. From the hydrodynamical models,
the neutrino luminosities for the individual neutrino flavors,
$L_{\nu_e}(t)$, $L_{\bar{\nu}_e}(t)$ and the corresponding values for
muon and tau neutrinos and antineutrinos, are available as functions of 
time as well as the average energies of the emitted neutrinos,
$\langle\epsilon_{\nu_e}(t)\rangle$, 
$\langle\epsilon_{\bar{\nu}_e}(t)\rangle$, 
and $\langle\epsilon_{\bar{\nu}_x}(t)\rangle$ for $\nu_{\mu}$,
$\bar\nu_{\mu}$, $\nu_{\tau}$, and $\bar\nu_{\tau}$.
We found that 
the typical radial size $R(t)$ of the neutrino emitting object 
during the phase where by far most of the $\nu\bar\nu$ annihilation
happens is not very
strongly time-dependent because the wobblings and oscillations
change the shape and size of the collision remnant only on smaller
scales but not globally. Therefore instead of
${\cal F}(t)$ from Eq.~(\ref{eq:abbrf}) we use in 
Eq.~(\ref{eq:Eapprox}) the simpler expression
\begin{eqnarray}
{\cal F}^{\ast}(t)\,\equiv\,
     & & L_{\nu_e}(t)\,L_{\bar{\nu}_e}(t)\,\eck{
     \ave{\epsilon_{\nu_e}(t)}  +
     \ave{\epsilon_{\bar{\nu}_e}(t)} } + \nonumber \\
     & & 2\cdot
     L_{\nu_{\mu}}(t)\,L_{\bar{\nu}_{\mu}}(t)\,\eck{
     \ave{\epsilon_{\nu_{\mu}}(t)}  +
     \ave{\epsilon_{\bar{\nu}_{\mu}}(t)} } \,.
\label{eq:abbrfbar}
\end{eqnarray}
Computing $E_{\rm ann}$ from Eq.~(\ref{eq:Eapprox}) instead of
Eq.~(\ref{eq:Eexact}) involves the approximation that the term
abbreviated by ${\cal F}(t)$ or ${\cal F}^{\ast}(t)$
contains the main time dependence of the integral in 
Eq.~(\ref{eq:Eexact}). Ideally, the ratio 
$L_{\rm ann}(t)/{\cal F}^{\ast}(t)$ would have to be constant.
Since this is not the case, we decided to employ an average value
for a small number $N$ of time points where the spatial integral of 
Eq.~(\ref{eq:Lann}) was evaluated. It turned out that the variation 
of $L_{\rm ann}(t)/{\cal F}^{\ast}(t)$ during the most interesting
phase of the evolution is less than a factor 2.

\subsection{Initial conditions\label{sec:initial}}

We started our simulations with two identical Newtonian
neutron stars, each having a baryonic mass of about 1.63~$M_\odot$
and a radius of 15~km, which were placed at a center-to-center 
distance of 42~km. 
The distributions of density $\rho$ and electron fraction 
$Y_e\equiv n_e/n_b$ (with $n_e$ being the number density of electrons 
minus that of positrons, and $n_b$ the baryon number density) were
taken from a one-dimensional model of a cold, deleptonized neutron
star in hydrostatic equilibrium and were the same as in RJS. 
For numerical reasons the 
surroundings of the neutron stars cannot be treated as completely
evacuated. The density of the ambient medium was set to less than
$10^8$~g/cm$^3$, more than six orders of magnitude smaller than 
the central densities of the stars. The total mass on the whole
grid, associated with this finite density is less than 
$10^{-3}\,M_{\odot}$.

The neutron stars were given the free-fall velocity at their respective
initial positions $(x,y,z) = (-21\,{\rm km},0,0)$ and
$(x,y,z) = (21\,{\rm km},0,0)$. The angle between the velocity vectors
and the vector connecting the stellar centers was varied to produce a
head-on collision for Models~h, H,
and~${\cal H}$, and an off-center collision for Models~o and~O.
The impact parameter of the latter was chosen to be one neutron star
radius. This impact parameter is the minimum distance that two
point masses reach along their orbits. A compilation of all models
together with their characterizing grid parameters, initial 
parameter settings, and some results of the numerical simulations
is given in Tables~\ref{tab:models1} and \ref{tab:models2}.

In degenerate matter variations of the temperature lead only to minor
changes of the internal energy and pressure (both are dominated
by degeneracy effects) or, inversely, the temperature is extremely
sensitive to small variations of the total internal energy.
Therefore any small fluctuation caused for example by small numerical 
errors in the calculation of the energy density, 
will be amplified and reflected in temperature fluctuations.
Subsequently, the neutrino emission, which scales with a high 
power of the temperature $T$, will be very noisy.
For this reason we did not start our simulations with cold
($T = 0$) or ``cool'' ($T \la 10^8$~K) neutron stars as suggested
by the investigations of Kochanek (1992), Bildsten \& Cutler (1992),
and Lai (1994). Instead, we constructed initial temperature distributions
inside the neutron stars by assuming thermal energy densities of 
about 3\% of the degeneracy energy density for a given density $\rho$
and electron fraction $Y_e$. The corresponding central temperature 
was around 7~MeV, the surface temperature less than half an MeV, 
and the average temperature was a few MeV. Because of the small
contribution of thermal effects to the pressure, these temperatures
are unimportant for the neutron star structure, and the rapid 
and violent hydrodynamical evolution ensures that the results are
essentially unaffected by the assumed finite initial temperatures.

The simulations were performed on a Cray-YMP~4/64.
The models with 64 zones needed about 24~MWords of main memory and took
approximately 160~CPU-hours each, models with 32 zones roughly a factor 
of 10 less. Movies were generated for every model.

\begin{figure*}
 \begin{tabular}{cc}
  \epsfxsize=8.8cm \epsfclipon \epsffile{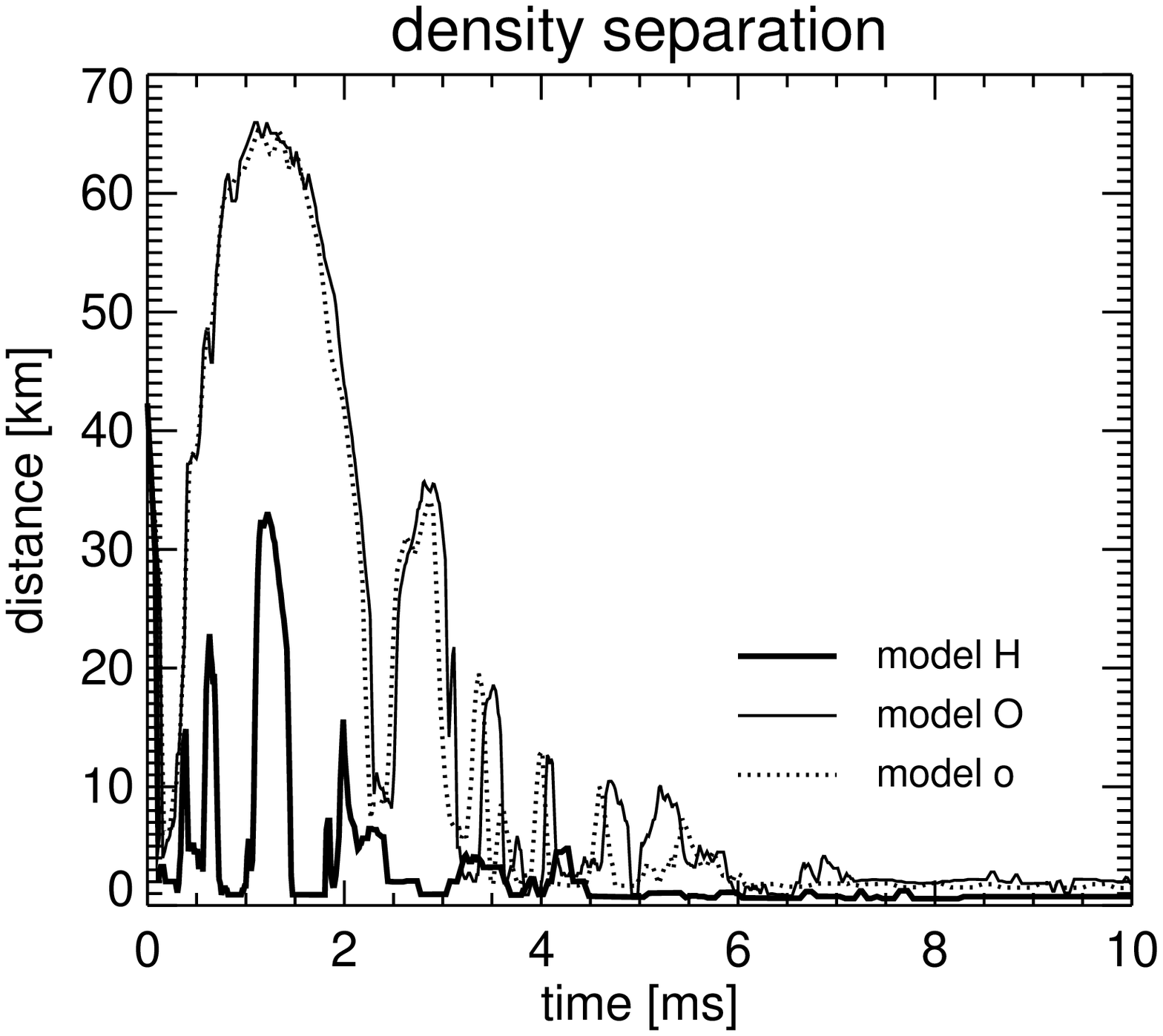} &
  \epsfxsize=8.8cm \epsfclipon \epsffile{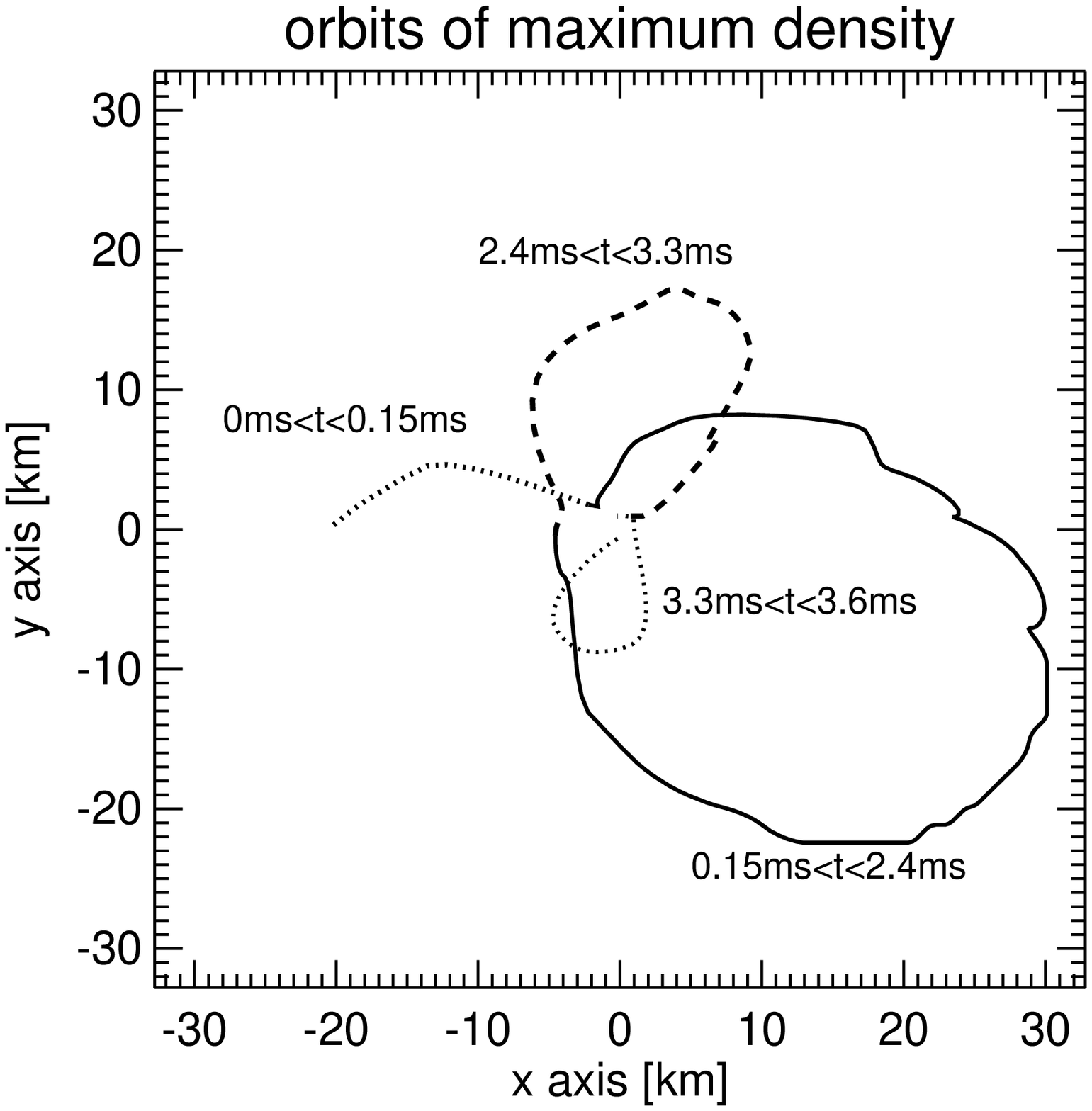}\\
  \parbox[t]{8.8cm}{\caption[]{The separation of the density maxima
  of the two neutron stars as a function of time for the three 
  Models~H, O, and o.}
  \label{fig:sepden}} &
  \parbox[t]{8.8cm}{\caption[]{Trajectory described by the density 
  maximum of one of the neutron stars in a Eulerian frame for the
  off-center collision, Model~O. Different line styles denote
  different orbits or phases between closest approaches.
  The kinks are numerical and due to the fact that the position of
  the maximum density is given by the integers corresponding to the 
  indices of the numerical grid cells.}
  \label{fig:O2orbit}} \\[15ex]
  \epsfxsize=8.8cm \epsfclipon \epsffile{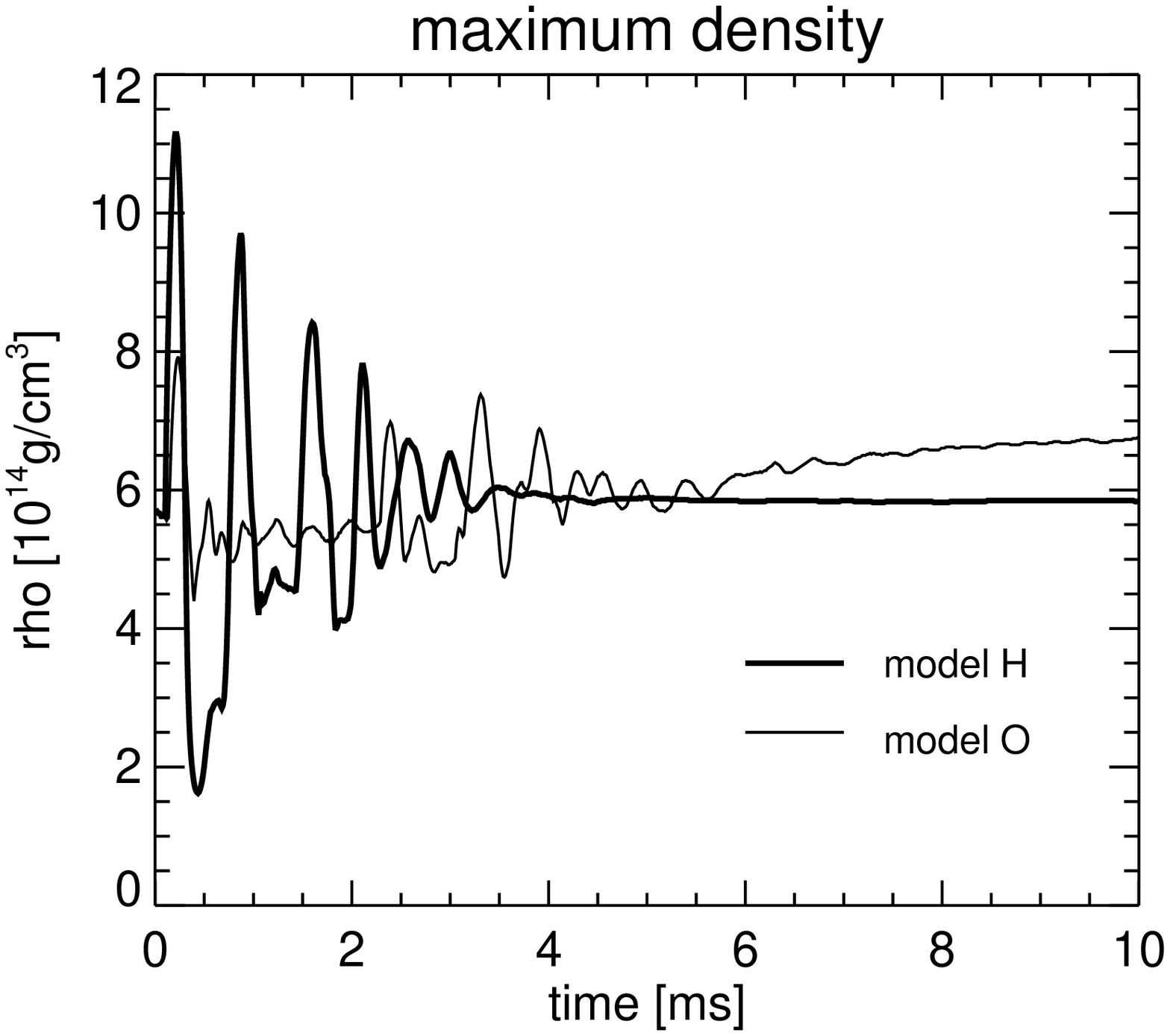} &
  \epsfxsize=8.8cm \epsfclipon \epsffile{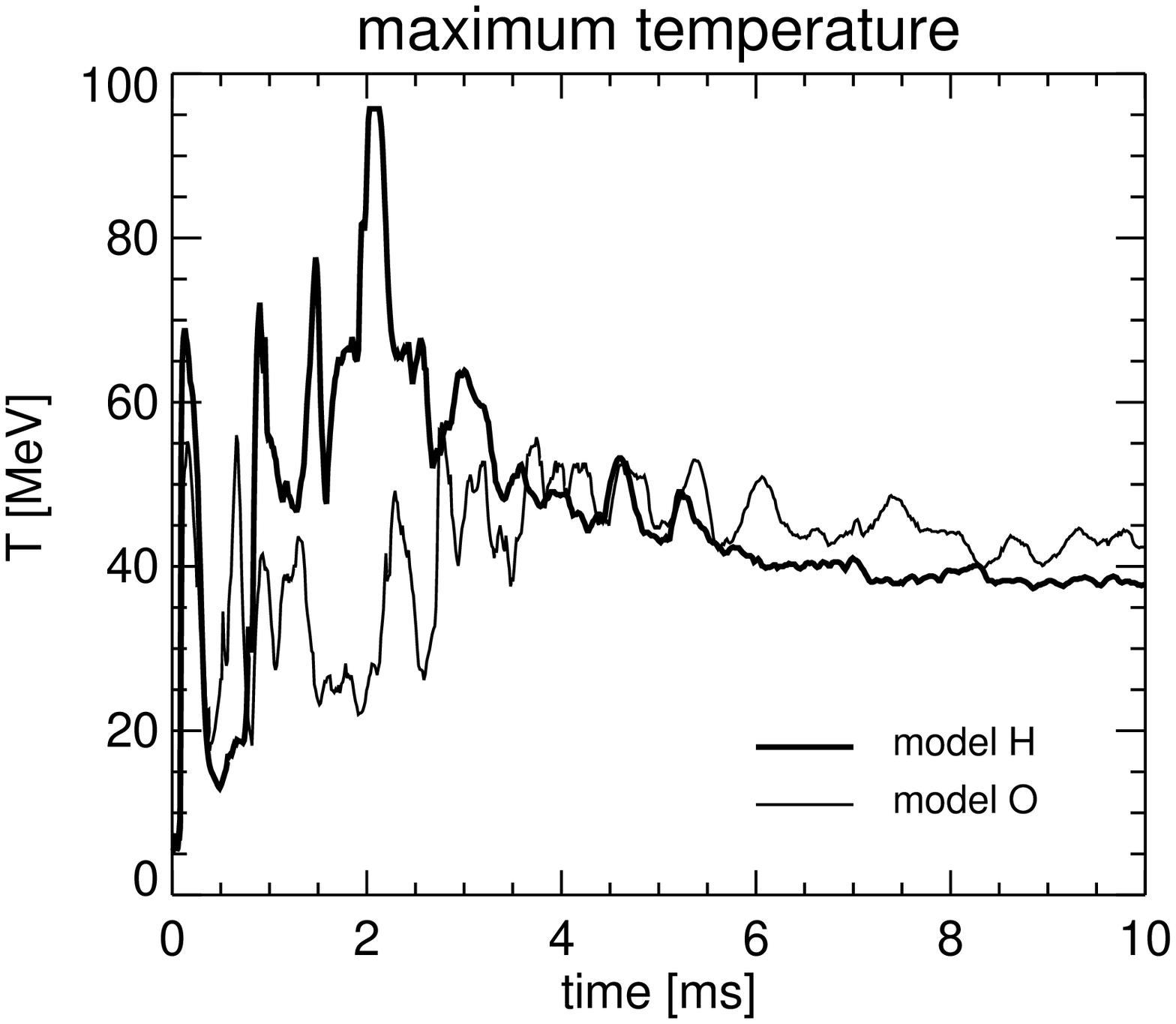}\\
  \parbox[t]{8.8cm}{\caption[]{The maximum density on the grid as a
  function of time for the two Models~H and O.}
  \label{fig:maxrho}} &
  \parbox[t]{8.8cm}{\caption[]{The maximum temperature on the grid as a
  function of time for the two Models~H and O.}
  \label{fig:maxtemp}}  \\
 \end{tabular}
\end{figure*}

\section{Hydrodynamical and thermodynamical evolution\label{sec:hydevol}}

\subsection{Head-on collision}

Figures~\ref{fig:Hcont1} and \ref{fig:Hcont2} show the density $\rho$, 
temperature $T$, electron fraction $Y_e$, and entropy $s$ at two 
different moments of the collision: At $t = 0.23\,{\rm ms}$ after
the start of the simulation the largest compression is reached with
a maximum density of $1.1\cdot 10^{15}\,{\rm g\,cm}^{-3}$
(see Fig.~\ref{fig:maxrho}), associated
with a prominent peak of the gravitational wave luminosity
(see Figs.~\ref{fig:lggrergH} and \ref{fig:grcompH}), and at 
$t = 2.87\,{\rm ms}$ 
when the collision remnant has performed several cycles of 
oscillatory motions before it begins to settle into a more quiet
state, and the neutrino emission starts to decrease from its
most powerful phase (Figs.~\ref{fig:neuterg} and \ref{fig:neutradH}). 
For the earlier time, data from the high-resolution Model~${\cal H}$
are plotted in Figs.~\ref{fig:Hcont1} and \ref{fig:Hcont2}, whereas
for the later moment only data from the 64$^3$ Model~H are available. 

Shortly after the surfaces of the neutron stars have touched during their
head-on collision (Figs.~\ref{fig:Hcont1}a and~\ref{fig:Hcont1}c),
a strong shock wave is generated at the common surface by the abrupt
deceleration of the matter. This shock propagates back into the as yet
practically undisturbed neutron star matter. The temperatures directly 
behind this shock reach values of more than 45$\,$MeV, while ahead of
the shock they were around 5$\,$MeV. The entropy in the initially cool 
neutron star ($s \ll 1\,k$/nucleon) is increased to values between
1$\,k$/nucleon and 2$\,k$/nucleon (Fig.~\ref{fig:Hcont2}c). 
At the same time, matter is being squeezed out perpendicularly 
to the collision axis and expands behind a very strong shock 
(postshock entropies near 30$\,k$/nucleon). This shock-heated matter
emits electron antineutrinos in large numbers 
(cf.~Fig.~\ref{fig:neutradH}) and quickly develops
from initially neutron-rich conditions to a much more symmetric nuclear
state characterized by an electron fraction $Y_e\ga 0.4$ 
(Fig.~\ref{fig:Hcont2}a).
In contrast, in the interior of the colliding bodies the composition
remains essentially unchanged because of the long neutrino diffusion 
timescales in the hot neutron star matter.

\begin{figure*}
 \begin{tabular}{cc}
  \epsfxsize=8.8cm \epsfclipon \epsffile{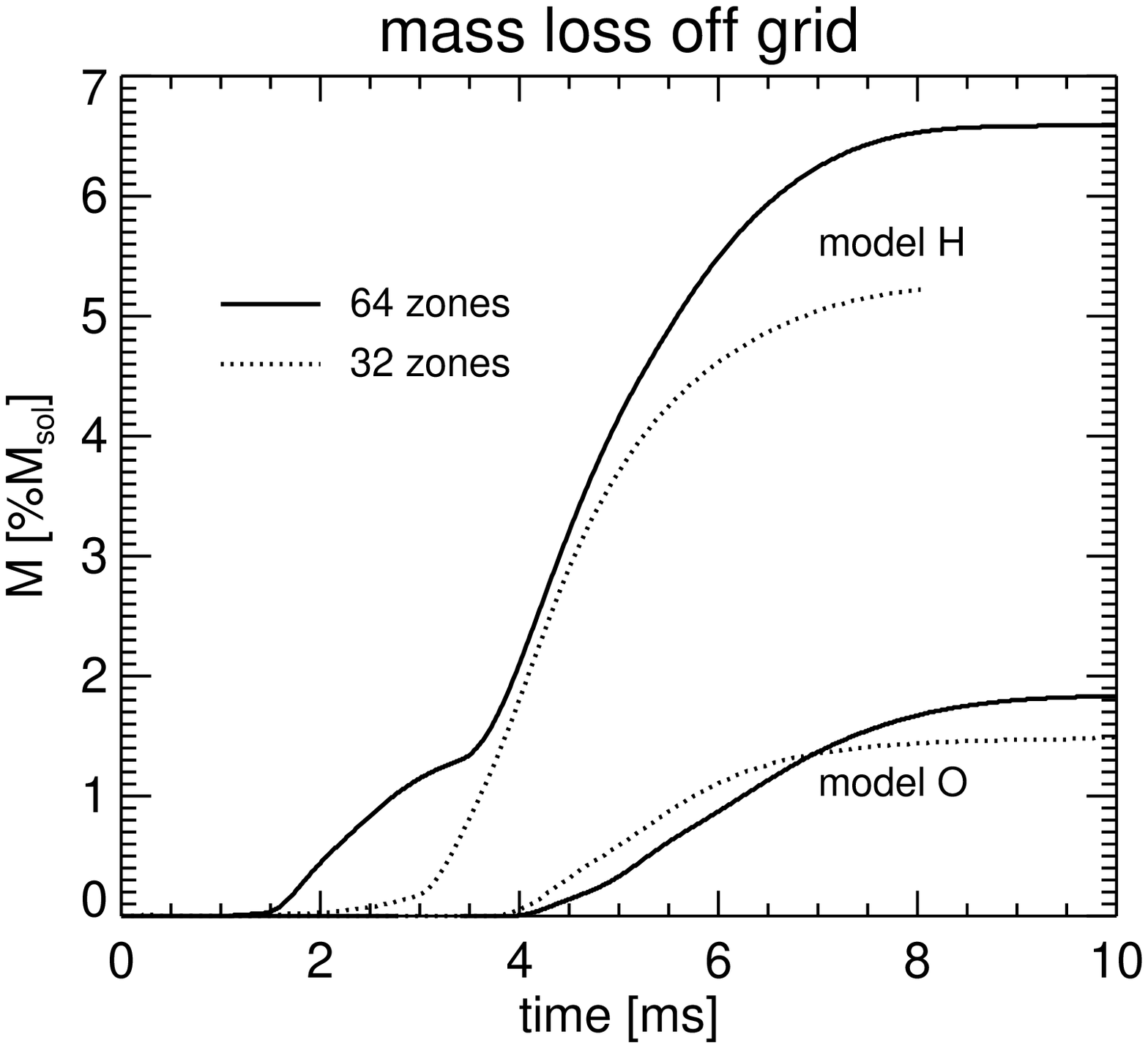} &
  \epsfxsize=8.8cm \epsfclipon \epsffile{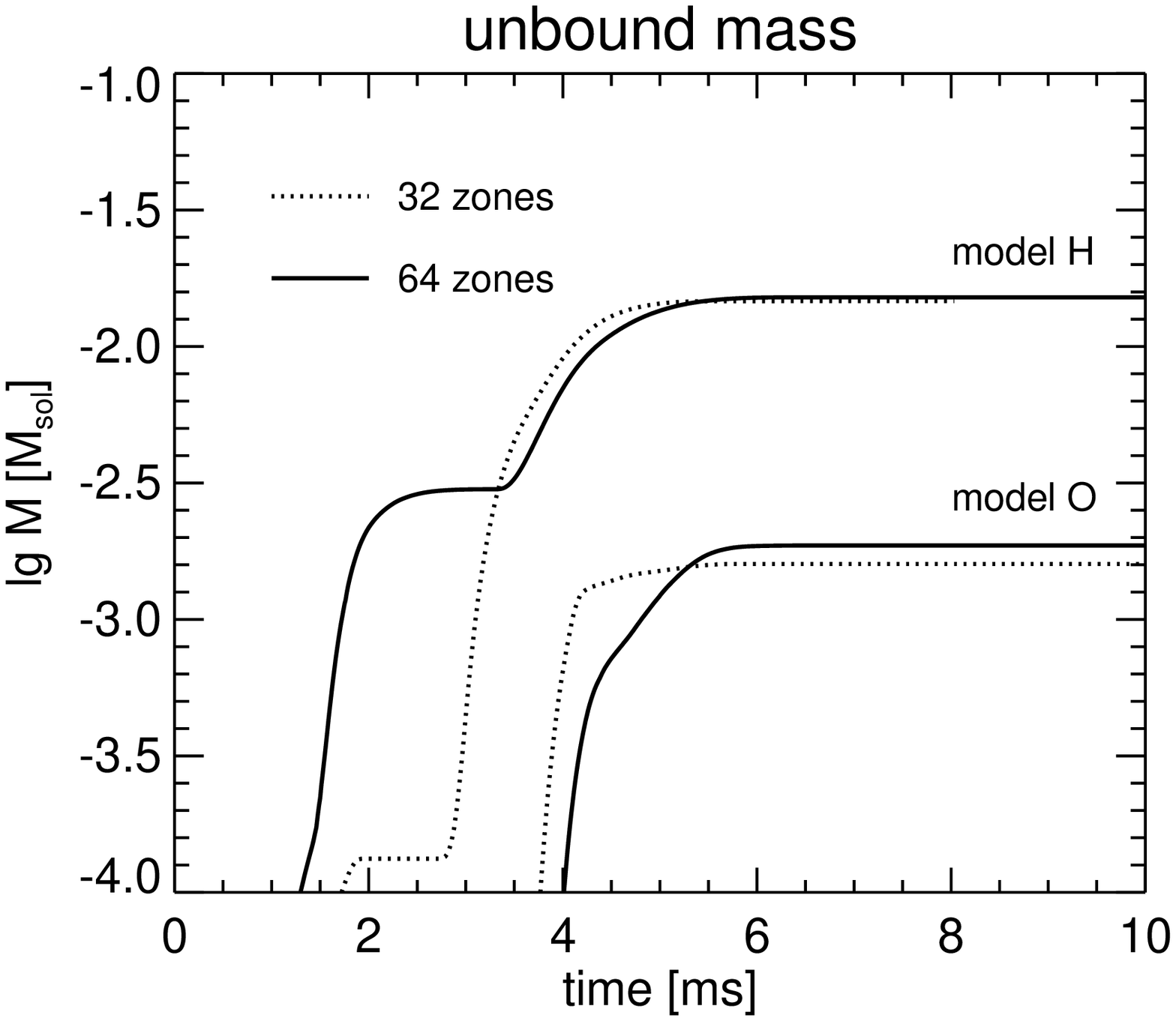} \\
  \parbox[t]{8.8cm}{\caption[]{Cumulative amount of matter flowing
  off the numerical grid (measured in $10^{-2}\,M_{\odot}$) as a 
  function of time from the beginning of the simulations.}
  \label{fig:massloss}} &
  \parbox[t]{8.8cm}{\caption[]{Cumulative amount of matter that 
  is unbound when it leaves the numerical grid (in units of 
  $M_{\odot}$, plotted logarithmically) as a function of time
  for Models~o, O, h, and H. The unbound mass is determined by
  the criterion that the sum of the specific gravitational,
  kinetic, and internal energies is positive.}
  \label{fig:unmass}} \\[15ex]
  \epsfxsize=8.8cm \epsfclipon \epsffile{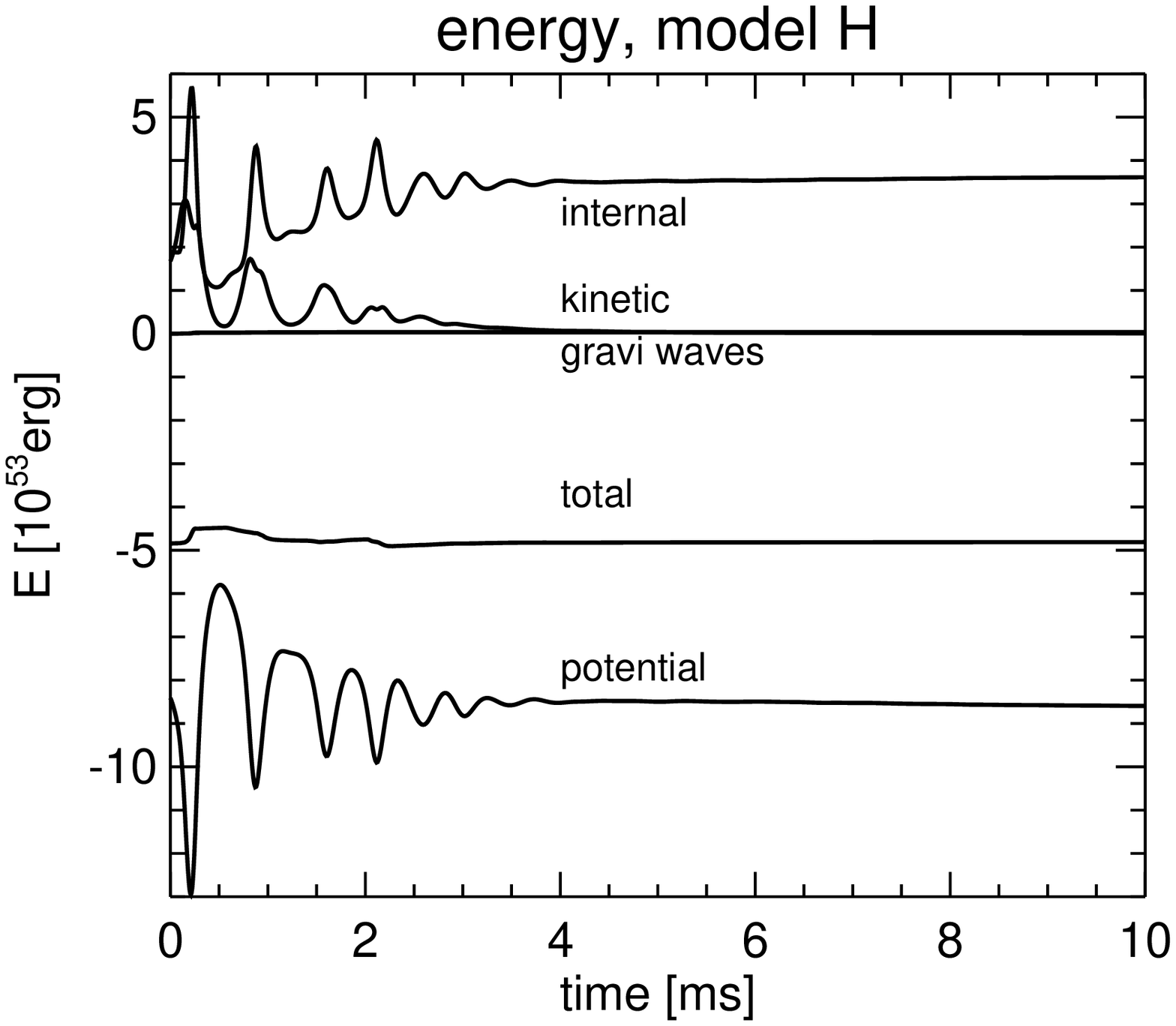} &
  \epsfxsize=8.8cm \epsfclipon \epsffile{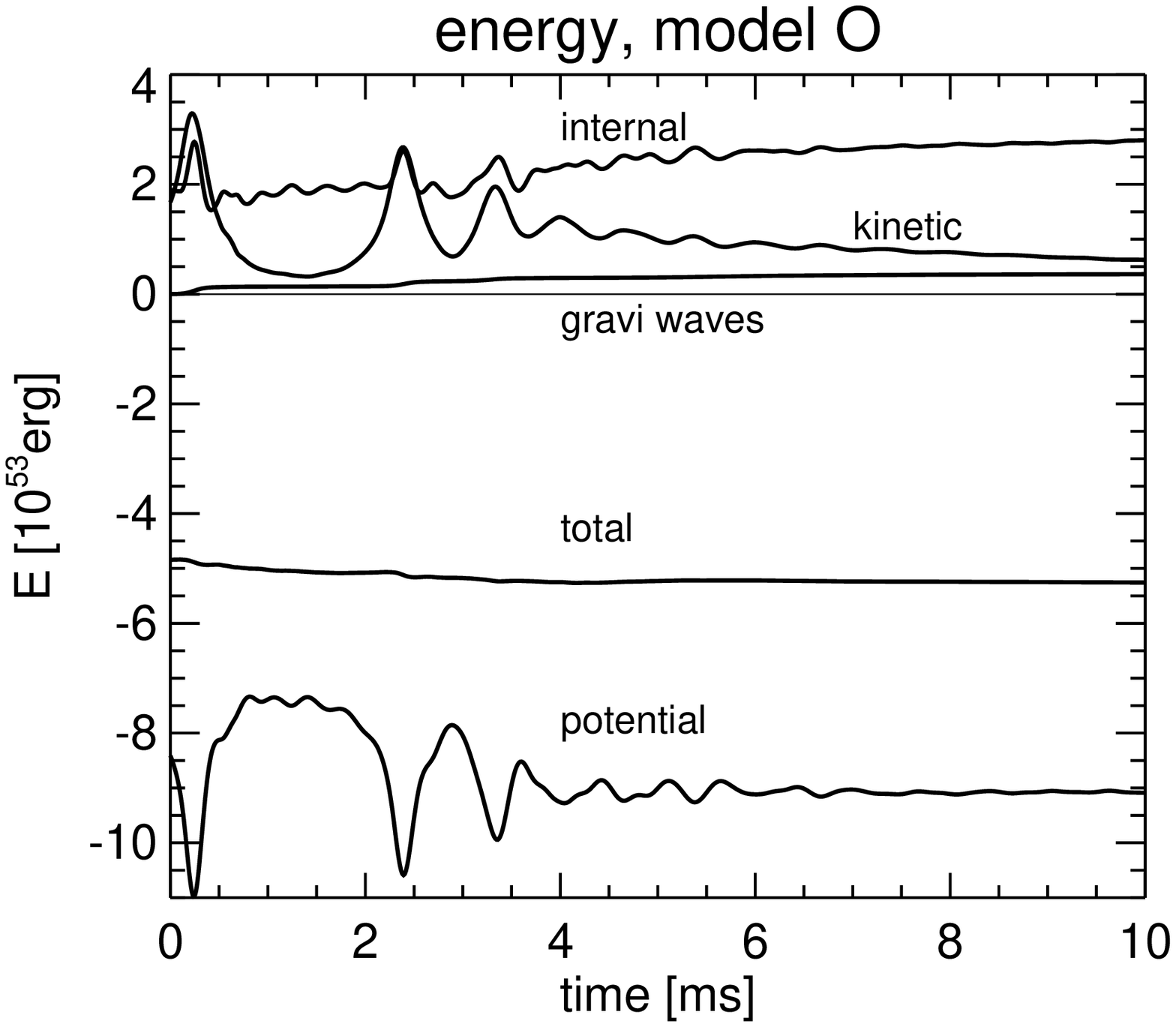} \\
  \parbox[t]{8.8cm}{\caption[]{The internal, kinetic, and potential
  energies of the matter on the grid and the sum of these
  energies as functions of time for Model~H. 
  Also plotted is the cumulative energy carried away in gravitational 
  waves.}
  \label{fig:energyH}} &
  \parbox[t]{8.8cm}{\caption[]{Same as Fig.~\ref{fig:energyH} but
  for Model~O. The thin solid line marks the $E = 0$ level.}
  \label{fig:energyO}} \\
 \end{tabular}
\end{figure*}

At the collision interface a thin ``pancake'' like layer with very 
high temperatures up to about 70$\,$MeV (Figs.~\ref{fig:Hcont1}c and
\ref{fig:maxtemp}) occurs. This sheet is dynamically unstable due to
shear motions. First indications of a break-down of the mirror symmetry
relative to the $y$-$z$-plane can be already seen at $t = 0.23\,$ms
in Figs.~\ref{fig:Hcont1}c and \ref{fig:Hcont2}a. Only a short
moment later, when the merged bodies bounce back and the oblate shape
changes into a prolate form, this flat pancake-like layer folds
asymmetrically and breaks up on a millisecond timescale
(Fig.~\ref{fig:Hcont1}d). Within 3$\,$ms the density distribution has
already smoothed into a nearly spherical shape (Fig.~\ref{fig:Hcont1}b)
and most of the kinetic energy of the impact has been dissipated 
by shocks into thermal energy or is carried away by ejected matter
(see Fig.~\ref{fig:energyH}). The collision has lead to an increase 
of the entropy to values near 2$\,k$/nucleon in the merged object
(Fig.~\ref{fig:Hcont2}c), whereas the shock heated gas that forms
a very extended, nearly spherical cloud around the dense central
body, has entropies between 6$\,k$/nucleon and 10$\,k$/nucleon
(Fig.~\ref{fig:Hcont2}d). Ejected clumps of matter with even higher 
entropy ($s \sim 20\,k$/nucleon) can be identified, and positron 
captures onto neutrons and $\bar\nu_e$ production in the hot gas
($T\sim$ several MeV) leads to a rapid increase of the electron fraction
to values $Y_e\sim 0.3$--0.4 and higher in the expanding debris. 

The formation of shock waves at the moment of the impact in Model~H
(Fig.~\ref{fig:Hcont1}) clearly indicates that the head-on 
collision is strongly inelastic and the dissipation of kinetic energy
happens very efficiently. Therefore the two neutron stars are not
able to separate again after the first compression and reexpansion,
however, it takes several (4--6) violent oscillations 
until all the kinetic energy is dissipated into heat. 
The reexpansions produce peaks of the separation of the density
maxima in Fig.~\ref{fig:sepden}, while the compression phases are
reflected in a sequence of very large density maxima in 
Fig.~\ref{fig:maxrho} and temperature maxima in Fig.~\ref{fig:maxtemp}.
The steady decrease of the density maxima in Fig.~\ref{fig:maxrho}
indicates that the oscillations come to a rest within about 4$\,$ms.
In contrast, the maximum temperature increases from one compression
to the next (Fig.~\ref{fig:maxtemp}) because of the dissipative 
heating of the stellar plasma. The most extreme temperatures that are
reached in Model~H during the dynamical phase, $0< t\la 4\,$ms, are 
close to 100$\,$MeV. After settling into a static state $t > 4\,$ms, 
the maximum temperatures in the collision remnant is around 
40--50$\,$MeV.

\subsection{Off-center collision}

The motion of the two neutron stars is rather complicated in 
case of the {\it off-center} collision, Model~O. The first phase of
the infall ($t \la 0.1\,$ms) proceeds essentially
along point-mass binary orbits (Fig.~\ref{fig:O2orbit}),
until the stars start to touch and orbital energy and angular
momentum are converted into neutron star spin and are consumed by
the acceleration of matter which is flung off the neutron star 
surfaces. The corresponding loss of orbital angular momentum and
kinetic energy leads to a transformation of the initially parabolic
orbits into elliptic ones. Even more orbital energy is transfered
to internal energy when the neutron stars come into contact and 
inelastic interaction sets in. 

After the first closest approach,
visible in Fig.~\ref{fig:sepden} as minimum distance 
$d_{\rm min}\approx 3\,$km of the density maxima of the two 
neutron stars at $t\approx 0.15\,$ms, the positions of the 
density maxima describe three nearly elliptic orbits between
moments of closest approach at $t \approx 0.15\,$ms, $2.4\,$ms, 
$3.3\,$ms, and $3.6\,$ms (see also Fig.~\ref{fig:sepden}).
In Fig.~\ref{fig:O2orbit} these three orbits, represented by 
the $x$-$y$-trajectory of the density maximum of one of the 
neutron stars, are discerned  
by different line styles. The diameters of the orbits (and thus
the maximum separation of the density maxima) become successively
smaller due to the inelasticity of the contact between the stars
during their close encounters. The major axis of the first ellipse
has a length of more than 30$\,$km, corresponding to an apastron 
separation of the density maxima on the grid of about 65$\,$km at
$t \approx 1.2\,$ms (Fig.~\ref{fig:sepden}), much larger than 
the initial distance of the neutron star centers which was $42\,$km.
This means that the neutron stars separate again after their first
encounter, but fall back towards each other again. Even during 
the second apastron at $t\approx 2.9\,$ms, the density maxima of
the two stars have a distance of about 35$\,$km, which is larger
than the sum of their initial radii $2R_{\rm ns,i} = 30\,$km.
During the third and the subsequent quasi-elliptic orbits the 
neutron stars are not able to separate again. While the first
bound orbit has a period of about 2.2$\,$ms, the second, smaller
one has only $\sim 0.9\,$ms, and the following are even shorter
(Figs~\ref{fig:sepden} and \ref{fig:O2orbit}).

In Figs.~\ref{fig:Ocont1} and~\ref{fig:Ocont2} the density distribution,
temperature, electron fraction, and entropy per nucleon are plotted
for Model~O in the orbital plane at two different stages during the 
off-center collision. The left panels show the results for a time
close to the apastron of the first orbital ellipse, i.e., a little
more than a millisecond after the first closest approach.
The neutron stars are tidally strongly deformed and gas has
been swept into the surroundings during the first direct contact
and interaction (Fig.~\ref{fig:Ocont1}a). There is a dense gas 
bridge ($\rho \sim 10^{12}$--$10^{13}\,{\rm g\,cm}^{-3}$) between the
stars which continue to wobble and oscillate along their orbits.
The temperature has climbed to nearly 40$\,$MeV in distinct
hot spots where the gas bridge hits the denser cores of the stars 
(Figs.~\ref{fig:Ocont1}c and \ref{fig:maxtemp}), whereas the 
extended cloud of gas surrounding the orbiting bodies has a 
temperature of 1--5$\,$MeV. In this ambient gas, $\bar\nu_e$ 
production by positron capture onto neutrons has raised the electron
fraction from initially less than about 0.05 to maximum values 
around 0.2 (Fig.~\ref{fig:Ocont2}a). The maximum entropy values of
$s \sim 11\,k$ per nucleon are produced by bow shocks in front of
the rather dilute ($\rho \sim 10^9$--$10^{10}\,{\rm g\,cm}^{-3}$)
clouds reaching outward from the two neutron stars. Clumps of
gas with entropy $s\sim 5$--$6\,k$/nucleon are scattered in the
surroundings of the collision site. Note that in contrast to the
head-on collision, there is no shock heating of the interior
of the neutron stars. Up to the end of the simulation the 
high-density cores of the neutron stars retain their low initial
entropy, even after they have merged into one body. 

The right panels in Figs.~\ref{fig:Ocont1} and~\ref{fig:Ocont2}
show snapshots at a time near the end of the simulation 
($t = 9.27\,$ms). The distributions of density, entropy, and
electron fraction have become roughly circular in the 
$x$-$y$-plane: A compact,
dense central body ($\rho > 10^{12}\,{\rm g\,cm}^{-3}$) with
$T\sim 5$--10$\,$MeV (outside of two distinct hot spots
where $T\sim 40\,$MeV), $Y_e\sim 0.04$--0.1, and $s\la 7\,k$/nucleon
is surrounded by an extended envelope with somewhat larger
$Y_e$ and $s$ (but lower temperature), which is rapidly rotating
and which is stabilized by centrifugal forces. Therefore the
vertical extension of the gas envelope is significantly smaller
than its diameter in the orbital plane; the density contour 
corresponding to $\rho = 10^{10}\,{\rm g\,cm}^{-3}$ extends to
a radius of about 130$\,$km in the orbital plane, whereas its
butterfly shape has a maximum vertical height of roughly
70$\,$km (Fig.~\ref{fig:O2densv}). Even the compact core is 
rotationally deformed with an axis ratio of 1:1.5. Nevertheless,
there is a considerable amount of gas at large heights $|z|$ above and
below the orbital plane. By spatial integration we find 
$3.8\times 10^{-2}\,M_{\odot}$ for $|z| \ge 40\,{\rm km}$,
$2.2\times 10^{-2}\,M_{\odot}$ for $|z| \ge 50\,{\rm km}$, and still
$6.0\times 10^{-3}\,M_{\odot}$ for $|z| \ge 80\,{\rm km}$.

\subsection{Comparison of head-on and off-center collisions}

The different dynamical evolutions of the head-on collision,
Model~H, and the off-center collision, Model~O, are reflected
in the different time histories of the maximum density
(Fig.~\ref{fig:maxrho}) and the maximum temperature 
(Fig.~\ref{fig:maxtemp}) on the grid. Model~H shows very large
amplitudes of the maximum density at the moments of strongest
compression with up to a factor of $\sim 10$ larger values 
compared to the return points of expansion phases.
These density fluctuations are damped in a sequence of 5--6
strong oscillations, and a stationary value is reached after
about 4$\,$ms. The temperature
evolution reveals spikes and valleys, respectively, at the same 
moments, but there is a general trend of the temperature to
increase during the dissipative oscillatory compressions and 
reexpansions. Although in Model~O the variations of the separation
of the density maxima are much more pronounced than in Model~H 
(Fig.~\ref{fig:sepden}), the maximum density and temperature
on the grid show much less extreme fluctuations because
the neutrons stars describe orbits around each other and do
not crash violently into each other. One can
recognize peaks of the maximum density correlated with the 
moments of closest approaches, $t\approx 0.15\,$ms, 2.4$\,$ms,
3.3$\,$ms, and 3.6$\,$ms (compare Fig.~\ref{fig:maxrho}
and Fig.~\ref{fig:O2orbit}). The whole evolution of Model~O is
much less violent than that of Model~H. Nevertheless, on a longer
timescale $t \ga 4\,$ms, both models settle to roughly the same
maximum temperature of 40--50$\,$MeV. The maximum density in
Model~O becomes even somewhat larger than in Model~H, because the
latter has been heated to higher entropies and therefore thermal 
pressure causes an expansion of the collision remnant. In addition, 
the compact core of the remnant of Model~H is less massive as a
result of the larger mass loss during the collision.


\begin{figure}
\epsfxsize=8.8cm \epsfclipon \epsffile{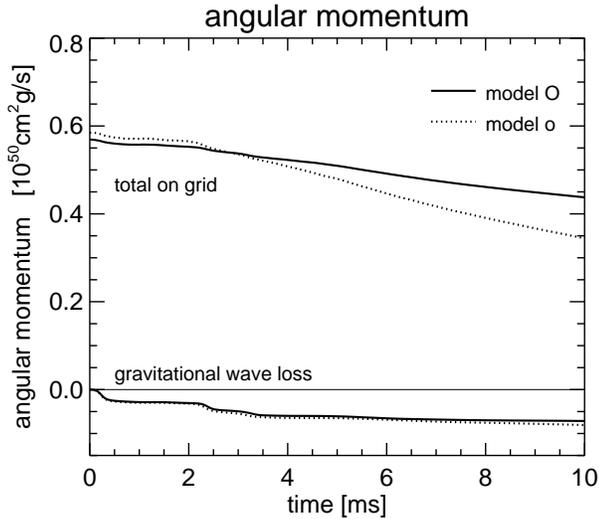}
\caption[]{Total angular momentum ($z$-component) of the matter
  on the grid (upper curves) and cumulative angular momentum loss
  by gravitational wave emission (lower curves) as functions of
  time for Models~O and o. Gravitational waves carry away about 13\% 
  of the initial angular momentum of the colliding neutron stars
  and another non-negligible amount ($\sim 3.5\%$)
  is taken away by the matter flowing off the numerical grid.
  }
\label{fig:angmom}
\end{figure}

The more violent collision and therefore higher temperatures 
in Model~H push more matter off the grid than in Model~O
(Fig.~\ref{fig:massloss}). Also a larger fraction of this
matter gets unbound (Fig.~\ref{fig:unmass}) which is the
case when the total specific energy of the gas, defined as the sum 
of the specific kinetic, internal, and gravitational potential
energies, becomes positive. In Model~H about $1.5\times 10^{-2}\,M_{\odot}$ 
are able to escape the gravitational potential of the collision 
remnant, whereas it is little more than one tenth of this amount
in Model~O (cf.~Table~\ref{tab:models1}). Obviously, the angular
momentum of Model~O (see Fig.~\ref{fig:angmom}) and the associated
centrifugal shredding of material can hardly compete with the 
ejection of gas in the strong shock waves occurring in Model~H.
We note in passing that Rasio \& Shapiro (1992) (RS) found a significantly
larger amount of mass loss (up to about 5\% of the total mass) in 
their simulations of head-on collisions between two identical 
$\gamma = 2$ polytropes,
compared to only $\sim 0.46$\% that can escape from the system in
our Model~H. The difference is presumably caused by a combination of 
reasons, the use of different equations of state and correspondingly
different stellar structure and mass (Lattimer \& Swesty nuclear
EOS here vs.~$\gamma = 2$ adiabatic EOS by RS), the inclusion of 
gravitational wave back-reaction on the hydrodynamical flow in our
simulations, and last but not least the use of different numerical 
schemes (Eulerian PPM here vs.~SPH by RS) in combination with 
possibly different criteria to determine the unbound mass.

The temporal evolutions of the internal, kinetic, and gravitational 
potential energies (Figs.~\ref{fig:energyH} and \ref{fig:energyO})
show structures that correspond to the stages of the dynamical 
interactions of the colliding stars. The internal energy of Model~H
oscillates strongly with maxima at the moments of strongest 
compression (compare Figs.~\ref{fig:maxrho} and \ref{fig:energyH})
which coincide with maxima of the kinetic and minima of the potential
energy. There is a general trend for the internal energy to increase
with time. This corresponds to a decrease of the kinetic energy 
(while the potential energy fluctuates around an essentially constant
level) and thus reflects the action of dissipative forces. 
In the off-center collision of Model~O the kinetic energy is much 
less efficiently converted into internal energy. The latter exhibits
a continuous increase with superimposed, but much less 
dramatic, local peaks at the instants of closest approach, also
coinciding with maxima of the kinetic and minima of the potential 
energy. This reflects the dynamical transformation between the 
different energy forms. Near the end of the computed evolution,
practically all of the kinetic energy of Model~O is rotational energy, 
$E_{\rm kin} = E_{\rm rot}$. Using the values from Fig.~\ref{fig:energyO},
we determine the following ratio of kinetic energy to the gravitational
binding energy: $\beta = E_{\rm kin}/E_{\rm rot} \approx 0.08$. 

The total
energies of Model~H and Model~O (defined as the sum of kinetic, 
internal, and potential energies of all gas on the grid) are also
shown in Figs.~\ref{fig:energyH} and \ref{fig:energyO}, respectively.
In Model~H minor variations of the total energy between 
$t\approx 0.2\,$ms and $t\approx 2.2\,$ms are numerical because of
the extremely violent collision. Since the energy carried away by
gravitational waves and mass loss off the numerical grid is negligible,
the total energy at the beginning and end of the simulation are 
nearly equal. In contrast, in Model~O the energy emitted in 
gravitational waves leads to a gradual decrease of the total energy
of the gas on the grid. 

Fig.~\ref{fig:angmom} shows the total angular momentum 
($z$-com\-po\-nent) of the gas on the grid and the cumulative value of
the angular momentum carried away by the emitted
gravitational waves as a function of time for Models~O and o. 
There is very good agreement of both calculations concerning the
angular momentum loss in gravitational waves. This indicates that
the overall mass distribution in the neutron stars (which enters
the calculation of the mass quadrupole moment needed for the 
evaluation of the grav\-ita\-tional-\-wave source terms) is sufficiently
well represented even on the coarser grid of Model~o. Most of the
decrease of the gas angular momentum is explained by the gravitational
wave emission. An additional effect comes from the mass loss off the
computational grid at $4\,{\rm ms}\la t \la 8\,{\rm ms}$
(Fig.~\ref{fig:massloss}) which removes an angular momentum of about
$\Delta J_z \approx M_{\rm g}r_{\rm g}v_{\rm g}\approx 
2\times 10^{48}\,{\rm g\,cm^2\,s^{-1}}$ (with $M_{\rm g}$
taken from Table~\ref{tab:models1} and
$r_{\rm g}\approx 160\,{\rm km}$ being the grid radius and 
$v_{\rm g}\approx 3.5\times 10^9\,{\rm cm\,s^{-1}}$ the 
nearly tangential velocity of the gas when it leaves the grid)
or 3.5\% of the initial angular momentum. However, although the
gravitational wave loss and the mass flowing off the grid are
very similar in both models, Model~o exhibits a steeper decrease
of the total angular momentum at $t\ga 3\,{\rm ms}$ than Model~O.
This difference is purely numerical and caused by the coarser
grid resolution of Model~o. Even in the better resolved calculation,
Model~O, about 7\% of the initial angular momentum are destroyed
by numerical effects at the end of the simulation. 

The relativistic rotation parameter is defined as 
$a\equiv Jc/(GM_{\rm tot}^2)$
where $M_{\rm tot}$ is the total mass of the system ($M_{\rm tot} = 2M$
initially with $M$ being the mass of one of the neutron stars) and
$J$ is the total angular momentum relative to the center of mass.
We have an initial value of $a = 0.60$ and find a value of $a \ga 0.47$
(in Model~O) at the end of the simulation after angular momentum
has been removed from the system by gravitational waves and ejected
matter. Since $a < 1$ rotation seems unable to prevent the 
collapse of the collision remnant to a black hole if the remnant mass 
exceeds the maximum stable mass of the employed equation of state
(see also RJS and Rasio \& Shapiro, 1992, and references therein).
Thermal pressure can increase this stable mass limit only 
insignificantly (cf.~Goussard et al.~1997) and, if so, 
only during the transient period of neutrino cooling (note that
the interior of the neutron stars retains its initial
low entropy in the off-center collision, see  
Figs.~\ref{fig:Ocont2}c and d, and thus the temperature remains 
fairly low), and also rotation leads to an
increase of the upper stability limit on the baryon mass by
only $\la 20\%$ (Friedman et al.~1986; Friedman \& Ipser 1987).

\begin{figure*}
 \begin{tabular}{cc}
  \epsfxsize=8.8cm \epsfclipon \epsffile{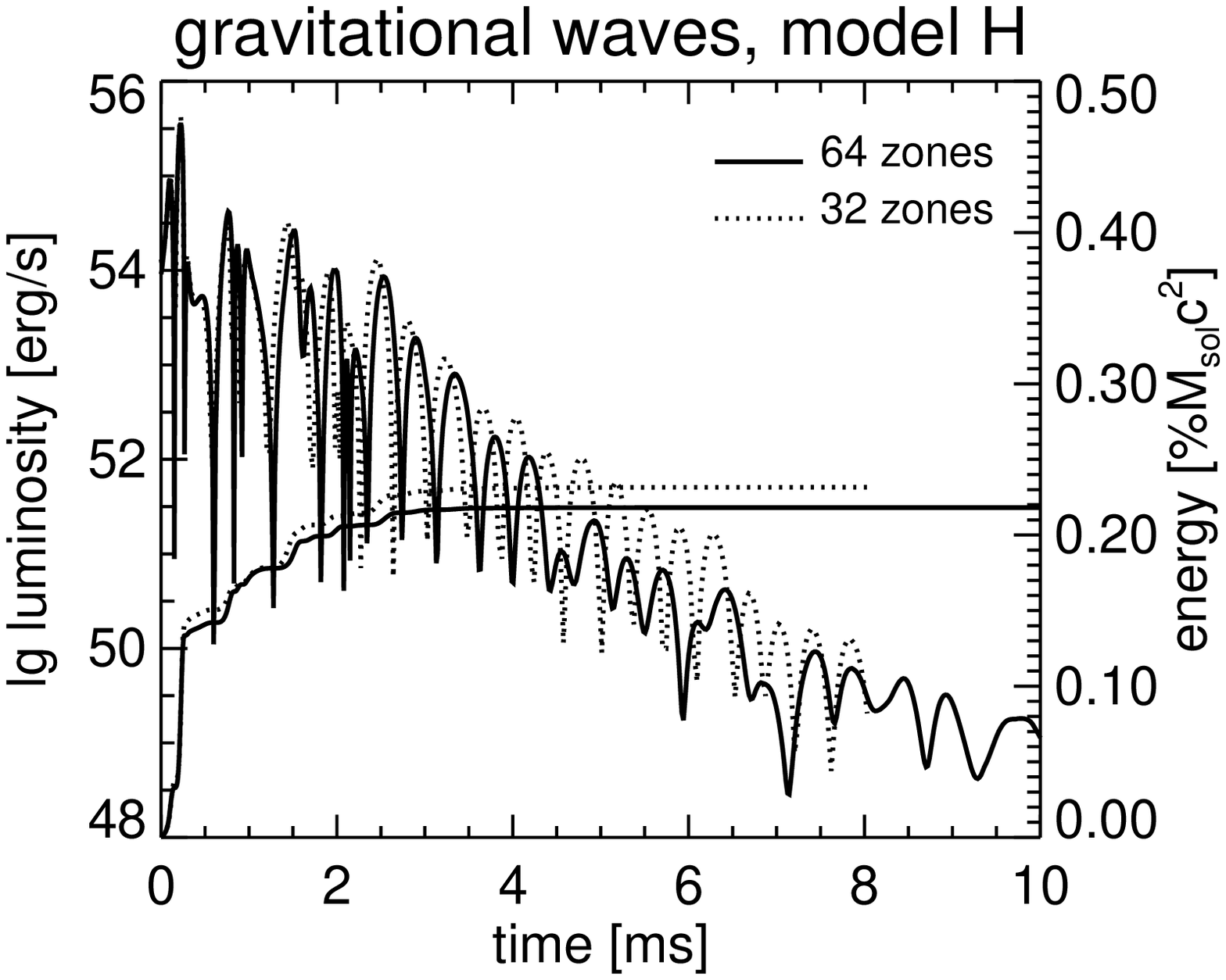} &
  \epsfxsize=8.8cm \epsfclipon \epsffile{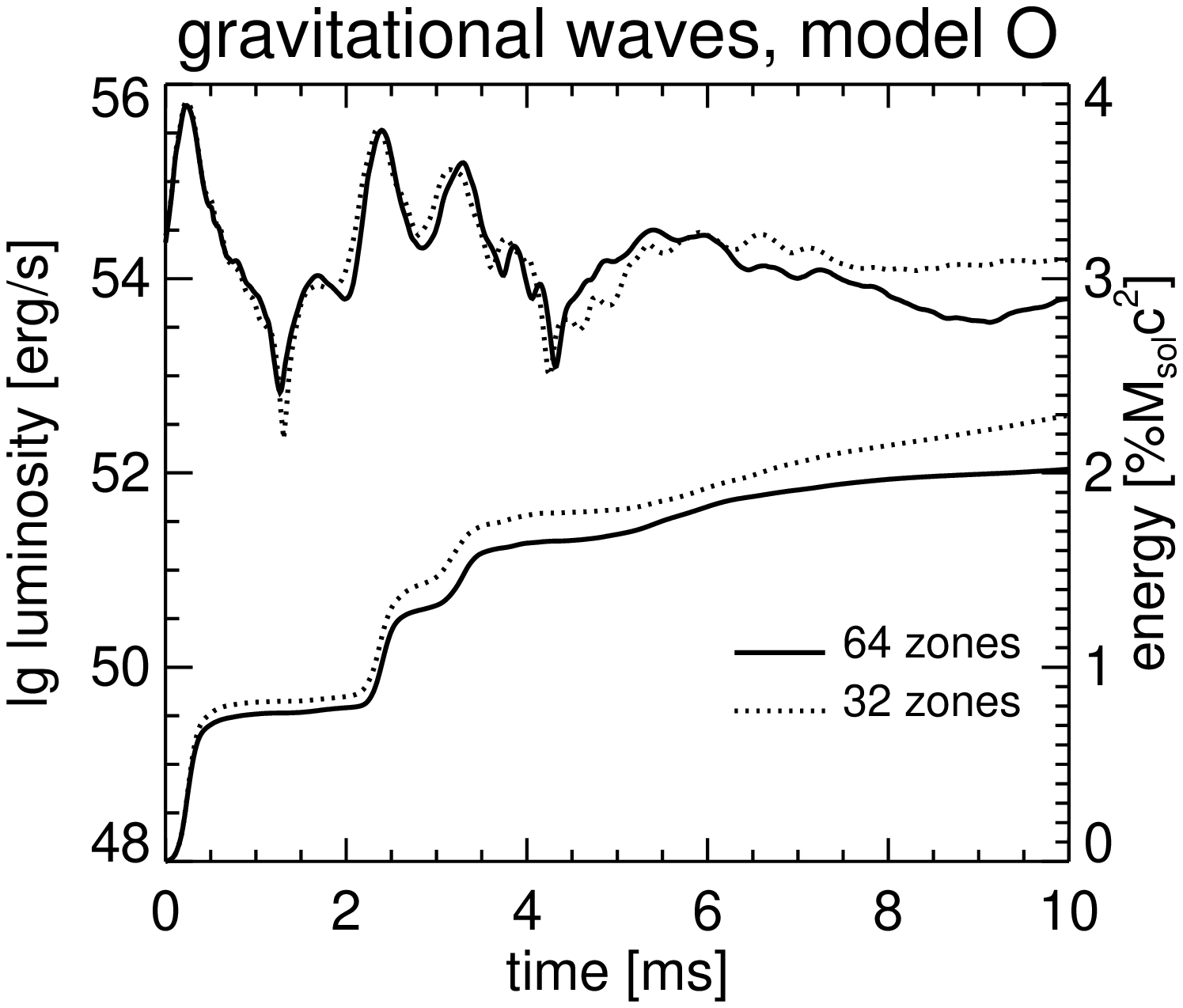} \\
  \parbox[t]{8.8cm}{\caption[]{Gravitational-wave luminosity and
  cumulative energy emitted in gravitational waves
  (measured in units of $10^{-2}\,M_{\odot}c^2$) as functions of
  time for Models~H and h.}
  \label{fig:lggrergH}} &
  \parbox[t]{8.8cm}{\caption[]{Same as Fig.~\ref{fig:lggrergH}
  but for Models~O and o.}
  \label{fig:lggrergO}} \\[15ex]
  \epsfxsize=8.8cm \epsfclipon \epsffile{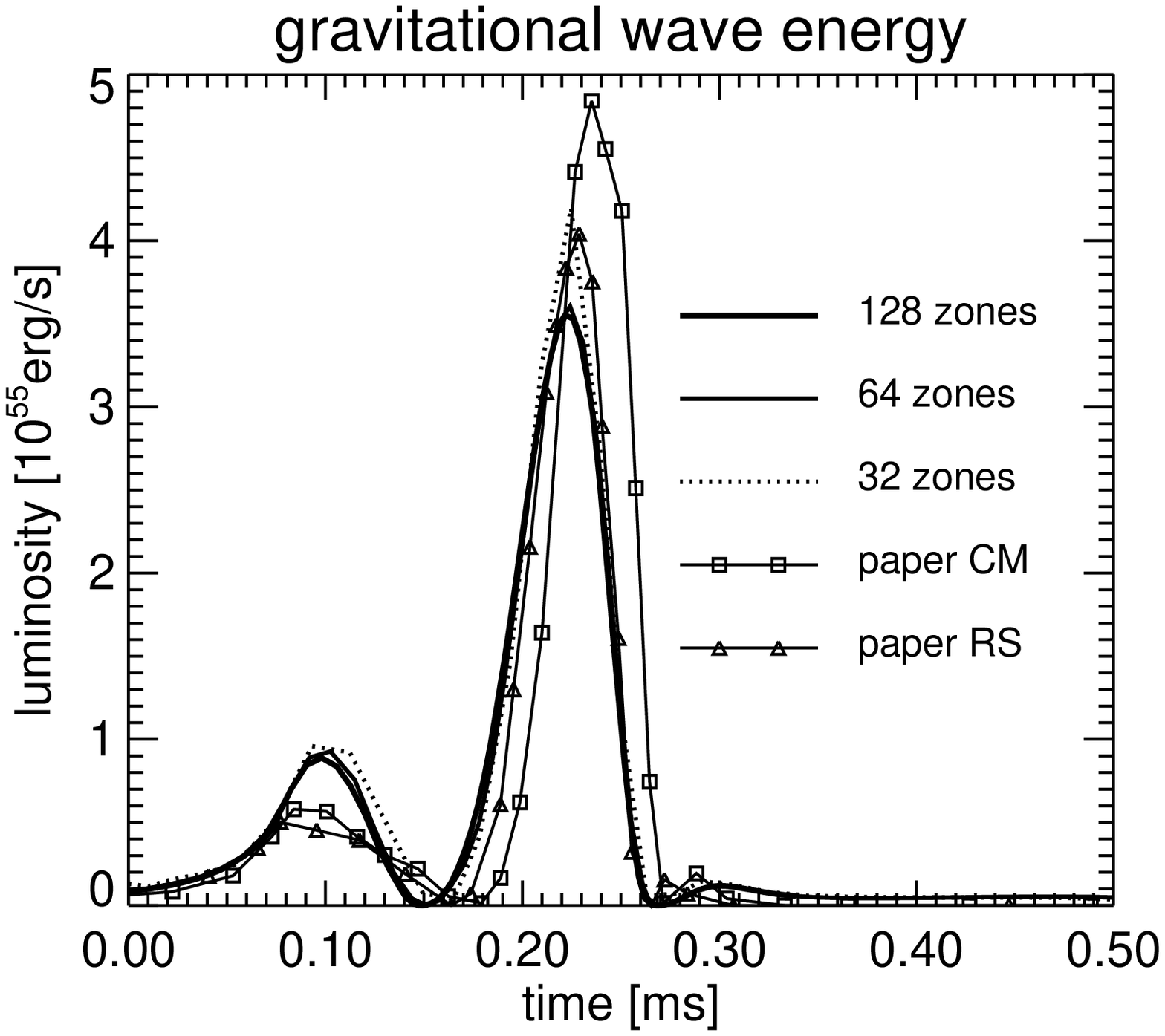} &
  \epsfxsize=8.8cm \epsfclipon \epsffile{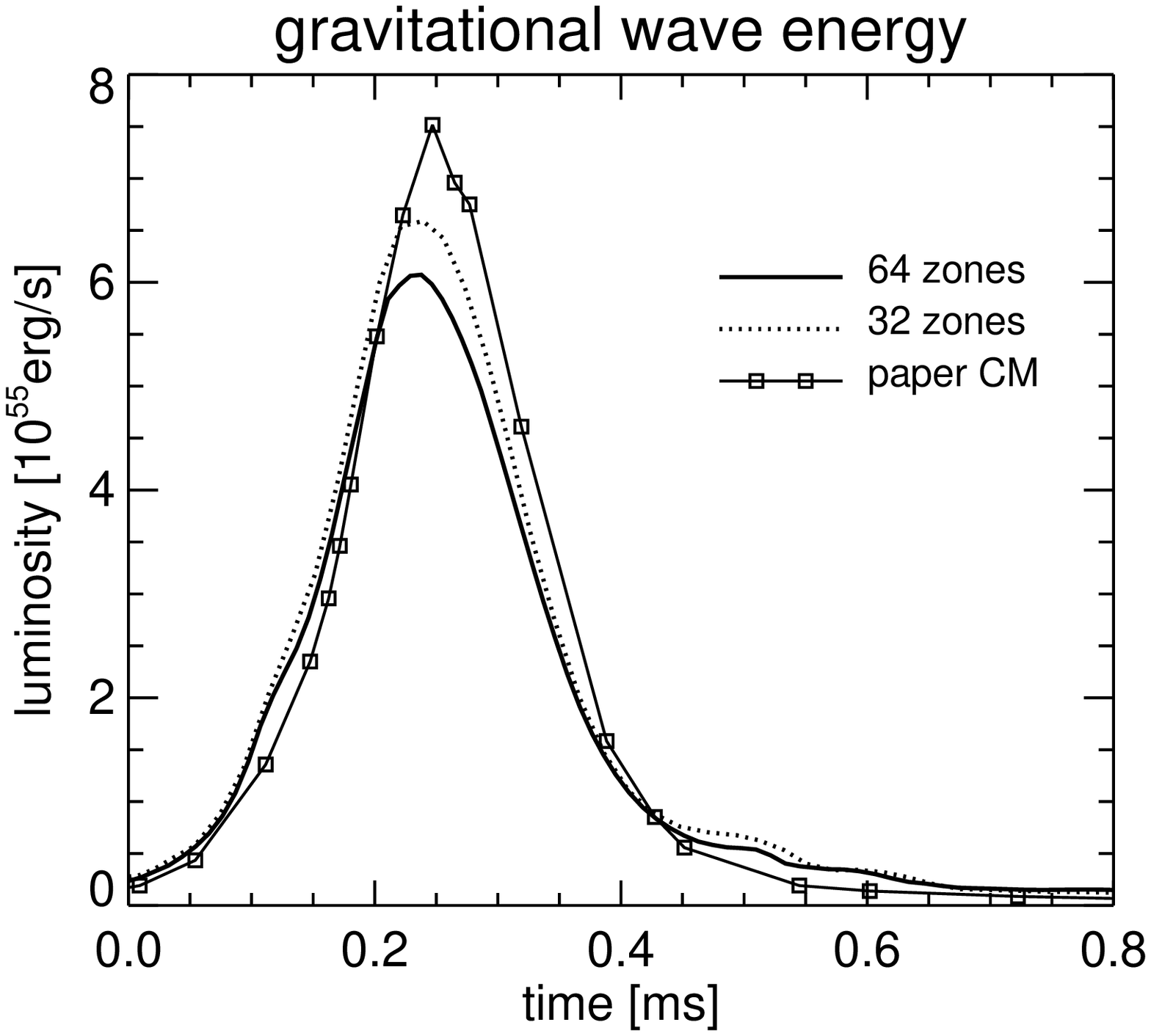} \\
  \parbox[t]{8.8cm}{\caption[]{Gravitational-wave luminosity
  as a function of time for the {\it head-on} collision
  Models~${\cal H}$, H and h, compared 
  with the results from Centrella \& McMillan
  (CM, 1993, Fig.~8) and from Rasio \& Shapiro (RS, 1992, Fig.~9).}
  \label{fig:grcompH}} &
  \parbox[t]{8.8cm}{\caption[]{Gravitational-wave luminosity
  as a function of time for the {\it off-center} collision 
  Models~O and o, compared with the result from Centrella \& McMillan 
  (CM, 1993, Fig.~11).}
  \label{fig:grcompO}} \\
  \end{tabular}
\end{figure*}

\section{Gravitational waves\label{sec:gwem}}

\begin{figure}
\epsfxsize=8.8cm \epsfclipon \epsffile{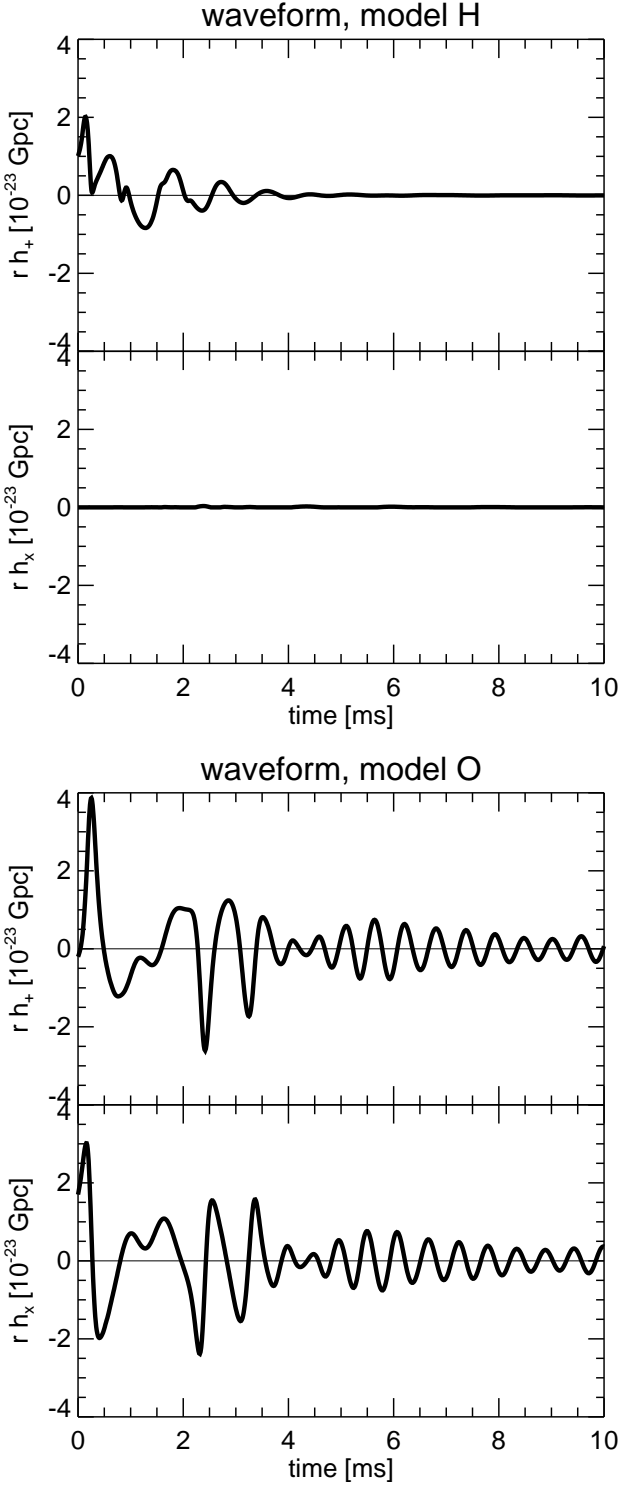}
\caption[]{Gravitational waveforms, $h_+$ and $h_\times$,
for Models~H and~O, respectively, as functions of time.
Note that the gravitational-wave field $h_{\times}$ with
the cross polarization is practically zero in case of 
the head-on collision, Model~H,
because there are only very small deviations from the mirror symmetry
relative to the $y$-$z$ plane when the hot contact layer of the
two neutron stars becomes unstable against shear motions
(compare Figs.~\ref{fig:Hcont1} and \ref{fig:Hcont2}).
}
\label{fig:waveform}
\end{figure}

The gravitational-wave luminosities and the cumulative energy loss
in gravitational waves as functions of time for Models~H and h 
and Models~O and o are shown in Figs.~\ref{fig:lggrergH} and
\ref{fig:lggrergO}, respectively, and the corresponding gravitational
waveforms $h_+$ and $h_\times$ for Models~H and O are plotted in 
Fig.~\ref{fig:waveform}. 

In Model~H the most prominent luminosity spike is created at the moment 
when the two neutron stars crash into each other and the gas flow
is abruptly decelerated and redirected by the recoil shocks 
($t \approx 0.22\,{\rm ms}$, cf.~Fig.~\ref{fig:Hcont1}) which
leads to a rapid change of the mass quadrupole moment.
The peak luminosity reaches about $3.7\times 10^{55}\,$erg/s
(see also Fig.~\ref{fig:grcompH}). A precursor with about 1/4 of the
maximum luminosity is caused by the increasing tidal deformation of the
neutron stars as they approach each other. After this initial outburst the
gravitational-wave luminosity continues to oscillate regularly with
a period between two maxima of roughly half a millisecond but with
peaks at least one order of magnitude below the maximum luminosity.
This indicates that the bulk of the matter quickly adopts a more
or less spherical distribution. Within about 8$\,$ms the luminosity
falls by more than 5 orders of magnitude to less than 
$10^{50}\,$erg/s. This dramatic drop is 
reflected in the gravitational waveforms of Model~H which indicate 
that after $\sim 4\,{\rm ms}$ the activity has essentially ceased.
This coincides with the complete dissipation of the kinetic energy
at that time (see Fig.~\ref{fig:energyH}). More than 50\% of the 
total energy emitted in gravitational waves are contained in the 
luminosity spike and after about $3\,{\rm ms}$ only insignificant
further contributions are added (cf.~cumulative energy loss in 
Fig.~\ref{fig:lggrergH}). From Figs.~\ref{fig:lggrergH} and
\ref{fig:grcompH} one learns that the coarser resolution of Model~h
leads only to minor differences compared to Models~H and ${\cal H}$
with the tendency to overestimate the gravitational-wave luminosity
and emitted energy. Models~H and ${\cal H}$ are hardly distinguishable 
in Fig.~\ref{fig:grcompH} and the computations seem to be converged.

Model~O is an approximately ten times more energetic source of 
gravitational waves than Model~H and emits a total energy of
$2\times 10^{-2}\,M_{\odot}c^2$. On the one hand the first 
luminosity maximum at $t\approx 0.22\,{\rm ms}$ is nearly twice as high 
($\sim 6\times 10^{55}\,$erg/s) as in Model~H and 
nearly four times longer (half width about $0.2\,{\rm ms}$ 
compared to $0.05\,{\rm ms}$) (Fig.~\ref{fig:grcompO})
On the other hand Model~O continues
its strong emission of gravitational waves for the whole computation
period of 10$\,$ms during which the luminosity on average stays 
around $10^{54}\,$erg/s and hardly ever drops below 
$10^{53}\,$erg/s. This can be explained by the rapid change of the quadrupole 
moment of the system as the two neutron stars repeatedly come close
and separate again on their quasi-elliptic orbits around each
other (cf.~Fig.~\ref{fig:O2orbit}) and by the large kinetic energy retained
as rotational energy of the collision remnant at the end of the
simulation (Fig.~\ref{fig:energyO}). Several peaks of the gravitational-wave
luminosity can be correlated with the moments of closest approach of
the density maxima of the two neutron stars (compare Figs.~\ref{fig:lggrergO}
and \ref{fig:sepden}), and two strong, short minima of the luminosity (at 
$t\approx 1.2\,{\rm ms}$ and $t\approx 4.1\,{\rm ms}$) coincide with 
instants of maximal separation. In Model~O only about 30\% of the total
energy emitted in gravitational waves is contained in the first luminosity
spike, another $\sim 50\%$ are added during the second and third 
periastrons ($t\approx 2.4\,{\rm ms}$ and $t\approx 3.3\,{\rm ms}$),
and a non-negligible fraction ($\sim 20\%$) comes at later times. 
The waveforms of Model~O (Fig.~\ref{fig:waveform}) exhibit a rather
irregular structure during the first 4--5$\,$ms of the evolution with
a peak amplitude at $t \approx 0.22\,{\rm ms}$ which is about twice as high
as in Model~H. They continue with a very regular, slowly damped sinusoidal 
modulation until the end of the simulation at $t \approx 10\,{\rm ms}$.

Finally, in Figs.~\ref{fig:grcompH} and~\ref{fig:grcompO} we compare 
the first large spike of the gravitational-wave luminosity with the
corresponding results obtained by Centrella \& McMillan (1993) and
Rasio \& Shapiro (1992). In order to do that we rescale and renormalize 
their dimensionless quantities to physical units for our chosen neutron
star parameters by using the dynamical timescale
\begin{equation}
  t_{\rm D}\equiv \left(\frac{R^3}{GM}\right)^{1/2}\approx 0.125~{\rm ms}
\end{equation}
with $M=1.63\,M_\odot$ as the mass and $R=15\,$km as the radius
of the neutron stars, and by the scaling the luminosity with the factor
\begin{equation}
 L_0 \equiv \frac{1}{G} \left( \frac{GM}{cR} \right)^5  
     \approx  3.86\cdot 10^{55}~{\rm erg/s} \, .
\end{equation}
Moreover, there is a time shift between their calculations and ours
because of the different initial center-to-center distances of the
neutron stars. While it was $a_0 = 42\,$km in our models, 
Rasio \& Shapiro (1992) and Centrella \& McMillan (1993) assumed 
$a_1=4R=60\,$km for their head-on collisions, and Centrella \& McMillan 
(1993) took $a_1\approx 4.53R\approx 68\,$km for the off-center case.
The time lag for the parabolic infall trajectories of the two masses
with different initial separations can be determined as the 
difference of the times to reach the minimum distance (periastron),
to be (Roy, 1982, Eqs.~(4.82) and (4.85))
\begin{equation}
\Delta t = \frac{\sqrt{2}(A_1-A_0)}{3\sqrt{GM_{\rm t}}}
\end{equation}
with $A_i \equiv (a_i+2R_{\rm p})\sqrt{a_i-R_{\rm p}}$ ($i = 0,\,1$).
$M_{\rm t}=M_1+M_2=2M$ is the total mass of the system and
the periastron distance $R_{\rm p}$ is the separation of two point
particles of mass $M$ at closest approach, which is set to zero for 
the head-on collision and to $R_{\rm p}=1R$ for the off-center case.

Taking into account these aspects in 
Figs.~\ref{fig:grcompH} and~\ref{fig:grcompO}, we find good overall
agreement between the gravitational-wave luminosities from the 
different calculations, both in shape and magnitude. The remaining
minor discrepancies can be attributed to the use of the Lattimer \&
Swesty (1991) EOS in our calculations instead of an adiabatic EOS with
constant index $\gamma = 2$, the possible influence from the inclusion of 
gravitational-wave back-reactions in our models, and the effects
resulting from different numerical schemes and different resolution.

\begin{figure*}
\tabcolsep=2.0mm
 \begin{tabular}{cc}
  \epsfxsize=8.8cm \epsfclipon \epsffile{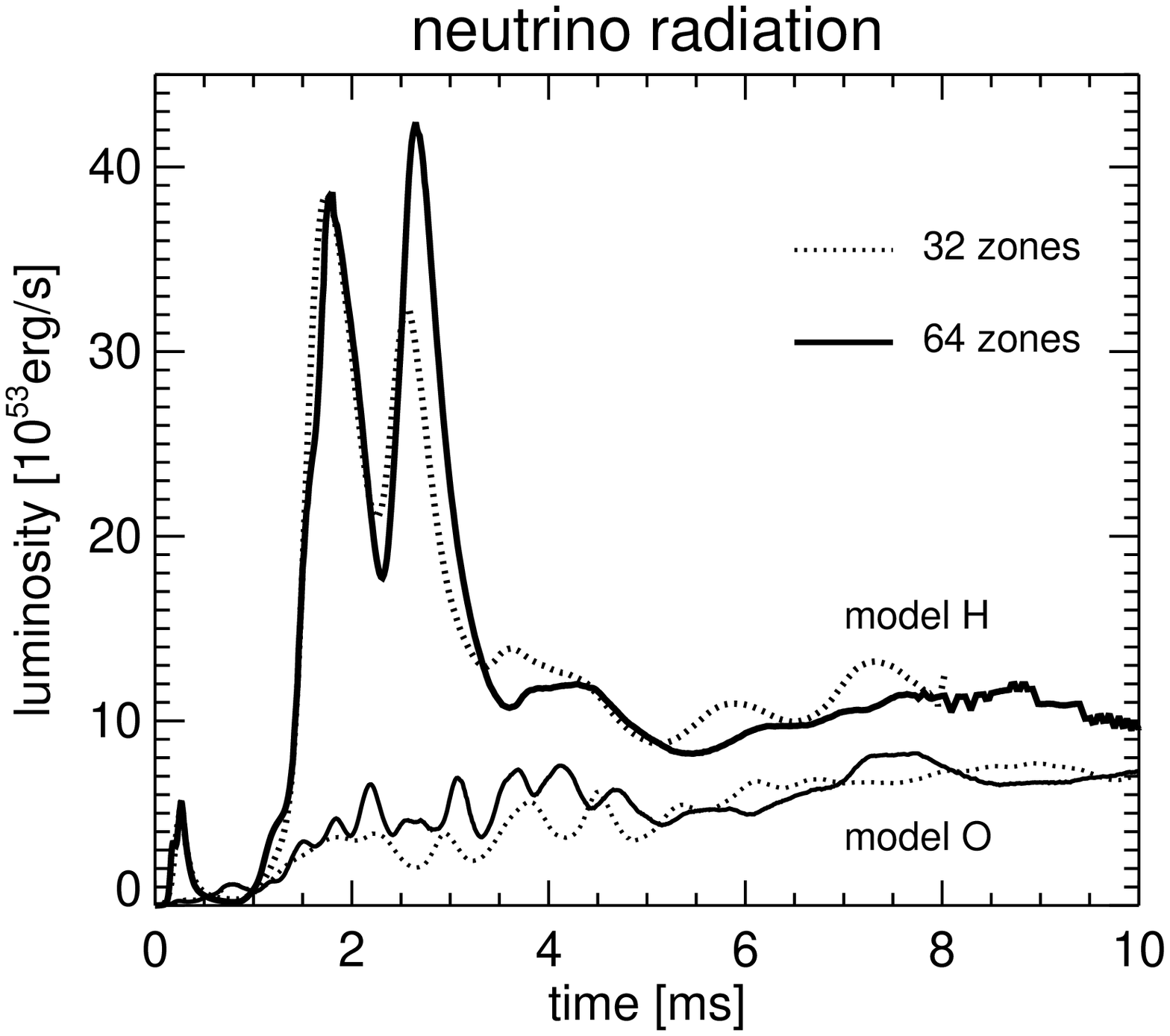} &
  \epsfxsize=8.8cm \epsfclipon \epsffile{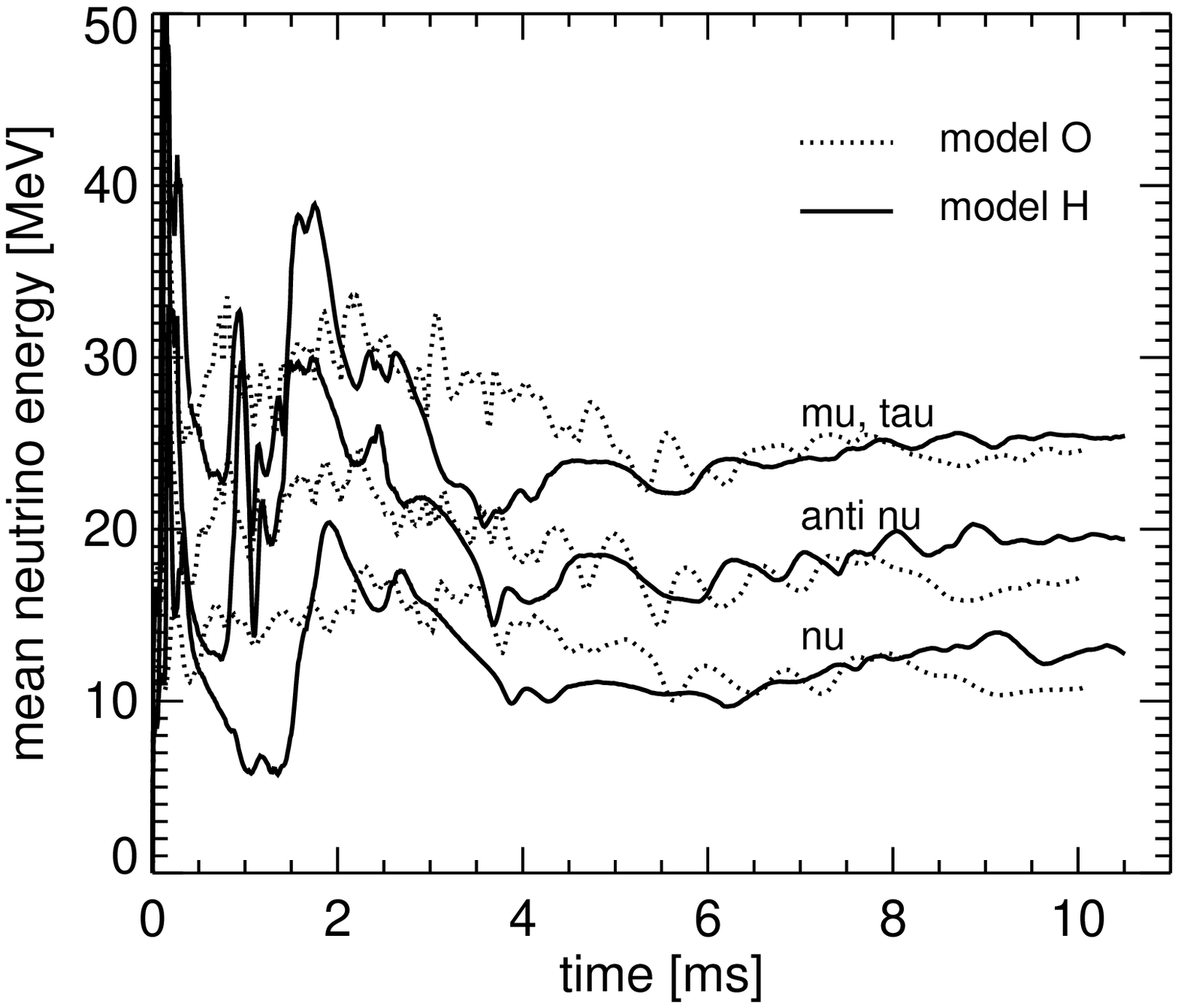} \\
  \parbox[t]{8.7cm}{\caption[]{Total neutrino luminosities (sums of 
         the contributions from all neutrino and antineutrino flavors) as
	 functions of time for Models~H and h and Models~O and o.}
         \label{fig:neuterg}} &
  \parbox[t]{8.7cm}{\caption[]{Average energies of emitted neutrinos 
	 $\nu_e$, $\bar\nu_e$, and $\nu_x$ ($\equiv \nu_{\mu}$, 
         $\bar\nu_{\mu}$, $\nu_{\tau}$, $\bar\nu_{\tau}$) 
         as functions of time for Models~H and O.}\label{fig:mene}} \\[15ex]
  \epsfxsize=8.8cm \epsfclipon \epsffile{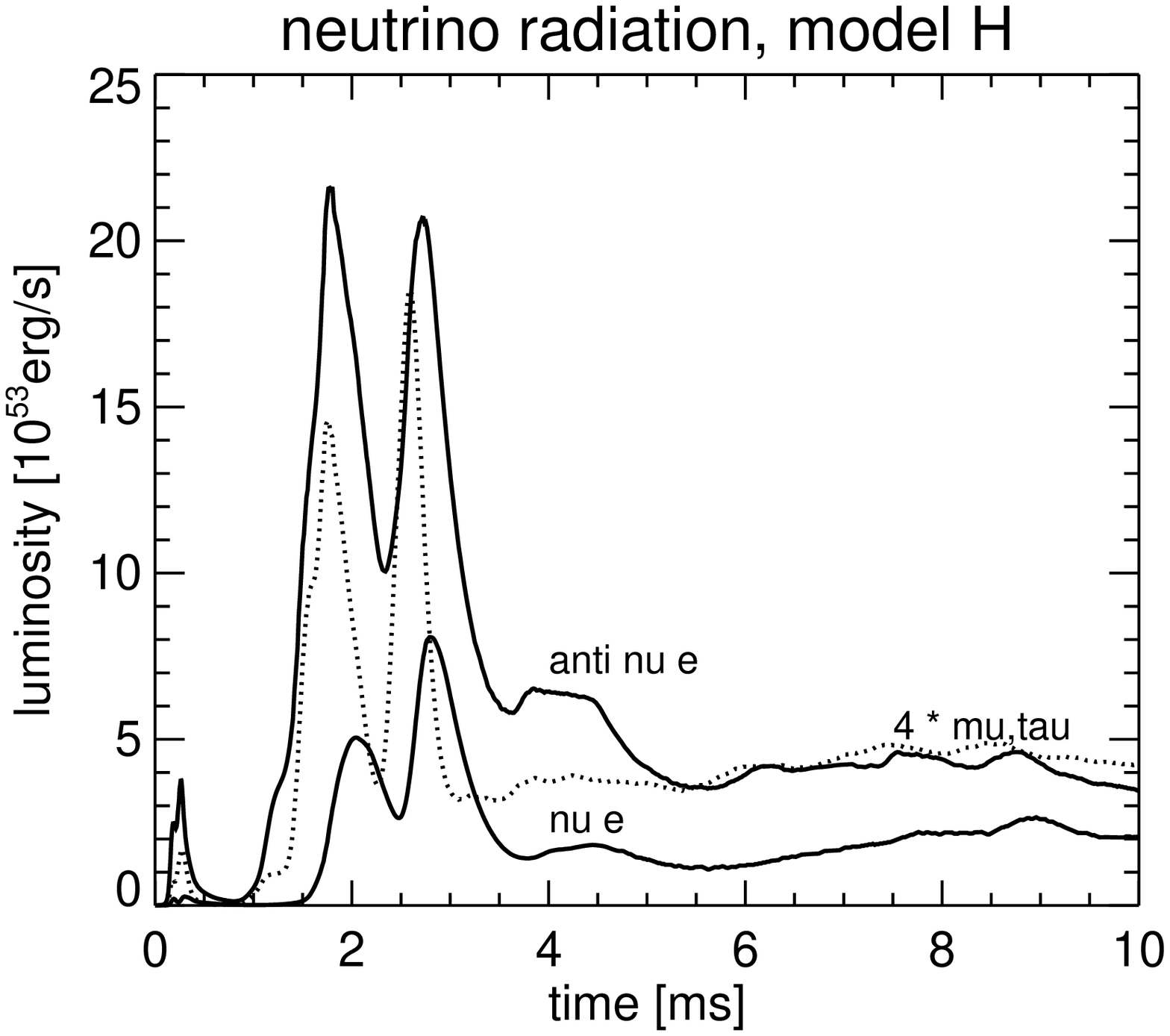} &
  \epsfxsize=8.8cm \epsfclipon \epsffile{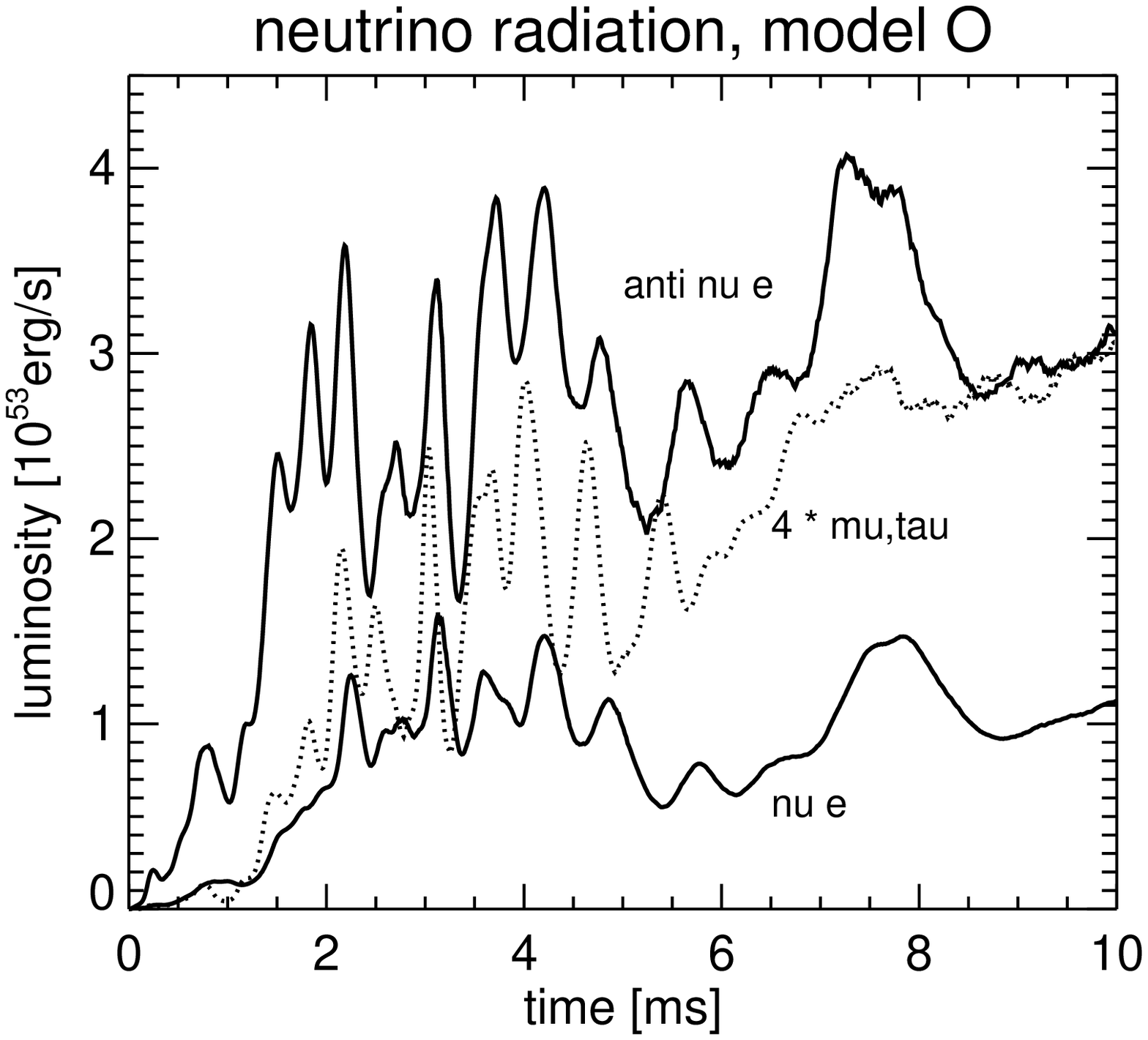} \\ 
  \parbox[t]{8.7cm}{\caption[]{Luminosities of the individual neutrino and
	 antineutrino flavors ($\nu_e$, $\bar\nu_e$, and the sum of all 
	 $\nu_x$) as functions of time for Model~H.}\label{fig:neutradH}} &
  \parbox[t]{8.7cm}{\caption[]{Same as Fig.~\ref{fig:neutradH} but for 
	 Model~O.}\label{fig:neutradO}}\\
 \end{tabular}
\end{figure*}

\section{Neutrino emission\label{sec:nuem}}

Because of the different dynamical evolution, the head-on and 
off-center collisions also show distinctive differences in the
neutrino emission. The total neutrino luminosities
for Models~H and h and Models~O and o as functions of time are
plotted in Fig.~\ref{fig:neuterg}. 

\begin{figure}
 \epsfxsize=8.8cm \epsfclipon \epsffile{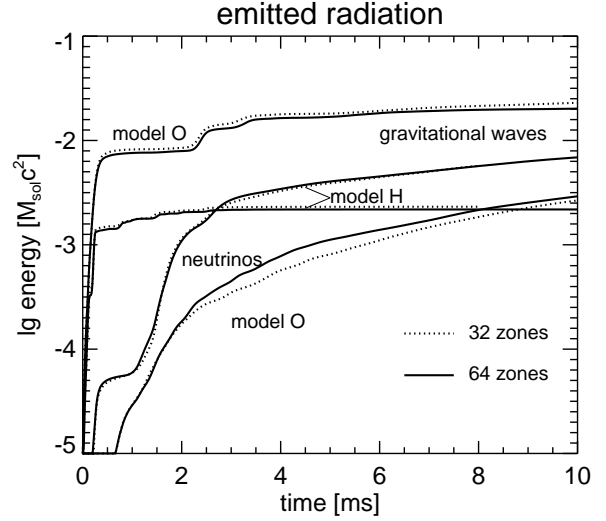}
 \caption[]{Comparison of the cumulative energies emitted in neutrinos
  and gravitational waves (in units of $M_{\odot}c^2$) as
  functions of time for the four Models~H, h, O, and o.}
 \label{fig:neugra}
\end{figure}

One can see that in the head-on collision (Models~H and h)
a first very luminous burst of neutrinos with a peak flux
of more than $5\times 10^{53}\,$erg/s is produced at the moment
when the two stars crash into each other and the neutron
star matter is shock heated and hot gas is squeezed out perpendicular
to the collision axis (see Fig.~\ref{fig:Hcont1}a and c). A second and
third luminosity maximum, however, with about 8 times larger 
peak fluxes of more than $4\times 10^{54}\,$erg/s are present in the time
interval $1.5\,{\rm ms}\la t \la 3.5\,{\rm ms}$ which is the time of
maximum temperatures in the collision remnant (Fig.~\ref{fig:maxtemp})
when the hot matter starts to expand and
to spread out over a larger volume 
(Figs.~\ref{fig:Hcont1}b and \ref{fig:Hcont1}d). A luminosity of
$L_{\nu} = 4\times 10^{54}\,{\rm erg/s}\approx
4\pi R_{\nu}^2 {c \over 4}(3\cdot {7\over 8}a_{\rm rad}T_{\nu}^4)$
corresponds to a neutrinosphere with radius $R_{\nu}\approx 50\,$km
which radiates neutrinos and antineutrinos of all flavors as black 
body with temperature $T_{\nu}\approx 8.5\,$MeV. 
Figures~\ref{fig:Hcont1}b and d confirm that the massive, nearly 
spherical central body of the collision remnant, which is embedded
in a cloud of less dense and cooler gas, has a radial size
and ``surface'' temperature in this estimated range. The duration 
of this extremely luminous burst is only about 2$\,$ms, after which
the total luminosity settles down to a much lower but still sizable
value around $10^{54}\,$erg/s. During the $\sim 10\,$ms of 
simulation time, Models~H and h emit an energy of 
$6.7\times 10^{-3}\,M_{\odot}c^2$ or $1.2\cdot 10^{52}\,$erg
in neutrinos, of
which 50\% are radiated during the double peak of the 
luminosity. Model~H is a stronger source of neutrinos than of
gravitational waves; the latter carry away only about
$2.2\times 10^{-3}\,M_{\odot}c^2$ (Fig.~\ref{fig:neugra}).
Nevertheless, despite of the extremely high luminosity
the total energy radiated in neutrinos by Model~H is still one order
of magnitude below the estimates by Katz \& Canel (1996).

\begin{figure*}
\tabcolsep=2.0mm
\begin{tabular}{cc}
  \epsfxsize=8.8cm \epsfclipon \epsffile{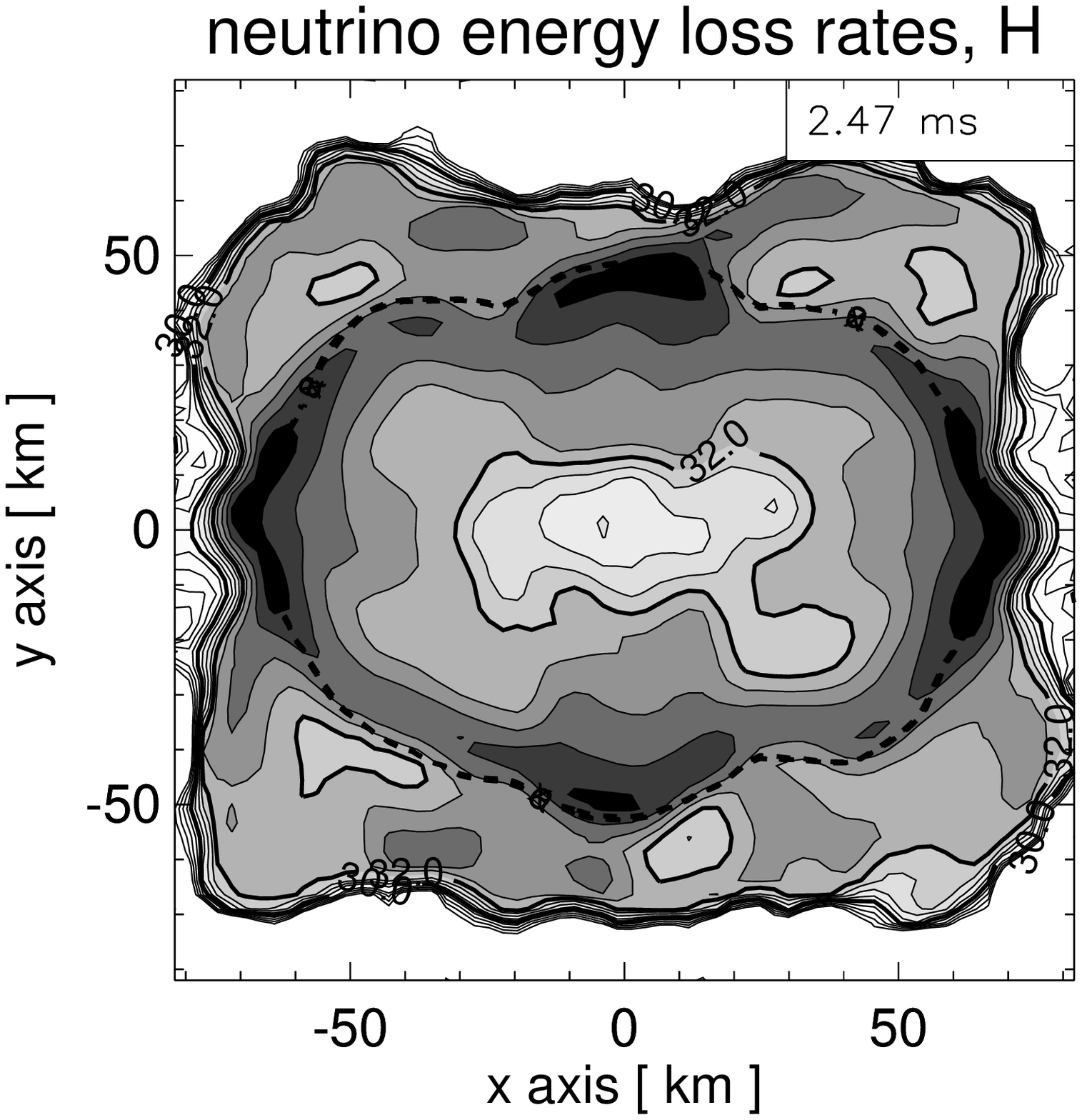} &
  \epsfxsize=8.8cm \epsfclipon \epsffile{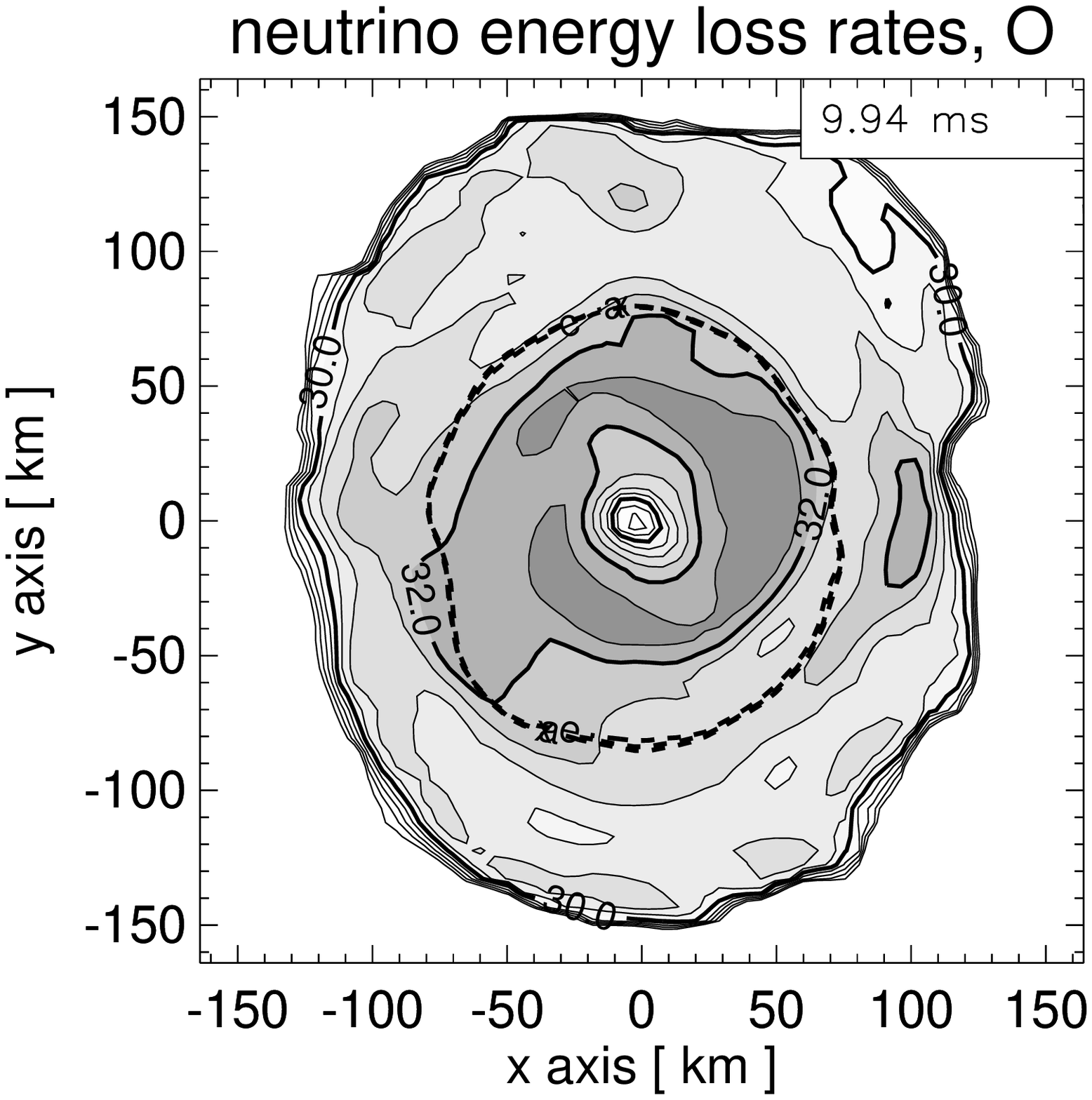} \\[-4ex]
\end{tabular}
\caption[]{Contour plots of the local energy loss
	 rates due to the emission of neutrinos and antineutrinos of all
	 flavors. The left plot shows a cut in the $z = 0$ plane for Model~H
	 at time $t = 2.47\,$ms, the right plot displays a cut in the orbital
	 plane of Model~O at time $t = 9.94\,$ms. Note that the visualized
	 region is larger in the latter figure.
         The contours are spaced logarithmically in steps of 
	 0.5~dex, the bold lines are labeled with their corresponding
	 values in units erg$\,$cm$^{-3}\,$s$^{-1}$. The dashed lines
	 indicate the approximate positions of the neutrinospheres
	 (defined where the optical depths for $\nu_e$, $\bar\nu_e$,
	 and $\nu_x$, respectively, are approximately unity).}
\label{fig:emmap} 
\end{figure*}

During the break-out of the bounce shock at $t\approx 0.2\,$ms 
neutrinos with average energies in excess of $50\,$MeV are emitted
from Model~H (Fig.~\ref{fig:mene}). A second phase of very high 
mean energies coincides with the two luminosity spikes and
therefore the phase of highest temperatures in Model~H. As
is well known from type II supernovae (see, e.g., Janka 1993), 
neutron-rich, hot neutron
star matter is more opaque to $\nu_e$ than to $\bar\nu_e$
because of frequent captures of the $\nu_e$ on the abundant
free neutrons. Heavy-lepton neutrinos ($\nu_x \equiv \nu_{\mu}$,
$\bar\nu_{\mu}$, $\nu_{\tau}$, $\bar\nu_{\tau}$) are even
less strongly coupled to the stellar medium since their
opacity is dominated by neutral-current neutrino-nucleon scatterings
but they do not interact with nucleons via charged-current reactions.
For these reasons $\bar\nu_e$ decouple energetically from the hot plasma
at higher densities and thus usually higher temperatures than 
$\nu_e$, and $\nu_x$ at even higher densities and temperatures.
This explains why typically the mean energy of the emitted $\nu_e$
is lower than that of $\bar\nu_e$ which in turn is below
the average energy of $\nu_x$ (see Fig.~\ref{fig:mene}). After
$t \ga 4\,$ms we obtain mean energies of 
$\ave{\epsilon_{\nu_e}}\approx 10$--13$\,$MeV,
$\ave{\epsilon_{\bar\nu_e}}\approx 15$--20$\,$MeV, and
$\ave{\epsilon_{\nu_x}}\approx 20$--25$\,$MeV which is in the range
of values found during the neutrino cooling phase of newly formed
neutron stars in type II supernovae. Despite of this generic ranking 
of the mean energies, the $\nu_e$ luminosity of the collision remnant
is larger than the luminosity in each individual type of $\nu_x$, and the
$\bar\nu_e$ luminosity of the neutron-rich, hot neutron star matter
dominates the $\nu_e$ luminosity (Fig.~\ref{fig:neutradH}).
The difference in ranking between neutrino luminosities and mean 
energies of emitted neutrinos reflects the fact that the
emission is not like an ideal black-body, but non-equilibrium 
effects play a role. Moreover, Fig.~\ref{fig:emmap} shows that
there is an extended region (with a broad range of temperatures and
densities) where the neutrino fluxes are fed by local
neutrino energy losses. Therefore the neutrino emission can 
hardly be characterized by the conditions of thermodynamical 
equilibrium at a well defined neutrino emitting surface
(``neutrinosphere'') associated with each type of neutrino
or antineutrino.

In the off-center collision, Models~O and o, the neutrino luminosity
reveals a steady increase and does not have such pronounced maxima as
seen in Model~H (Fig.~\ref{fig:neuterg}), although some fluctuations
up to a factor of 2 are present. At the end of our simulations 
($t \approx 10\,$ms), Model~O has a total neutrino luminosity of
about $7\times 10^{53}\,$erg/s which is only 30\% less than in Model~H.
Because of the lack of a phase of extremely high neutrino emission and
the strong production of gravitational waves during its whole 
evolution, Model~O loses only half the energy in neutrinos as Model~H
but is a much stronger gravitational-wave source (Fig.~\ref{fig:neugra}).
A comparison of the two plots in Fig.~\ref{fig:emmap} shows that
in Model~O the neutrino emitting gas cloud (at $t\approx 10\,$ms) is
more extended than in Model~H (at $t\approx 2.5\,$ms) but the peak
values of the local energy loss rate in neutrinos are only around
$3\times 10^{32}\,$erg$\,$cm$^{-3}\,$s$^{-1}$ in Model~O whereas they
are above $10^{34}\,$erg$\,$cm$^{-3}\,$s$^{-1}$ in case of Model~H.
The mean energies of the emitted neutrinos (Fig.~\ref{fig:mene})
are very similar in Models~O and H, and also the relative contributions
of the different neutrino types to the energy loss are roughly similar,
as can be seen from a comparison of the relative sizes of the individual
neutrino luminosities in Figs.~\ref{fig:neutradO} and \ref{fig:neutradH}.

\section{Neutrino-antineutrino annihilation\label{sec:annihi}}

The rate of energy deposition by neutrino-antineutrino annihilation 
increases, roughly, with the product of the local neutrino and 
antineutrino energy densities times the mean energy of these neutrinos
times a factor that accounts for the angular distribution of the 
neutrinos (the process is very sensitive to the angle at which
neutrinos and antineutrinos collide, for details, see RJST). In 
the spherically
symmetric situation this can be converted into a product of the 
neutrino and antineutrino luminosities times the sum of the mean 
neutrino and antineutrino energies times a normalized factor which
results from the phase space integration over the local neutrino 
distribution functions and which depends on the geometry of the
considered problem. This was used to arrive at the approximate
description summarized in Eqs.~(\ref{eq:doteann})--(\ref{eq:abbrfbar}) 
of Sect.~\ref{sec:evalann} employed here to evaluate Model~H for the 
energy deposition by $\nu\bar\nu$ annihilation in the surroundings 
of the collision remnant.

Neutrino-antineutrino annihilation is considered as a
mechanism to pump energy into a fireball consisting of $e^+$, $e^-$, 
photons, and a small number of baryons. This fireball was
suggested to be a possible source of
gamma-ray bursts from neutron star collisions at cosmological 
distances (see, e.g., Katz \& Canel 1996) if the energy in the
fireball is sufficiently large, $E_{\rm fb}\approx E_{\gamma}
\approx E_{\rm ann} \ga 10^{51}\delta\Omega/(4\pi)\,$erg, and if the
baryon loading of the fireball is sufficiently small, 
$M_{\rm fb}\la 10^{-5}\,M_{\odot}$ (for a canonical energy 
$E_{\rm fb}\sim 10^{51}\,$erg) 
so that the fireball can expand relativistically with a Lorentz
factor $\Gamma_{\rm fb}= E_{\rm fb}/(M_{\rm fb}c^2)\ga 100$.
Two questions arise from this suggestion. First, is the conversion
of neutrino-antineutrino energy into electron-positron pairs 
enough efficient to provide the desired energy, and, second, how
large is the baryon mass contained in the fireball created through
$\nu\bar\nu$ annihilation?

In the following we attempt to give answers to these two questions 
on grounds of our hydrodynamical collision models. Here we only report
on the evaluation of the head-on collision Model~H. We
concentrate on this model for two reasons. On the one hand, the
efficiency of $\nu\bar\nu$ annihilation increases strongly with
the neutrino luminosities and with the mean energies of the emitted 
neutrinos. Therefore Model~H with its larger neutrino emission 
appeared to us as the more interesting
one. On the other hand, our simulations have demonstrated that even
in the off-center collision the interaction of the two neutron stars
is so dramatic that a lot of matter is ejected perpendicularly to 
the orbital plane. Therefore, despite of the large angular momentum 
in the system, the axis region is polluted with baryons
(see Fig.~\ref{fig:O2densv}), and both the remnants of the head-on and 
off-center collisions adopt more or less spherical shapes after
the dynamic interactions, with a central massive object 
being surrounded by a less dense, extended cloud of hot baryonic 
gas (Figs.~\ref{fig:Hcont1} and \ref{fig:Ocont1}).
From this point of view Model~O did not seem to offer better 
perspectives for the emergence of a relativistic fireball from
a baryon-depleted region near the collided neutron stars.

\begin{figure}
 \epsfxsize=8.8cm \epsfclipon \epsffile{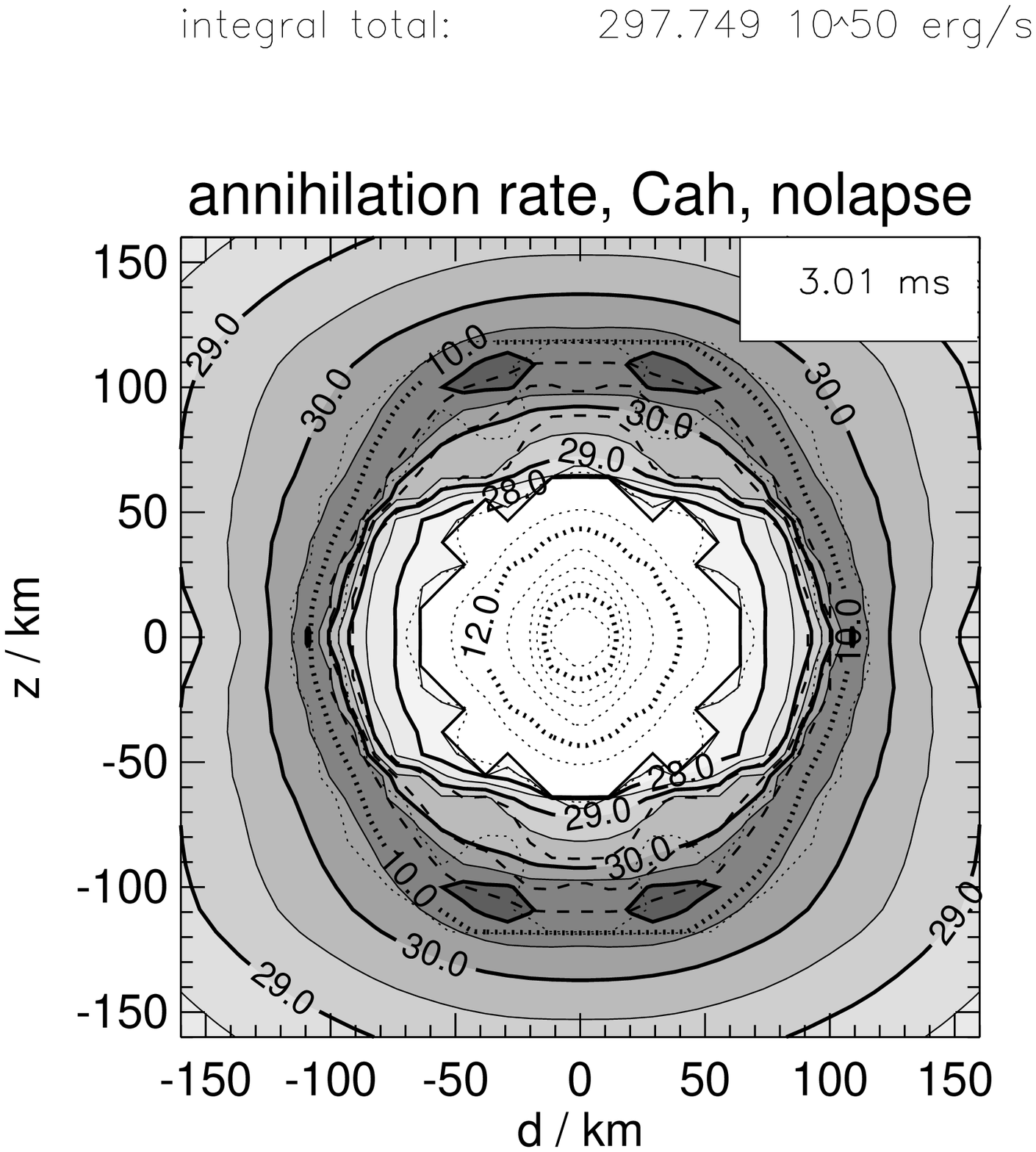}
\caption[]{Map of the local energy deposition rates 
(in~erg$\,$cm$^{-3}\,$s$^{-1}$) by $\nu\bar\nu$ annihilation into
$e^+e^-$ pairs in the vicinity of the merger for Model~H at
time $t = 3$~ms after the start of the simulation. 
The values are obtained as averages over the azimuthal angle
around the $z$-axis, $d$ measures the distance from the grid
center in the $x$-$y$-plane. The corresponding solid contour lines 
are logarithmically spaced in steps of 0.5 dex, the grey shading
emphasizes the levels with dark grey meaning high energy
deposition rate. The dashed lines mark the (approximate) positions 
of the neutrinospheres of $\nu_e$, $\bar\nu_e$, and $\nu_x$
(from outside inward), defined by the requirement that the
optical depths in $z$-direction are $\tau_{z,\nu_i} = 1$.
The dotted contours indicate levels of the
azimuthally averaged density, also logarithmically spaced with
intervals of 0.5 dex. The energy deposition rate was
evaluated only in that region around the merged object, where
the mass density is below $10^{11}$~g$\,$cm$^{-3}$. The integral
value of the energy deposition rate at the displayed time
is $3.4\times 10^{52}\,{\rm erg\,s}^{-1}$.
}
\label{fig:annmap}
\end{figure}
 
\begin{figure}
 \epsfxsize=8.8cm \epsfclipon \epsffile{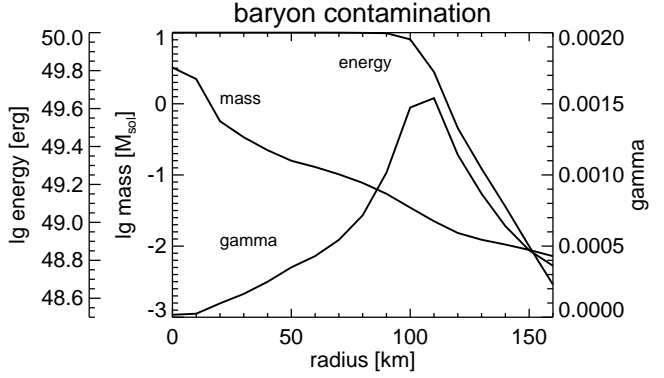}
\caption[]{Cumulative mass $M(r\ge R)$ and annihilation energy 
$E_{\nu\bar\nu}(r\ge R)$ outside of radius $R$, as functions of $R$
for Model~H at time~3$\,$ms (same time as in Fig.~\ref{fig:annmap}). 
The corresponding relativistic Lorentz factor
$\Gamma(R)\equiv E_{\nu\bar\nu}(r\ge R)/\eck{M(r\ge R)c^2}$ is also
plotted.
}
\label{fig:anngam}
\end{figure}

Figure~\ref{fig:annmap} gives a map of the energy deposition
rate by $\nu\bar\nu$ annihilation into $e^+e^-$ pairs
(averaged over the azimuthal angle around the $z$-axis according
to Eq.~(\ref{eq:doteann}))
in the surroundings of the collision remnant in
Model~H at time $t = 3\,$ms which is inside the double peak 
structure of the neutrino luminosity of Fig.~\ref{fig:neuterg}.
One can see that the highest energy deposition rates of the
order of 
$3\times 10^{30}\,{\rm erg\,cm^{-3}\,s^{-1}}$ to $10^{31}\,{\rm 
erg\,cm^{-3}\,s^{-1}}$ occur immediately outside the 
neutrinospheres but in layers with densities still above and
around $10^{10}\,$g$\,$cm$^{-3}$. At the displayed time, the 
integral energy deposition rate in matter with density below 
$10^{11}\,$g$\,$cm$^{-3}$ (evaluated according to 
Eq.~(\ref{eq:Lann})) is $3.4\times 10^{52}\,{\rm erg\,s}^{-1}$.

Since the neutrino luminosities and mean energies of the 
emitted neutrinos show significant variation during the computation
time (see Figs.~\ref{fig:neuterg}, \ref{fig:mene}, and
\ref{fig:neutradH}), we compute the time integral of the 
energy deposition rate, Eq.~(\ref{eq:Eexact}), by employing the
approximate treatment summarized in 
Eqs.~(\ref{eq:Eapprox})--(\ref{eq:abbrfbar}). The phase 
between $t\approx 1.5\,{\rm ms}$ and $t\approx 3.5\,{\rm ms}$ has 
by far the highest neutrino luminosity and therefore yields the
largest contribution to the time integral. For this reason we calculate
the temporal average of the term $L_{\rm ann}(t)/{\cal F}^\ast(t)$
in Eq.~(\ref{eq:Eapprox}) by summing over $N =3$ time points in 
this interval: $t_1 = 2.47\,$ms, $t_2 = 3.01\,$ms, and 
$t_3 = 3.59\,$ms. We obtain
\begin{equation}
\frac{1}{N} \sum_{i=1}^{N}
\left( \frac{L_{\rm ann}(t_i)}{{\cal F}^{\ast}(t_i)} \right)
\approx 2\times 10^{-57}\,{\rm MeV^{-1}\,erg^{-1}\,s} \,.
\label{eq:part1}
\end{equation}
Because the three terms of the sum are different by less than
a factor 2, we think that the splitting of the time integral
of Eq.~(\ref{eq:Eexact}) which led to the approximate form
of Eq.~(\ref{eq:Eapprox}) was justified. 
Taking the data for the individual neutrino luminosities 
$L_{\nu_e}(t)$, $L_{\bar{\nu}_e}(t)$, and $L_{\nu_x}(t)$
(Fig.~\ref{fig:neutradH}) and the average neutrino energies
$\langle\epsilon_{\nu_e}(t)\rangle$, 
$\langle\epsilon_{\bar{\nu}_e}(t)\rangle$, and
$\langle\epsilon_{\nu_x}(t)\rangle$ (Fig.~\ref{fig:mene}),
we further find
\begin{equation}
 \int_0^{10\,{\rm ms}} {\cal F}^{\ast}(t)\,{\rm d}t \approx
 5\times 10^{106}\,{\rm MeV\,erg^2\,s^{-1}}\,.
 \label{eq:part2}
\end{equation}
Multiplying the results of Eqs.~(\ref{eq:part1}) and (\ref{eq:part2})
we end up with a total energy deposition of
$E_{\rm ann} \approx 10^{50}\,{\rm erg}$
within our simulation interval of 10$\,$ms for Model~H 
(see also Fig.~\ref{fig:anngam}). The 
corresponding average energy deposition rate by $\nu\bar\nu$ annihilation
of $\sim 10^{52}\,{\rm erg\,s}^{-1}$ is very large and means 
a conversion efficiency of $\nu\bar\nu$ energy to $e^+e^-$ pairs
of the order of 1\%. Most of this energy is liberated during the
2$\,$ms interval between $t \approx 1.5\,$ms and $t \approx 3.5\,$ms
after the start of the simulation because of the enormous neutrino
luminosity shortly after the collision of the neutron stars.

The energy of approximately $10^{50}\,$erg in $e^+e^-$ pairs and photons
is released nearly isotropically (Fig.~\ref{fig:annmap}) and is only
one order of magnitude below the canonical fireball energy 
$E_{\rm fb}\approx E_{\gamma}\sim 10^{51}\delta\Omega/(4\pi)\,$erg.
It may therefore be sufficient to account for the shorter and weaker
bursts whose energy is estimated to be typically more than a factor 
of 10 below the mean energy of the longer and more powerful bursts
(Mao et al.~1994).
However, most of the $\nu\bar\nu$ energy deposition happens very 
close to the neutrinospheres and thus in a high-density region
(Fig.~\ref{fig:annmap}). Because of this, the baryon loading
of the $e^+e^-$-photon fireball is a serious problem. This is obvious
from Fig.~\ref{fig:anngam} where we give the energy from 
$\nu\bar\nu$-annihilation and the baryonic mass, integrated from 
outside inward to the radial position $R$ given on the abscissa.
In the region where $10^{50}\,$erg are deposited, one has a baryon
mass of $M\ga 5\times 10^{-2}\,M_{\odot}$ which is about
5 orders of magnitude too large to allow for highly relativistic
expansion. The Lorentz factors which can be estimated as
$\Gamma(R)\equiv E_{\nu\bar\nu}(r\ge R)/\eck{M(r\ge R)c^2}$ are 
therefore around $10^{-3}$ instead of 100. For this reason, the 
huge and nearly isotropic baryon pollution of the $\nu\bar\nu$ 
energy deposition region seems to rule out the
possibility that neutrinos from colliding neutron stars produce 
gamma-ray bursts.

\begin{figure}
 \epsfxsize=8.8cm \epsfclipon \epsffile{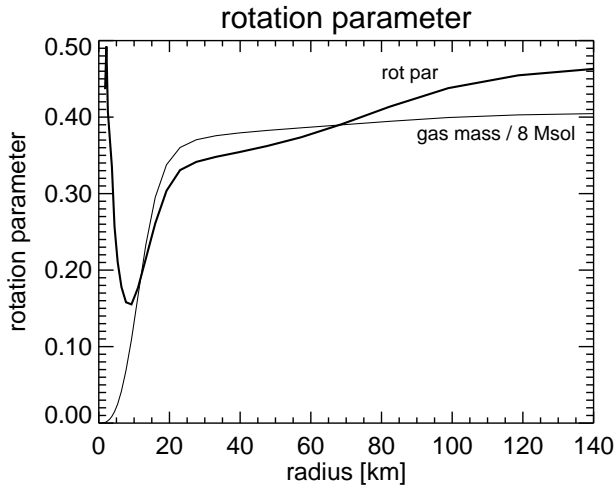}
\caption[]{The thick solid line gives the relativistic rotation parameter 
$a(r) \equiv J(r)c/(GM_{\rm gas}^2(r))$ for Model~O as function 
of radius $r$. $J(r)$ is the angular momentum (in $z$-direction)
inside $r$, $M_{\rm gas}(r)$ is the gas mass inside $r$ which is
plotted as thin solid line, given in units of $M_{\odot}$ and scaled 
down by a factor 8.
}
\label{fig:rotparO}
\end{figure}

\section{Summary and discussion\label{sec:end}}

We have reported results of three-dimensional Newtonian hydrodynamical 
simulations of the collision along parabolic orbits of two identical, 
non-rotating neutron stars with a baryonic mass of about $1.6\,M_{\odot}$.
The simulations were done with a Eulerian PPM code,
employing nested grids, using a physical nuclear EOS, and 
taking into account gravitational-wave emission and its
back-reaction on the hydrodynamical flow as well as the emission
of neutrinos from the heated neutron star matter. We have studied
two different cases, each with varied resolution on the finest grid, a 
head-on collision and an off-center collision with a periastron
distance of one stellar radius ($\approx 15\,$km). Our simulations
allow us to draw the following main conclusions:

\noindent
1.) {\it Dynamical evolution:} During the head-on collision of the 
two neutron stars, the collision remnant is heated by recoil shocks
and the initial kinetic energy is efficiently shock-dissipated to 
thermal energy within a few violent pulsations on a timescale of
about 3--4$\,$ms, after which the collision remnant has an
essentially spherical mass distribution. The neutron star
matter is transiently heated to peak temperatures close to $100\,$MeV.
Average temperatures are around 40--50$\,$MeV, corresponding to 
entropies between 2$\,k$/nucleon and 10$\,k$/nucleon. The violent 
crash of the stars into each other leads to the dynamical ejection
of about 0.5\% of the system mass. In contrast, the off-center
collision is much ``milder'' and the ejected mass is about a 
factor of 10 smaller, although the collision remnant retains 
a large fraction (about 30\%) of the initial kinetic energy as
rotational energy even after 10$\,$ms at the end of our simulations.
Only for extreme assumptions about the nuclear EOS will the 
$\ga 3\,M_{\odot}$ remnant of the head-on collision not form
a black hole on a dynamical timescale. We also believe that the 
remnant of the off-center collision will probably not be able to escape
the collapse to a black hole within a few milliseconds. A mass of
$3\,M_{\odot}$ is distributed within a radius of 30$\,$km and
$2.8\,M_{\odot}$ are within 20$\,$km (Fig.~\ref{fig:rotparO})
which is only about twice the Schwarzschild radius of a 
$3\,M_{\odot}$ black hole. Thermal pressure can only
insignificantly raise the maximum stable mass of neutron stars
(Goussard et al.~1997) and rotation is able to increase the 
stable mass limit by only $\la 20\%$ (Friedman et al.~1986).
The interior $\sim 2.8\,M_{\odot}$ of the remnant rotate
nearly uniformly and the relativistic rotation parameter of 
this mass is only $a(r=20\,{\rm km})=\eck{Jc/(GM^2)}_{20\,{\rm km}}
\approx 0.3$ (cf.~Fig.~\ref{fig:rotparO}), which is smaller than 
that of the maximally rotating,
maximum-mass models constructed by Friedman et al.~(1986).
For all but one extreme EOS tested by Friedman et al.~(1986) such
a configuration is unstable. Therefore we conclude that it is very
likely that gravitational instability will set in as soon as the two 
stars have merged into one massive body within $t\ga 3$--$4\,$ms. 
General relativistic effects will certainly influence the 
transformation of the infall orbit of the neutron stars into 
a bound one; this needs to be studied with relativistic simulations.

\noindent
2.) {\it Gravitational waves and neutrinos:} The gravitational-wave
signal will certainly depend on general relativistic effects and
our basically Newtonian models have only a limited ability to make 
predictions of a possibly measurable pulse. Moreover, the
duration of the gravitational-wave and neutrino emission from the 
hot collision remnant will depend on the timescale of the delay until
black hole formation. Our simulations yield a maximum 
amplitude of the gravitational waveform that is 
$h_{\rm max}\approx 2$--$4\times 10^{-23}$ for neutron star collisions
happening at a distance of $1\,$Gpc. The off-center collision is the
stronger gravitational-wave source due to the larger quadrupole
moment of the rotating
system and the longer duration of the emission which
might last for 10--20 wave periods in the 1000--2000$\,$Hz range.
The gravitational-wave strain
would be close to the lower sensitivity limit of the new
generation of gravitational-wave interferometers which are currently 
under construction and will start operation within the next years. Of
course, neutron star collisions are very rare and very short 
events and therefore the chance to catch a signal is rather small.
The head-on collision is the more powerful neutrino source of the 
two investigated cases and
emits an energy about 3 times larger in neutrinos
than in gravitational waves. The peak neutrino luminosity reaches
$4\times 10^{54}\,{\rm erg\,s^{-1}}$ and the total energy
radiated in neutrinos within a few milliseconds is around 
$10^{52}\,$erg.

\noindent
3.) {\it Gamma-ray bursts:} Because of the larger energy output in
neutrinos and the very high neutrino luminosity 
($L_{\nu} \ga 10^{54}\,{\rm erg\,s}^{-1}$) as well as high 
mean energies of the
emitted neutrinos ($\ave{\epsilon_{\nu}}\la 40\,$MeV), the head-on 
collision provides more favorable conditions for producing
gamma-ray bursts from
$e^+e^-$-photon fireballs created by $\nu\bar\nu$ annihilation. 
We calculate a conversion efficiency of neutrino energy into 
$e^+e^-$-pairs of about 1\% and find an integral value for the 
energy deposited in the vicinity of the collision remnant
of $10^{50}\,$erg within only 10$\,$ms.
However, most of this energy is deposited in the immediate 
neighbourhood of the neutrinospheres where the density is still
higher than about $10^{10}\,{\rm g\,cm}^{-3}$. Therefore the 
baryon loading of the $e^+e^-$-photon fireball is at least 
5 orders of magnitude too high and instead of having values 
above 100 the relativistic Lorentz factor is estimated to be only
around $10^{-3}$. Dynamically ejected material together with a flow
of baryonic matter driven by neutrino-energy transfer to the
surface layers of the collision remnant are
therefore a harmful poisonous combination which prevents relativistic 
expansion of the pair-plasma fireball even though
the latter seems to obtain an interesting amount of energy from
$\nu\bar\nu$-annihilation. 

Strong shock heating of the neutron stars
during their violent collision (or during their merging as suggested
by post-Newtonian simulations, see Oohara \& Nakamura 1997) can indeed
raise the neutrino luminosities significantly and can thus enhance
the energy deposition by $\nu\bar\nu$ annihilation. The associated
dynamical ejection of gas and the neutrino-driven wind caused by 
the intense neutrino fluxes, however, impede the emergence of 
gamma-rays from this scenario. Fireballs powered by neutrinos from
colliding neutron stars should therefore be ruled out
as possible sources of cosmological gamma-ray bursts.
Also, we do not see the formation of highly relativistic shocks 
during the collisions
by which a fraction of 0.1--1\% of the kinetic energy at impact
($\ga 10^{53}\,$erg) might be hydrodynamically focused into a small
amount of mass ($\sim 10^{-6}$ to $10^{-5}\,M_{\odot}$). The
expansion velocity of the collision shocks is at most a few
tenths of the speed of light and the masses ejected in the
presented models are more than 3 orders of magnitude too large
to allow for ultrarelativistic motion.


\begin{acknowledgements}
M.R.~is grateful to Sabine Schindler for her
patience in their common office.
We would like to thank M.~Camenzind, W.~Hillebrandt,
F.~Meyer, M.~Rees, and S. Woos\-ley for educative discussions.
MR acknowledges support by the PPARC, HTJ by the 
``Sonderforschungsbereich 375-95 f\"ur Astro-Teilchenphysik''
der Deutschen Forschungsgemeinschaft.
The calculations were performed at the Rechenzentrum Garching on a
Cray-YMP 4/64.
\end{acknowledgements}

\end{document}